\pdfoutput=1

\documentclass[11pt,twoside,a4paper,cmspaper,final,collab]{cms-tdr}
\def\svnVersion{227782}\def\cmsCernNo{2013-188}\def\cmsCernDate{\today}\def\cmsMessage{Published in Physical Review D as \href{http://dx.doi.org/10.1103/PhysRevD.89.012003}{\doi{10.1103/PhysRevD.89.012003.}}}

\begin{document}\cmsNoteHeader{HIG-13-012}

\hyphenation{had-ron-i-za-tion}
\hyphenation{cal-or-i-me-ter}
\hyphenation{de-vices}

\newcommand\HBB{\ensuremath{\PH\to\bbbar}\xspace}
\newcommand\mH{\ensuremath{m_\PH}\xspace}
\newcommand{\Vvar}{\ensuremath{\cmsSymbolFace{V}}\xspace}
\newcommand\VH{\ensuremath{{\Vvar}\PH}\xspace}
\newcommand\WH{\ensuremath{\PW\PH}\xspace}
\newcommand\WZ{\ensuremath{\PW\cPZ}\xspace}
\newcommand\ZH{\ensuremath{\cPZ\PH}\xspace}
\newcommand\WmnH{\ensuremath{\PW(\Pgm\cPgn)\PH}\xspace}
\newcommand\WenH{\ensuremath{\PW(\Pe\cPgn)\PH}\xspace}
\newcommand\WlnH{\ensuremath{\PW(\ell\cPgn)\PH}\xspace}
\newcommand\WtnH{\ensuremath{\PW(\Pgt\cPgn)\PH}\xspace}
\newcommand\ZmmH{\ensuremath{\cPZ(\Pgm\Pgm)\PH}\xspace}
\newcommand\ZeeH{\ensuremath{\cPZ(\Pe\Pe)\PH}\xspace}
\newcommand\ZnnH{\ensuremath{\cPZ(\cPgn\cPgn)\PH}\xspace}
\newcommand\ZllH{\ensuremath{\cPZ(\ell\ell)\PH}\xspace}
\newcommand\ptV {\ensuremath{\pt(\Vvar)}\xspace}
\newcommand\ptH {\ensuremath{\pt(\PH)}\xspace}
\newcommand\dphiVH {\ensuremath{\Delta\phi(\Vvar,\PH)}\xspace}
\newcommand\WtoLN {\ensuremath{\PW\to\ell\cPgn}}
\newcommand\ZtoLL {\ensuremath{\cPZ\to\ell\ell}}
\newcommand\ZtoNN {\ensuremath{\cPZ\to\cPgn\cPagn}}
\newcommand\dphiMJ {\ensuremath{\Delta\phi(\MET, \text{jet})}}
\newcommand\dphiMtrk {\ensuremath{\Delta\phi(\MET,\MET\text{(tracks)})}}
\newcommand\Nal {\ensuremath{N_{\mathrm{a}\ell}}}
\newcommand\ptjj   {\ensuremath{{\pt}(\mathrm{jj})}}
\newcommand\ZllHbb  {\ensuremath{\cPZ(\ell\ell)\PH(\cPqb\cPqb)}}
\newcommand\Mjj     {\ensuremath{m(\mathrm{jj})}}
\newcommand\Naj {\ensuremath{N_{\mathrm{aj}}}}
\newcommand\dRJJ {\ensuremath{\Delta R(\mathrm{jj})}}
\newcommand\mtau       {\ensuremath{\tau}\xspace}
\newcommand\dphiMtkM {\ensuremath{\Delta\phi(\MET,{\MET}\text{(tracks)})}}
\newcommand\dPhiMETlep {\ensuremath{\Delta\phi(\MET,\ell)}}
\newcommand\dEtaJJ {\ensuremath{\abs{\Delta \eta(\mathrm{jj})}}}
\newcommand\dThPull {\ensuremath{\Delta\theta_{\text{pull}}}}
\newcommand\AddJetMaxCSV {\ensuremath{\mathrm{max}\mathrm{CSV}_{\mathrm{aj}}}}
\newcommand\AddJetMindR  {\ensuremath{\mathrm{min}\Delta R(\PH,\mathrm{aj})}}
\newcommand\Vudscg   {\ensuremath{\Vvar+\cPqu\cPqd\cPqs\cPqc\Pg}}
\newcommand\Voneb   {\ensuremath{\Vvar+\cPqb}}
\newcommand\Vtwob   {\ensuremath{\Vvar+\bbbar}}
\newcommand\Woneb   {\ensuremath{\PW+\cPqb}}
\newcommand\Wtwob   {\ensuremath{\PW+\bbbar}}
\newcommand\Wudscg   {\ensuremath{\PW+\cPqu\cPqd\cPqs\cPqc\Pg}}
\newcommand\Zoneb   {\ensuremath{\cPZ+\cPqb}}
\newcommand\Ztwob   {\ensuremath{\cPZ+\bbbar}}
\newcommand\Zudscg   {\ensuremath{\cPZ+\cPqu\cPqd\cPqs\cPqc\Pg}}

\RCS$Revision: 216721 $
\RCS$HeadURL: svn+ssh://svn.cern.ch/reps/tdr2/papers/HIG-13-012/trunk/HIG-13-012.tex $
\RCS$Id: HIG-13-012.tex 216721 2013-11-16 03:46:34Z alverson $
\newlength\cmsFigWidth
\ifthenelse{\boolean{cms@external}}{\setlength\cmsFigWidth{0.85\columnwidth}}{\setlength\cmsFigWidth{0.4\textwidth}}
\ifthenelse{\boolean{cms@external}}{\providecommand{\cmsLeft}{top\xspace}}{\providecommand{\cmsLeft}{left\xspace}}
\ifthenelse{\boolean{cms@external}}{\providecommand{\cmsRight}{bottom\xspace}}{\providecommand{\cmsRight}{right\xspace}}
\cmsNoteHeader{HIG-12-044} 
\title{Search for the standard model Higgs boson produced in association with a \PW\ or a \cPZ\ boson and decaying to bottom quarks}

\date{\today}

\abstract{A search for the standard model Higgs boson (\PH) decaying to \bbbar when produced in association with a weak
vector boson (\Vvar) is reported for the following channels: $\PW(\Pgm\Pgn)\PH$, $\PW(\Pe\Pgn)\PH$, $\PW(\Pgt\Pgn)\PH$, $\cPZ(\Pgm\Pgm)\PH$, $\cPZ(\Pe\Pe)\PH$, and $\cPZ(\Pgn\Pgn)\PH$. The search is performed in data samples corresponding to integrated luminosities of up to 5.1\fbinv at $\sqrt{s} = 7$\TeV and
up to 18.9\fbinv at $\sqrt{s} = 8$\TeV, recorded by the CMS experiment
at the LHC. An excess of events is observed above the expected background with a local significance of 2.1 standard deviations for a Higgs boson mass of 125\GeV, consistent with the expectation from the production of the standard model Higgs boson. The signal strength corresponding to this excess, relative to that of
the standard model Higgs boson, is  ${1.0\pm 0.5}$.
}

\hypersetup{%
pdfauthor={CMS Collaboration},%
pdftitle={Search for the standard model Higgs boson produced in association with a W or a Z  boson and decaying to bottom quarks},%
pdfsubject={CMS},%
pdfkeywords={CMS, physics, Higgs}}

\maketitle 

\section{Introduction}\label{sec:hbb_Introduction}

At the Large Hadron Collider (LHC), the ATLAS and CMS collaborations have reported the
discovery of a new
boson~\cite{Chatrchyan:2012ufa,Aad:2012tfa} with a mass, \mH, near
125\GeV and properties compatible with those of the
standard model (SM) Higgs
boson~\cite{Englert:1964et,Higgs:1964ia,Higgs:1964pj,Guralnik:1964eu,Higgs:1966ev,Kibble:1967sv}. To
date, significant signals have been observed in
channels where the boson decays into $\gamma \gamma$, $\cPZ\cPZ$, or $\PW\PW$.
The interaction of this boson with the massive \PW\ and \cPZ\  vector
bosons indicates that it plays a role in electroweak symmetry
breaking. The interaction with the fermions and whether the Higgs
field serves as the source of mass generation in the fermion sector,
through a Yukawa interaction, remains to be firmly established.

At $\mH\approx 125$\GeV  the standard model Higgs boson decays
predominantly into a bottom quark-antiquark pair (\bbbar) with a
branching fraction of ${\approx}58\%$~\cite{Dittmaier:2011ti}. The
observation and study of the \HBB\ decay, which involves the direct
coupling of the Higgs boson to down-type quarks, is therefore
essential in determining the nature of the newly discovered boson.
The measurement of the \HBB\ decay  will be the first direct test of whether the
observed boson interacts as expected with the quark sector, as the
coupling to the top quark has only been tested through loop effects.

In their combined search for the SM Higgs boson~\cite{PhysRevD.88.052014},
the CDF and D0 collaborations at the Tevatron  \Pp\Pap\ collider have reported evidence for
an excess of events in the 115--140\GeV mass range, consistent with the mass of the Higgs boson observed at the LHC.
In that search, the sensitivity below a mass of 130\GeV is dominated by the channels in
which the Higgs boson is produced in association with a weak vector boson and decaying to
\bbbar~\cite{PhysRevLett.109.071804}. The combined local significance
  of this excess is reported to be 3.0 standard deviations at
  $\mH=125$\GeV, while the expected local significance is 1.9 standard deviations.  At the LHC, a search for $\PH\to \bbbar$ by the ATLAS
experiment using data samples corresponding to an integrated
luminosity of $4.7$\fbinv at $\sqrt{s}=7\TeV$ resulted in exclusion limits on Higgs boson
production, at the 95\% confidence level (CL), of 2.5 to 5.5 times the
standard model cross section in the 110--130\GeV mass range~\cite{Aad:2012gxa}.

This article reports on a search at the Compact Muon Solenoid (CMS) experiment for the standard model Higgs
boson in the $\Pp\Pp\to \VH$ production mode, where \Vvar is either a \PW\ or a \cPZ\ boson and $\PH\to \bbbar$.   The previous Higgs boson search in
this production mode at CMS used
data samples corresponding to integrated
luminosities of up to $5.1$\fbinv at $\sqrt{s}=7\TeV$ and
up to $5.3$\fbinv at $\sqrt{s}=8\TeV$~\cite{Chatrchyan:2013lba}.
The results presented here combine the analysis of the 7\TeV data
sample in Ref.~\cite{Chatrchyan:2013lba}
with an updated analysis of the full 8\TeV data sample corresponding to a luminosity of up to $18.9$\fbinv.

The following six channels are
considered in the search: $\PW(\mu\nu)\PH$,
   $\PW(\Pe\nu)\PH$,
$\PW(\tau\nu)\PH$, $\cPZ(\mu\mu)\PH$, $\cPZ(\Pe\Pe)\PH$, and
   $\cPZ(\nu\nu)\PH$, all
with the Higgs boson decaying to \bbbar. Throughout this article the term ``lepton''
refers only to charged leptons and the symbol
$\ell$ is used to refer to both muons and electrons, but not
to taus. For the
$\PW(\tau\nu)\PH$ final state, only the 8\TeV data are
included and only taus with 1-prong
hadronic decays are explicitly considered; the $\tau$ notation
throughout this article refers exclusively to such decays. The leptonic
decays of taus in \WH\ processes are implicitly accounted for in the
$\PW(\mu\nu)\PH$ and $\PW(\Pe\nu)\PH$
channels.
Backgrounds arise from production of \PW\ and \cPZ\  bosons in association with
jets (from gluons and from light- or heavy-flavor quarks), singly and pair-produced top quarks (\ttbar), dibosons, and
quantum chromodynamics (QCD) multijet processes.

Simulated samples of signal and background events are used to provide
guidance in the optimization of the analysis. Control regions in data
are selected to adjust the event yields from simulation for the main
background processes in order to estimate their contribution in the signal
region. These regions also test the accuracy of the modeling of kinematic
distributions in the simulated samples.

Upper limits at the 95\% CL on the
$\Pp\Pp\to \VH$ production cross section times the $\PH\to
\bbbar$ branching fraction are
obtained for Higgs boson masses in the
110--135\GeV range. These limits are extracted by fitting the shape
of the output distribution of a boosted-decision-tree (BDT)
discriminant~\cite{Roe:2004na,Hocker:2007ht}. The results of the fitting procedure
allow to evaluate the presence of a Higgs boson signal over the
expectation from the background components. The
significance of any excess of events, and the corresponding event yield, is compared
with the expectation from a SM Higgs boson signal.

\section{Detector and simulated samples}\label{sec:hbb_Simulations}

A detailed description of the CMS detector can be found
elsewhere~\cite{Chatrchyan:2008aa}.
The momenta of charged particles are measured using a silicon pixel
and strip tracker that covers the pseudorapidity range
$\abs{\eta} < 2.5$ and is immersed in a 3.8\unit{T}
axial magnetic field. The pseudorapidity is defined as $\eta = -\ln[\tan(\theta/2)]$, where $\theta$
is the polar angle of the trajectory of a particle with respect to
the direction of the counterclockwise proton beam.
Surrounding the tracker are a crystal electromagnetic calorimeter
(ECAL) and a brass/scintillator hadron calorimeter (HCAL), both used to
measure particle energy deposits and consisting of a barrel assembly and two endcaps. The ECAL
and HCAL extend to a pseudorapidity range of $\abs{\eta} < 3.0$. A
steel/quartz-fiber Cherenkov forward detector extends the calorimetric
coverage to $\abs{\eta} < 5.0$. The outermost component of the CMS detector is the
muon system, consisting of gas-ionization detectors placed in the
steel return yoke of the magnet
to measure the momenta of muons traversing through the detector. The two-level CMS trigger system selects events of interest for
permanent storage. The first trigger level,
composed of custom hardware processors, uses information from the
calorimeters and muon detectors to select events in less than 3.2\mus.
The high-level trigger software algorithms, executed on a farm of
commercial processors, further reduce the
event rate using information from all detector subsystems. The
variable $\Delta R = \sqrt {(\Delta\eta)^2 +(\Delta\phi)^2}$ is used to
measure the separation between reconstructed objects in the detector,
where $\phi$ is the angle (in radians) of the trajectory of the object in the
plane transverse to the direction of the proton beams.

Simulated samples of signal and background events are produced using
various Monte Carlo (MC) event generators, with the CMS detector response modeled
with \GEANTfour~\cite{GEANT4}. The Higgs boson signal samples are
produced using the {\POWHEG}~\cite{POWHEG} event generator.
The {\MADGRAPH 5.1}~\cite{Alwall:2011uj} generator is used for the diboson, \PW+jets, \cPZ+jets, and \ttbar\ samples.
The single-top-quark samples, including the \cPqt\PW-, $t$-, and $s$-channel processes, are produced with {\POWHEG} and the QCD
multijet samples with {\PYTHIA 6.4}~\cite{Sjostrand:2006za}.  The production
cross sections for the diboson and \ttbar\ samples are rescaled to the
cross sections from the next-to-leading-order (NLO) {\MCFM} generator~\cite{Campbell:2010ff},
while the cross sections for the \PW+jets and \cPZ+jets samples are rescaled
to next-to-next-to-leading order (NNLO) cross sections calculated using the \textsc{fewz} program~\cite{Gavin:2010az,Li:2012wna,Gavin:2012sy}. The default set of parton distribution
functions (PDF) used to produce the NLO {\POWHEG} samples is
the NLO {MSTW2008} set~\cite{Martin:2009iq}, while the leading-order (LO)
CTEQ6L1 set~\cite{Pumplin:2002vw} is used for the other samples.
For parton showering and hadronization the {\POWHEG} and {\MADGRAPH} samples are interfaced with
 {\HERWIG++}~\cite{Bahr:2008pv} and  {\PYTHIA},
 respectively. The {\PYTHIA} parameters for the underlying event description are set to the Z2
tune for the 7\TeV samples and to the Z2$^*$ tune for the 8\TeV
samples~\cite{Chatrchyan:2011id}. The {\TAUOLA}~\cite{Jadach1991275} library is used to simulate tau decays.

During the 2011 data-taking period the LHC instantaneous luminosity reached up to $3.5\times 10^{33}\percms$ and
the average number of $\Pp\Pp$ interactions per bunch crossing was
approximately nine. During the 2012 period the LHC instantaneous luminosity
reached $7.7\times 10^{33}\percms$ and
the average number of $\Pp\Pp$ interactions per bunch crossing was
approximately twenty-one. Additional simulated $\Pp\Pp$ interactions overlapping with the event of interest in the same bunch crossing,
denoted as pileup events, are therefore added in the simulated samples to reproduce the pileup
distribution measured in data.

\section{Triggers}\label{sec:hbb_Triggers}

Several triggers are used to collect events consistent with
the signal hypothesis in the six channels under consideration.

For the \WmnH\ and  \WenH\ channels, the trigger paths consist of several single-lepton
triggers with tight lepton identification. Leptons are also required
to be isolated from other tracks and calorimeter energy deposits to maintain an acceptable trigger
rate. For the \WmnH\ channel and for the 2011 data, the trigger
thresholds for the muon transverse momentum, \pt, are in the range of 17 to 24\GeV. The higher
thresholds are used for the periods of higher instantaneous
luminosity. For the 2012 data the muon trigger \pt\ threshold for the single-isolated-muon trigger is
set at 24\GeV. For both the 2011 and 2012 data,
a single-muon trigger with a 40\GeV \pt\ threshold, but without any
isolation requirements, is also used for this channel.
The combined single-muon trigger efficiency is ${\approx}$90\% for
\WmnH\ events that pass all offline requirements that are described
in Section~\ref{sec:hbb_Event_Selection}.

For the \WenH\ channel and for the 2011 data, the electron \pt\ threshold ranges from 17 to 30\GeV.
To maintain acceptable trigger
rates during the periods of high instantaneous luminosity,
the lower-threshold triggers also require two central ($\abs{\eta}<2.6$) jets, with a \pt\
threshold in the 25--30\GeV range, and a minimum
requirement on the value of an online estimate of the missing transverse
energy, \MET, in the 15--25\GeV range. \MET is defined online as the magnitude of the vector sum of the transverse momenta
of all reconstructed objects identified by a
particle-flow algorithm~\cite{CMS-PAS-PFT-09-001,CMS-PAS-PFT-10-002}. This
algorithm combines the information
from all CMS subdetectors to identify and reconstruct online individual
particles emerging from the proton-proton collisions: charged hadrons, neutral hadrons,
photons, muons, and electrons.
These particles are then used to reconstruct jets, \MET and hadronic $\tau$-lepton decays,
and also to quantify the isolation of leptons and photons. For the
2012 data, the electron trigger uses a 27\GeV threshold on the \pt\ and no other
requirements on jets or \MET are made.
The combined efficiency for these triggers for \WenH\ events
to pass the offline selection criteria is $>$95\%.

For the \WtnH\ channel trigger, a 1-prong hadronically-decaying tau is required.
The \pt\ of the charged track candidate coming from the tau decay is required to be above 20\GeV and the \pt\
of the tau (measured from all reconstructed charged and neutral decay
products) above 35\GeV.
Additionally, the tau is required to be isolated inside
an annulus with inner radius $\Delta R = 0.2$ and outer
radius $\Delta R = 0.4$, where no reconstructed charged
candidates with $\pt > 1.5$\GeV must be found. A further requirement of a minimum of 70\GeV is placed on the \MET.
The efficiency of this trigger for \WtnH\ events that
pass the offline selection criteria is $>$90\%.

The \ZmmH\ channel uses the same single-muon triggers as the \WmnH\
channel. For the \ZeeH\ channel, dielectron triggers with lower \pt\
thresholds, of 17 and 8\GeV, and tight isolation requirements are used.
These triggers are nearly 100\% efficient for all \ZllH\  signal events that pass the
final offline selection criteria.

For the \ZnnH\ channel, combinations of several triggers are
used, all requiring \MET to be
above a given threshold. Extra requirements are added to
keep the trigger rates manageable as the instantaneous luminosity increased and to reduce
the \MET thresholds in order to improve signal
acceptance.
A trigger with $\MET >150$\GeV is used for the complete
data set in both 2011 and 2012. During 2011 additional
triggers that require the
presence of two central jets with $\pt >20\GeV$ and \MET thresholds
of 80 or 100\GeV, depending on the instantaneous luminosity, were used.
During 2012 an additional trigger that
required two central jets with $\pt >30\GeV$ and  $\MET>80$\GeV was used. This last trigger was discontinued when the instantaneous
luminosity exceeded $3\times 10^{33}\percms$
and was replaced by a trigger that required $\MET>100$\GeV,
at least two central jets with vectorial sum $\pt>100\GeV$
and individual \pt\ above 60 and 25\GeV, and no jet
with $\pt>40\GeV$ closer than 0.5 in azimuthal angle to the
\MET direction. In order to increase signal acceptance at lower values of \MET,
triggers that require jets to be identified as coming from \cPqb\ quarks
are used. For these triggers, two central jets with \pt\ above 20 or 30\GeV,
depending on the luminosity conditions, are required. It is also
required that at least one central jet with \pt\ above 20\GeV be
tagged by the online combined secondary vertex (CSV) \cPqb-tagging
algorithm described in
Section~\ref{sec:hbb_Event_Reconstruction}. This online
b-tagging requirement has an efficiency that is equivalent to that of the tight
offline requirement, $\mathrm{CSV} >0.898$, on the value of the output of the CSV discriminant. The \MET is required to
be greater than 80\GeV for these triggers. For \ZnnH\ events with $\MET >130$\GeV, the combined
trigger efficiency for \ZnnH\ signal events is near 100\% with
respect to the offline event reconstruction and selection, described
in the next sections. For events with \MET between 100 and 130\GeV the
efficiency is 88\%.

\section{Event reconstruction}\label{sec:hbb_Event_Reconstruction}

The characterization of \VH events, in the channels studied here,
requires the reconstruction of the following objects, all originating from a common interaction vertex:
electrons, muons, taus, neutrinos, and jets (including those originating from \cPqb\ quarks). The charged leptons
and neutrinos (reconstructed as \MET) originate
from the vector boson decays. The \cPqb-quark jets originate from the Higgs boson decays.

The reconstructed interaction vertex with the largest value of
$\sum_i {\pt}_i^2$, where ${\pt}_i$ is the transverse momentum of
the $i$th track associated with the vertex, is selected as the primary event vertex. This vertex is used as the
reference vertex
for all relevant objects in the event, which are reconstructed with
the particle-flow algorithm. The pileup interactions affect jet momentum
reconstruction, missing transverse energy reconstruction,
lepton isolation, and \cPqb-tagging efficiencies.
To mitigate these effects,  all charged hadrons that do not
originate from the primary interaction are identified by a
particle-flow-based algorithm and removed from consideration in the event.
In addition, the average neutral energy density  from pileup interactions is
evaluated  from particle-flow objects and
subtracted from the reconstructed jets in the
event and from the summed energy in the isolation cones used for
leptons, described below~\cite{Cacciari:subtraction}.
These pileup-mitigation procedures are applied on an event-by-event basis.

Jets are reconstructed from particle-flow objects using the
anti-\kt clustering algorithm~\cite{antikt}, with a distance parameter of 0.5,
as implemented in the \textsc{fastjet}
package~\cite{Cacciari:fastjet1,Cacciari:fastjet2}.  Each jet is required to
lie within $\abs{\eta} < 2.5$, to have at least two tracks associated with it,
and to have electromagnetic and hadronic energy fractions of at least
1\%. The last requirement removes jets originating from
instrumental effects. Jet energy corrections are applied as a function of pseudorapidity and
transverse momentum of the jet~~\cite{Chatrchyan:2011ds}. The
missing transverse energy vector is calculated offline as the negative
of the vectorial sum of transverse momenta of all particle-flow objects identified in the
event, and the magnitude of this vector is referred to as \MET in the
rest of this article.

Muons are reconstructed using two algorithms~\cite{Chatrchyan:2012xi}: one in which
tracks in the silicon tracker are matched to signals in the muon
detectors, and another in which a global track fit is performed, seeded by
signals in the muon systems.  The muon
candidates used in the analysis are required to be successfully reconstructed
by both algorithms.  Further identification criteria  are
imposed on the muon candidates to reduce the fraction
of tracks misidentified as muons. These include the number of measurements in the tracker and
in the muon systems,
the fit quality of the global muon track and its consistency with the primary
vertex. Muon candidates are
considered in the  $\abs{\eta} < 2.4$ range.

Electron reconstruction requires the matching
of an energy cluster in the ECAL with a track in the silicon
tracker~\cite{CMS-PAS-EGM-10-004}. Identification criteria based on the ECAL
shower shape, matching between the track and the ECAL cluster, and consistency with the
primary vertex are imposed. Electron identification relies on a
multivariate technique that combines observables sensitive to the
amount of bremsstrahlung along the electron trajectory, the
geometrical and momentum matching between the electron trajectory and
associated clusters, as well as shower-shape observables. Additional requirements are imposed to remove electrons
produced by photon conversions. In this analysis, electrons are
considered in the pseudorapidity range $\abs{\eta} < 2.5$,
excluding the  $1.44 < \abs{\eta}< 1.57$ transition
region between the ECAL barrel and endcap, where electron
reconstruction is suboptimal.

Charged leptons from the \PW\ and \cPZ\  boson decays are expected to be isolated
from other activity in the event. For each lepton candidate, a cone
is constructed around the track direction at the event vertex.  The scalar
sum of the transverse momentum of each reconstructed
particle compatible with the primary vertex and contained within the cone is calculated,
excluding the contribution from the lepton candidate itself. If this
sum exceeds approximately 10\% of the candidate \pt, the lepton is
rejected; the exact requirement depends on the lepton $\eta$, \pt,
and flavor. Including the isolation requirement, the total efficiency to reconstruct
muons is in the 87--91\% range, depending on \pt and $\eta$. The
corresponding efficiency for electrons is in the 81--98\% range.

The hadronically-decaying taus are reconstructed using the hadron plus strips (HPS) algorithm~\cite{Chatrchyan:2012zz} which
uses charged hadrons and neutral electromagnetic objects (photons) to
reconstruct tau decays. Reconstructed taus are required to be in the range $\abs{\eta} < 2.1$. In the first step of reconstruction, charged hadrons are reconstructed using the particle-flow algorithm.
Since neutral pions are often produced in hadronic tau decays, the HPS algorithm is optimized to reconstruct
neutral pions in the ECAL as objects called ``strips''. The strip reconstruction starts by centering one strip on the most
energetic electromagnetic particle and then looking for other particles in a window of 0.05 in $\eta$ and 0.20 in $\phi$.
Strips satisfying a minimum transverse momentum of $\pt(\text{strip})>1$\GeV are combined with the charged hadrons
to reconstruct the hadronic tau candidate.
In the final step of reconstruction, all charged hadrons and strips are required to be contained within a narrow cone size
of $\Delta R$ = 2.8/$\pt(\tau)$, where
$\pt(\tau)$ is measured from the reconstructed hadronic tau
candidate and is expressed in \GeVns.
Further identification criteria are imposed on the tau candidate to
reduce the fraction of electron and muons misidentified as taus.
These include the tau candidate passing an anti-electron
discriminator and an anti-muon discriminator. The isolation
requirement for taus is that the sum
of transverse momenta of particle-flow charged hadron and photon
candidates, with $\pt > 0.5$\GeV and within a cone of $\Delta R<0.5$, be less than 2\GeV. The tau reconstruction efficiency is approximately
50\% while the misidentification rate from jets is about 1\%.

Jets that originate from the hadronization of \cPqb\ quarks
are referred to as ``\cPqb\ jets''.  The CSV \cPqb-tagging
algorithm~\cite{Chatrchyan:2012jua} is used to identify such jets.
The algorithm combines
the information about track impact parameters and secondary
vertices within jets in a likelihood discriminant to provide separation
between \cPqb\ jets and jets originating from light quarks, gluons, or
charm quarks. The output of this CSV discriminant has values between zero
and one; a jet with a CSV value above a certain threshold is referred to
as being ``\cPqb\ tagged''. The efficiency to tag \cPqb\ jets and the rate of
misidentification of non-\cPqb\ jets depend on the
threshold chosen, and are typically parameterized as a function of the
\pt and $\eta$ of the jets.
These performance measurements are obtained directly from data in
samples that can be enriched in \cPqb\ jets, such as $\ttbar$ and multijet
events (where, for example, requiring the presence of a muon in the
jets enhances the heavy-flavor content of the events).
Several thresholds for the CSV output discriminant are used in this analysis.
Depending on the threshold used, the efficiencies to tag jets
originating from \cPqb\ quarks, \cPqc\ quarks, and light quarks or gluons are in
the 50--75\%, 5--25\%, and 0.15--3.0\% ranges, respectively.

Events from data and from the simulated samples are required to
satisfy the same trigger and event reconstruction requirements. Corrections
that account for the differences in the performance of these
algorithms between data and simulations are computed from data and used in
the analysis.

\section{Event selection}\label{sec:hbb_Event_Selection}

The background processes to \VH production with \HBB\ are the production of vector
bosons in association with one or more jets ({\Vvar}+jets), \ttbar production,
single-top-quark production, diboson production ({\Vvar}{\Vvar}),
and QCD multijet production. Except for dibosons, these processes have
production cross sections that are several orders of magnitude larger
than Higgs boson production. The production cross section for the ${\Vvar}\cPZ$ process, where
$\cPZ\to\bbbar$, is only a few times larger than the \VH
production cross section, and given the nearly identical final state this process provides a
benchmark against which the Higgs boson search strategy can be tested.

The event selection is based on the
reconstruction of the vector bosons in their leptonic decay modes and of the Higgs boson decay
into two \cPqb-tagged jets. Background events are substantially reduced by
requiring a significant boost of the \pt of the vector boson, \ptV, or
of the
Higgs boson~\cite{PhysRevLett.100.242001}.
In this kinematic region the \Vvar and $\PH$ bosons recoil away  from each other with a large azimuthal opening angle,  \dphiVH,
between them.  For each channel, different
\ptV\ boost regions are selected.
Because of different signal and background
content, each \ptV\ region has different sensitivity and
the analysis is performed separately in each region. The
results from all regions are then combined for each channel. The
{low-,} \mbox{intermediate-}, and high-boost regions
for the \WmnH\ and \WenH\ channels are defined by $100<\ptV<130$\GeV,  $130<\ptV<180$\GeV, and
$\ptV>180$\GeV.
For the \WtnH\ a single $\ptV>120$\GeV region is considered.
For the \ZllH\ channels, the low- and high-boost
regions are defined by $50<\ptV<100$\GeV
and $\ptV>100$\GeV.
For the \ZnnH\ channel \MET\ is used to define the low-, intermediate-, and high-boost
\ptV\ regions as $100<\MET<130$\GeV, $130<\MET<170$\GeV, and
$\MET>170$\GeV, respectively. In the rest of the article
the term ``boost region'' is used to refer to these \ptV\ regions.

Candidate \WtoLN\ decays are identified by requiring
the presence of a single-isolated lepton and additional missing
transverse energy. Muons are required to have $\pt>20$\GeV; the corresponding thresholds for electrons and taus are 30 and
40\GeV, respectively. For the \WlnH\ and \WtnH\ channels, \MET is
required to be $>$45 and $>$80\GeV, respectively,
to reduce contamination from QCD multijet processes. To further reduce
this contamination, it is also required for the \WlnH\
channels that the azimuthal angle between the \MET direction and the lepton be
${<}\pi/2$, and that the lepton isolation for the low-boost
region be tighter.

Candidate \ZtoLL\ decays are reconstructed by combining
isolated, oppositely-charged pairs of electrons or muons and requiring the dilepton invariant
mass to satisfy $75<m_{\ell\ell}<105$\GeV.  The
\pt\ for each lepton is required to be $>$20\GeV.

The identification of \ZtoNN\ decays
requires the \MET in the event to be within the boost regions
described above. The QCD multijet
background is reduced to negligible levels in this channel when requiring that the \MET does not originate from
mismeasured jets. To that end three event requirements are made. First,
for the high-boost region, a $\dphiMJ>0.5$ radians requirement is applied on the azimuthal angle
between the \MET direction and
the closest jet with $\abs{\eta}<2.5$ and $\pt>20$\GeV for the 7\TeV
analysis or $\pt>25$\GeV for the 8\TeV analysis (where more pileup
interactions are
present). For the low- and intermediate-boost regions the requirement is
tightened to $\dphiMJ>0.7$ radians. The second requirement is that the
azimuthal angle between the missing transverse energy direction as
calculated from charged tracks only (with $\pt>0.5\GeV$ and $\abs{\eta}<2.5$) and the \MET direction, \dphiMtrk, should be smaller
than 0.5 radians. The third requirement is made for the low-boost
region where the \MET significance (defined as the ratio between the \MET
and the square root of the total transverse energy in the
calorimeter, measured in \GeV) should be larger than 3.  To reduce background
events from \ttbar\ and \WZ\ production in the \WlnH, \WtnH, and \ZnnH\ channels,
events with an additional number of isolated leptons, $\Nal >0$, with $\pt>20\GeV$ are
rejected.

The reconstruction of the \HBB\ decay proceeds by selecting the pair
of jets in the event, each with $\abs{\eta}<2.5$ and \pt above a minimum
threshold, for which the value of the magnitude of the vectorial sum
of their transverse momenta, \ptjj, is the highest.
These jets are then also required to
be tagged by the CSV algorithm, with the value of the CSV discriminator
above a minimum threshold. The background from {\Vvar}+jets and
diboson production is reduced significantly
when the \cPqb-tagging requirements are applied and processes where the two jets originate from
genuine \cPqb\ quarks dominate the final selected data sample.

\begin{figure}[tbhp]
 \begin{center}
    \includegraphics[width=0.48\textwidth]{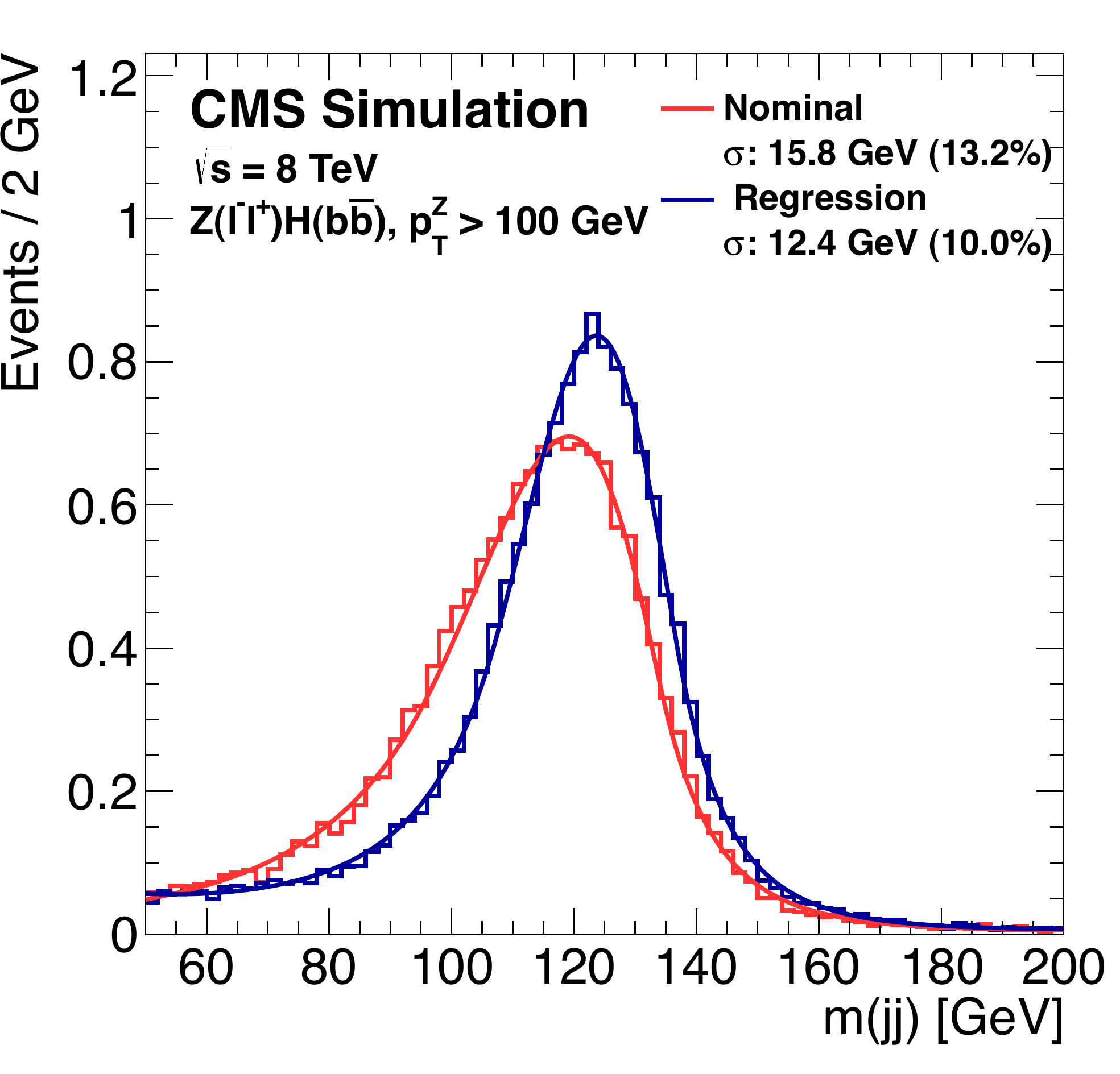}
    \caption{Dijet invariant mass distribution for simulated samples of \ZllHbb\ events ($m_{H} = 125\GeV$), before (red) and after (blue) the energy correction from the
regression procedure is applied. A Bukin function~\cite{Verkerke:2003ir} is fit to the distribution and the fitted width of the core of the distribution is displayed on the figure.
    }
    \label{fig:regression_VV_VH}
  \end{center}
\end{figure}

After all
event selection criteria described in this section are applied, the
dijet invariant-mass resolution of the two \cPqb\ jets from the Higgs decay is
approximately 10\%, depending on the \pt\ of the reconstructed Higgs boson,
with a few percent shift on the value of the mass peak. The Higgs boson mass resolution
is further improved by applying multivariate regression techniques similar to those used at
the CDF experiment~\cite{1107.3026}. An additional correction, beyond the
standard CMS jet energy corrections, is computed for
individual \cPqb\ jets in an attempt to recalibrate to the true \cPqb-quark
energy. For this purpose, a specialized BDT is trained on
simulated \HBB\ signal events with
inputs that include detailed  jet structure information
which differs in jets from \cPqb\ quarks from that of jets from light-flavor
quarks or gluons.
These inputs include variables related to several properties of the secondary vertex
(when reconstructed), information about tracks, jet constituents, and
other variables related to the energy reconstruction of the jet.
Because of semileptonic \cPqb-hadron decays, jets from \cPqb\ quarks contain, on average, more leptons and a larger
fraction of missing energy than jets from light quarks or gluons. Therefore,
in the cases where a low-\pt lepton is found in the jet or in its
vicinity, the following variables are also included in the
BDT regression: the \pt
of the lepton, the $\Delta R$ distance between the lepton and the jet
directions, and the
transverse momentum
of the lepton relative to the jet direction. For the \ZllH\ channels the \MET in the event and the azimuthal angle
between the \MET and each jet are also considered in the regression.
The output of the BDT regression is the corrected jet energy.
The average improvement on the mass resolution, measured on simulated
signal samples, when the corrected jet energies
are used is $\approx$15\%, resulting in an increase in the analysis sensitivity of
10--20\%, depending on the specific channel. This improvement is shown
in Fig.~\ref{fig:regression_VV_VH} for simulated samples of \ZllHbb\ events where the improvement in resolution is $\approx$25\%. The
validation of the regression technique in data is done with samples of
$\cPZ\to\ell\ell$ events with two \cPqb-tagged jets and in  $\ttbar$-enriched samples in the lepton+jets final state. In the $\cPZ\to\ell\ell$ case, when the jets are corrected by
  the regression procedure, the \pt\ balance distribution, between the
  $\cPZ$ boson, reconstructed from the leptons, and the \cPqb-tagged dijet
  system is improved to be better centered at zero and narrower than
  when the regression correction is not applied. In the  $\ttbar$-enriched case, the reconstructed top-quark mass distribution is closer to the nominal top-quark mass
  and also narrower than when the correction is not applied. In both
  cases the distributions for data and the simulated samples are
  in very good agreement after the regression correction is applied.

The signal region is defined by events that satisfy the vector boson and Higgs boson reconstruction criteria
described above together with the requirements listed in Table~\ref{tab:BDTsel}.
In the final stage of the analysis, to better separate signal from background under
different Higgs boson mass hypotheses, an event BDT discriminant is trained separately at
each mass value using simulated samples for signal and all background
processes. The training of this BDT is performed with all events in the signal region.
The set of event input variables used, listed in Table~\ref{tab:BDTvars},  is chosen by iterative optimization from a larger number of
potentially discriminating variables. Among the most discriminant
variables for all channels are the dijet invariant mass distribution
(\Mjj), the number of additional jets (\Naj), the value of CSV for the
Higgs boson daughter
with the second largest CSV value (CSV$_{\text{min}}$), and the
distance between Higgs boson daughters (\dRJJ). It has been suggested that variables related to techniques
that study in more detail the substructure of jets could help improve
the sensitivity of the \HBB\ searches~\cite{PhysRevLett.100.242001}. 
In this analysis, several combinations of such variables were
considered as additional inputs
to the BDT discriminant. However they did not yield significant gains in
sensitivity and  are not included in the final training used.

A fit is performed to the shape of the output distribution of the event BDT discriminant
to search for events resulting from Higgs boson production. Before testing all
events through this
final discriminant, events are classified based
on where they fall in the output distributions of several other background-specific BDT discriminants
that are trained to discern signal from individual background processes. This
technique, similar to the one used by the CDF collaboration~\cite{Aaltonen:2012id}, divides the samples into four distinct subsets that are
enriched in \ttbar, {\Vvar}+jets, dibosons, and \VH. The increase in the analysis sensitivity from using this technique in the
 \ZnnH\ and \WlnH\ channels is 5--10\%. For the  \ZllH\ channel the
improvement is not as large and therefore the technique is not used for that
 case. The technique is also not used in the \WtnH\ channel because of the
 limited size of the simulated event samples available for training
 multiple BDT discriminants.
The first background-specific BDT discriminant is trained to separate
\ttbar\ from \VH,
the second one is trained to separate {\Vvar}+jets from \VH, and the third one separates diboson events from \VH.
The output distributions of the background-specific BDTs are used to separate
events in four subsets: those that fail a requirement on the \ttbar\ BDT
are classified as \ttbar-like events,
those that pass the \ttbar\ BDT requirement but fail a requirement on the {\Vvar}+jets BDT
are classified as {\Vvar}+jets-like events,
those that pass the {\Vvar}+jets BDT requirement but fail the requirement on the diboson BDT
are classified as diboson-like events
and, finally, those that pass all BDT requirements are considered \VH-enriched events.
The events in each subset are then run through the final event BDT
discriminant and the resulting distribution, now composed of four
distinct subsets of events, is used as input to the fitting procedure.

As a validation of the multivariate approach to this analysis, these
BDT discriminants are also trained to find diboson signals ($\cPZ\cPZ$ and \PW\cPZ, with
$\cPZ\to \bbbar$) rather than the \VH signal. The event selection used in this case is identical
to that used for the \VH search.

\begin{table*}[tbp]
\topcaption{Selection criteria that define the signal region.
Entries marked with ``--'' indicate that the variable is not used in
the given channel.
If different, the entries in square brackets indicate the selection
for the different boost regions as defined in the first row of the table.
The \pt\ thresholds for the highest and second highest \pt\ jets are
$\pt(\mathrm{j}_1)$ and $\pt(\mathrm{j}_2)$, respectively.
The transverse momentum of the leading tau track is $\pt(\text{track})$.
The values listed for kinematic variables are in units of \GeV, and
for angles in units of radians.}
\label{tab:BDTsel}
\centering
\resizebox{\linewidth}{!}{
\begin{scotch}{ccccc}
Variable                     & \WlnH                           &     \WtnH             & \ZllH                 & \ZnnH                               \\\hline
\ptV                         & [100--130] [130--180] [$>$180]    &
[$>$120]             & [50--100] [$>$100]   & [100--130] [130--170] [$>$170]  \\ \hline
$m_{\ell\ell}$               & --                              &      --               & [75--105]            & --                                  \\
$\pt(\mathrm{j}_1)$                   & $>$30                           &     $>$30             & $>$20                 & $>$60                               \\
$\pt(\mathrm{j}_2)$                   & $>$30                           &     $>$30             & $>$20                 & $>$30                               \\
\ptjj                        & $>$100                          &
$>$120            & --                    & [$>$100] [$>$130] [$>$130]            \\
\Mjj                         & $<$250                          &
$<$250             & [40--250] $[<250]$   & $<$250                              \\
\MET                         & $>$45                           &   $>$80               & --                    & [100--130] [130--170] $[>170]$ \\
$\pt(\mtau)$               & --                              &      $>$40            & --                    & --                                  \\
$\pt(\text{track})$               & --                              &     $>$20             & --                    & --                                  \\
CSV$_{\text{max}}$         & $>$0.40                         &
$>$0.40          & [$>$0.50] [$>$0.244]    & $>$0.679                            \\
CSV$_{\text{min}}$         & $>$0.40                         &      $>$0.40          & $>$0.244              & $>$0.244                            \\
\Naj                         & --                              &
--              & --                    & $[<2]$ [--] [--]                      \\
\Nal                         & $=$0                            &      $=$0             & --                    & $=$0                                \\
\dphiVH                      & --                              &      --               & --                    & $>$2.0                              \\
\dphiMJ                      & --                              &
--               & --                    & [$>$0.7] [$>$0.7] [$>$0.5]           \\
\dphiMtkM                    & --                              &      --               & --                    & $<$0.5                              \\
\MET significance            & --                              &
--               & --                    & [$>$3] [--] [--]                     \\
\dPhiMETlep                  & $<\pi/2$                        &          --             & --                    & --                                  \\
\end{scotch}}
\end{table*}

\begin{table*}[tbp]
\topcaption{Variables used in the training of the event BDT discriminant. Jets are counted as additional jets if they
satisfy the following:  $\pt>20\GeV$ and $\abs{\eta} < 4.5$
for \WlnH,   $\pt>20\GeV$ and $\abs{\eta} < 2.5$
for \ZllH, and  $\pt>25\GeV$ and $\abs{\eta} < 4.5$
for \ZnnH. }
\label{tab:BDTvars}
\centering
\resizebox{\linewidth}{!}{
\begin{scotch}{l}
Variable \\\hline
$\pt(\mathrm{j}_1),\pt(\mathrm{j}_2)$: transverse momentum of each Higgs boson daughter       \\
\Mjj: dijet invariant mass                                \\
\ptjj: dijet transverse momentum                          \\
\ptV: vector boson transverse momentum (or \MET)          \\
\Naj: number of additional jets (see caption)                        \\
CSV$_{\text{max}}$: value of CSV for the Higgs boson daughter
with largest CSV value                                    \\
CSV$_{\text{min}}$: value of CSV for the Higgs boson daughter
with second largest CSV value                             \\
\dphiVH: azimuthal angle between \Vvar (or \MET) and dijet    \\
\dEtaJJ: difference in $\eta$ between Higgs boson daughters     \\
\dRJJ: distance in $\eta$--$\phi$ between Higgs boson daughters \\
\dThPull: color pull angle~\cite{Gallicchio:2010sw}  \\
\dphiMJ: azimuthal angle between \MET and the closest jet
(only for \ZnnH )                                         \\
\AddJetMaxCSV: maximum CSV of the additional jets in an
event (only for \ZnnH\ and \WlnH)                                   \\
\AddJetMindR: minimum distance between an additional jet
and the Higgs boson candidate (only for \ZnnH\ and \WlnH )                 \\
Invariant mass of the \VH system  (only for \ZllH )\\
Cosine of the angle between the direction of the \Vvar boson in the rest
frame of the \VH system and \\\hspace{0.2in}  the direction of the \VH system in the
laboratory frame  (only for \ZllH )\\
Cosine of the angle between the direction of one of the leptons in the
rest frame of the \cPZ\  boson and\\ \hspace{0.2in}  the direction of the \cPZ\  boson in the
laboratory frame  (only for \ZllH )\\
Cosine of the angle between the direction of one
of the jets in the rest frame of the reconstructed Higgs boson and\\\hspace{0.2in}  the
direction of the reconstructed Higgs boson in the laboratory frame  (only for \ZllH )\\
\end{scotch}}
\end{table*}

As a cross-check to the BDT-based analysis, a simpler analysis is
done by performing a fit to the shape of the dijet invariant mass distribution
of the two jets associated with the reconstructed Higgs boson, \Mjj. The event
selection for this analysis is more
restrictive than the one used in the BDT analysis and is optimized for
sensitivity in this single variable. Table~\ref{tab:MjjSel} 
lists the event
selection of the \Mjj\ analysis. Since the diboson background also
exhibits a peak in the \Mjj\ distribution from \cPZ\  bosons that decay into
b quark pairs, the distribution is also used to measure the
consistency of the diboson rate with the expectation from the
standard model. A consistent rate measurement would support the
validity of the estimate of the background processes in the Higgs
boson search.

\begin{table*}[tbp]
\topcaption{Selection criteria for the samples used in the \Mjj\ analysis
 in each channel. Entries marked with ``--'' indicate that the variable is not used in the given
channel.
If different, the entries in square brackets indicate the selection
for the different boost regions as defined in the first row of the table.
The \pt\ thresholds for the highest and second highest \pt\ jets are
$\pt(\mathrm{j}_1)$ and $\pt(\mathrm{j}_2)$, respectively.
The transverse momentum of the
leading tau track is $\pt(\text{track})$.
The values listed for kinematic variables are in units of \GeV, and
for angles in units of radians.}
\label{tab:MjjSel}
\centering
\resizebox{\linewidth}{!}{
\begin{scotch}{cccccc}
Variable       & \WlnH               & \WtnH  & \ZllH                        & \ZnnH                          \\ \hline
\ptV            & [100--150] [$>$150] (\Pe)  & [$<$250] & [50--100] [100--150] [$>$150] & [100--130] [130--170] [$>$170]\\
                  &  [100--130] [130--180] [$>$180] (\Pgm) & & \\\hline
$m_{\ell\ell}$ & --                  & --     & $75<m_{\ell\ell}<105$        & --                             \\
$\pt(j_1)$     & $>$30               &
$>$30  & $>$20                        & $[>60] [>60] [>80]$              \\
$\pt(j_2)$     & $>$30               & $>$30  & $>$20                        & $>$30                          \\
\ptjj          & $>$100              &
$>$120 & --                           & $[>110] [>140] [>190]$           \\
\Naj          & $=$0                & $=$0   & --                           & $=$0                           \\
\Nal           & $=$0                & $=$0  & --                           & $=$0                           \\
\MET            & $>$45               & $>$80     & $< 60$                      & --\\
$\pt(\mtau)$ & --                  & $>$40  & --                           & --                             \\
$\pt(\text{track})$ & --                  & $>$20  & --                           & --                             \\
CSV$_{\text{max}}$        & 0.898                & 0.898   & 0.679                         & 0.898                           \\
CSV$_{\text{min}}$            & $>$0.5              & $>$0.4 & $>$0.5                       & $>$0.5                         \\
\dphiVH         & $>$2.95             & $>$2.95& --                           & $>$2.95                        \\
\dRJJ           & --                  & --   & $[-] [-] [<1.6]$                & --                             \\
\dphiMJ          & --                  & --     & --                           & $[>0.7] [>0.7] [>0.5]$           \\
\dphiMtkM       & --                  & --     & --                           & $<$0.5                         \\
\dPhiMETlep    & $<\pi/2$            & --     & --                           & --                             \\
\end{scotch}}
\end{table*}

\section{Background control regions}\label{sec:hbb_Background_Control_Regions}

Appropriate control regions are identified in data and used to
validate the simulation modeling of the distributions used as input to the BDT
discriminants, and to obtain scale factors used to
adjust the simulation event yield
estimates for the most important background processes:
production of \PW\ and \cPZ\  bosons in association with jets and \ttbar\ production. For the \PW\ and \cPZ\  backgrounds the
control regions are defined such that they are enriched in either
heavy-flavor (HF) or light-flavor (LF) jets. Furthermore, these processes are split according to how many of the two
jets selected in the Higgs boson reconstruction
originate from \cPqb\ quarks, and separate scale factors are obtained for
each case. The notation used is: \Vudscg\ for the case where none of
the jets originate from a \cPqb\ quark, \Voneb\ for the case where only
one of the jets is from a \cPqb\ quark, and \Vtwob\ for the case where both
jets originate from \cPqb\ quarks.

To obtain the scale factors by which the simulated event
yields are adjusted, a set of binned likelihood fits is simultaneously performed to CSV distributions of
jets for events in the control regions. These fits are done separately
for each channel. Several other distributions are also fit to
verify consistency. These scale factors
account not only for cross section discrepancies, but also for potential residual
differences in physics object selection.
Therefore, separate scale factors are used for each background
process in the different channels. The uncertainties in the scale factor determination include two components: the statistical uncertainty due to the finite size of the samples and the systematic uncertainty.
The latter is obtained by subtracting, in quadrature, the statistical component from the full uncertainty which includes the effect of various sources of systematic uncertainty such as \cPqb-tagging, jet energy scale, and jet energy resolution.

Tables~\ref{tab:WlnControl}--\ref{tab:ZnnControl} list the selection criteria
used to define the control regions for the \WlnH, \ZllH, and \ZnnH\ channels,
respectively.  Because of the limited size of the simulated event
samples the scale factors obtained for the \WlnH\ channels are applied to the \WtnH\ channel.
Table~\ref{tab:SFs2012} summarizes the fit results
for all channels for the 8\TeV data. The scale factors are found to be
close to unity for all processes except for \Voneb\, for which the scale factors are consistently
found to be close to two. In this case, most of the excess occurs in
the region of low CSV$_{\text{min}}$ values in which events with two displaced vertices are found relatively close to each
other, within a distance $\Delta{\mathrm R}<0.5$ defined by the
directions of their displacement trajectories with respect to the
primary vertex. This discrepancy
is interpreted as arising mainly from mismodeling in the generator
parton shower of the process of gluon-splitting to \cPqb-quark pairs. In this
process the dominant contribution typically contains a low-\pt \cPqb\ quark that can end up not being reconstructed
as a jet above the \pt\ threshold used in the analysis, or that is merged with the jet from the more energetic \cPqb\ quark. These
discrepancies are consistent with similar observations in other
studies of the production of vector
bosons in association with heavy-flavor quarks by the
ATLAS and CMS experiments~\cite{Aad:2013vka,Chatrchyan:2013zja,Chatrchyan:2012vr}.

\begin{figure*}[tbh]
 \begin{center}
    \includegraphics[width=0.45\textwidth]{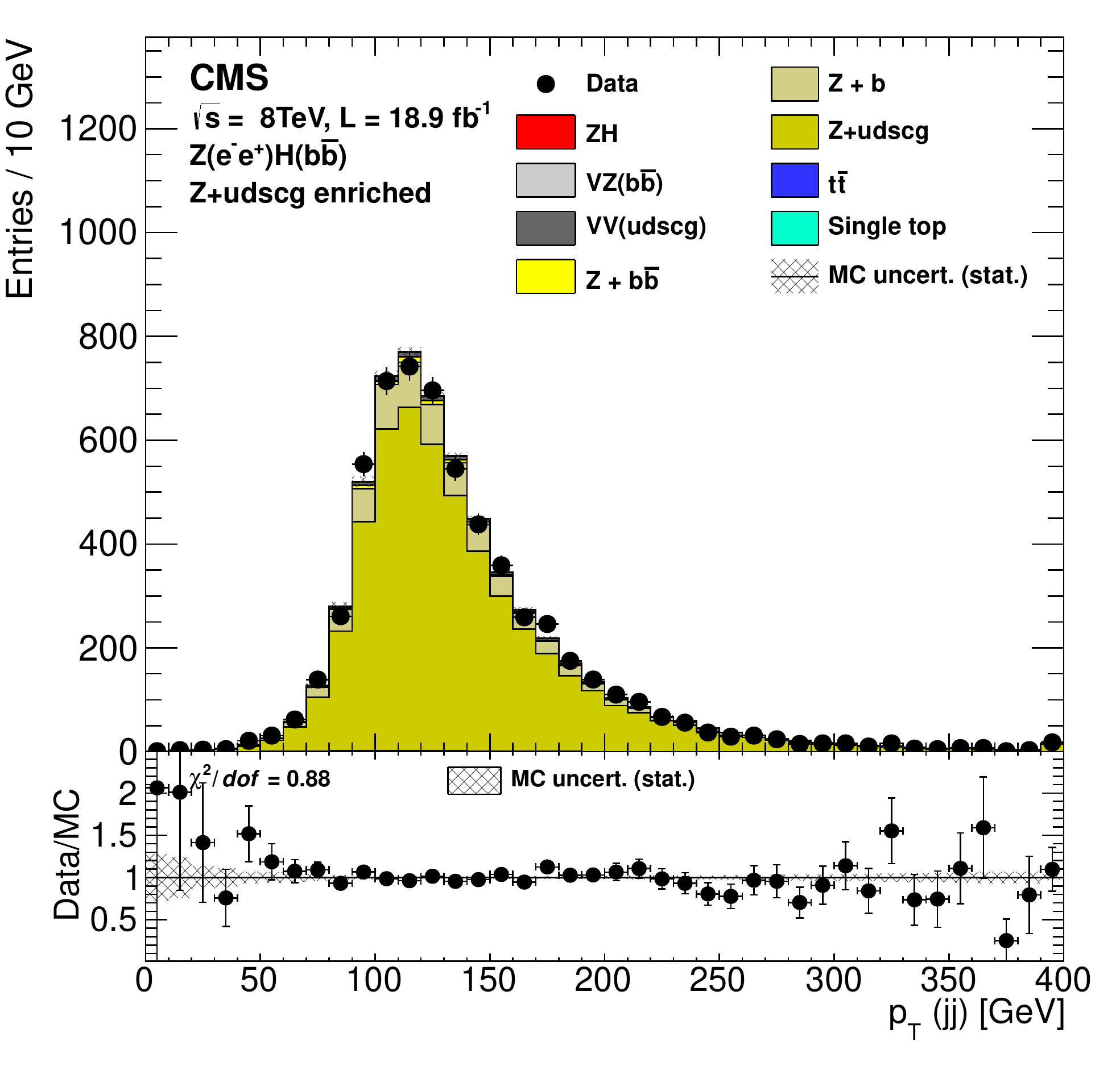}
    \includegraphics[width=0.45\textwidth]{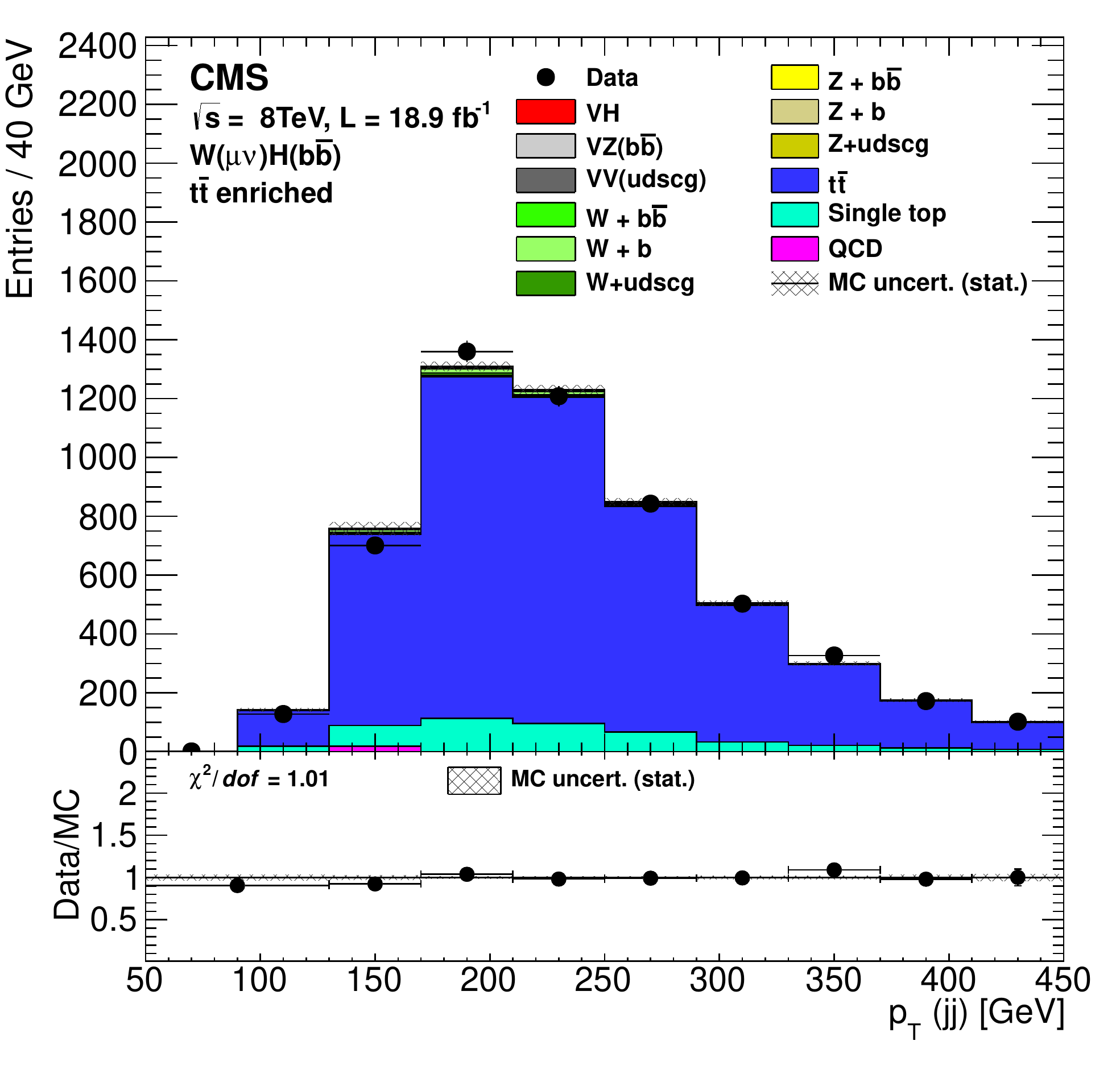}
    \includegraphics[width=0.45\textwidth]{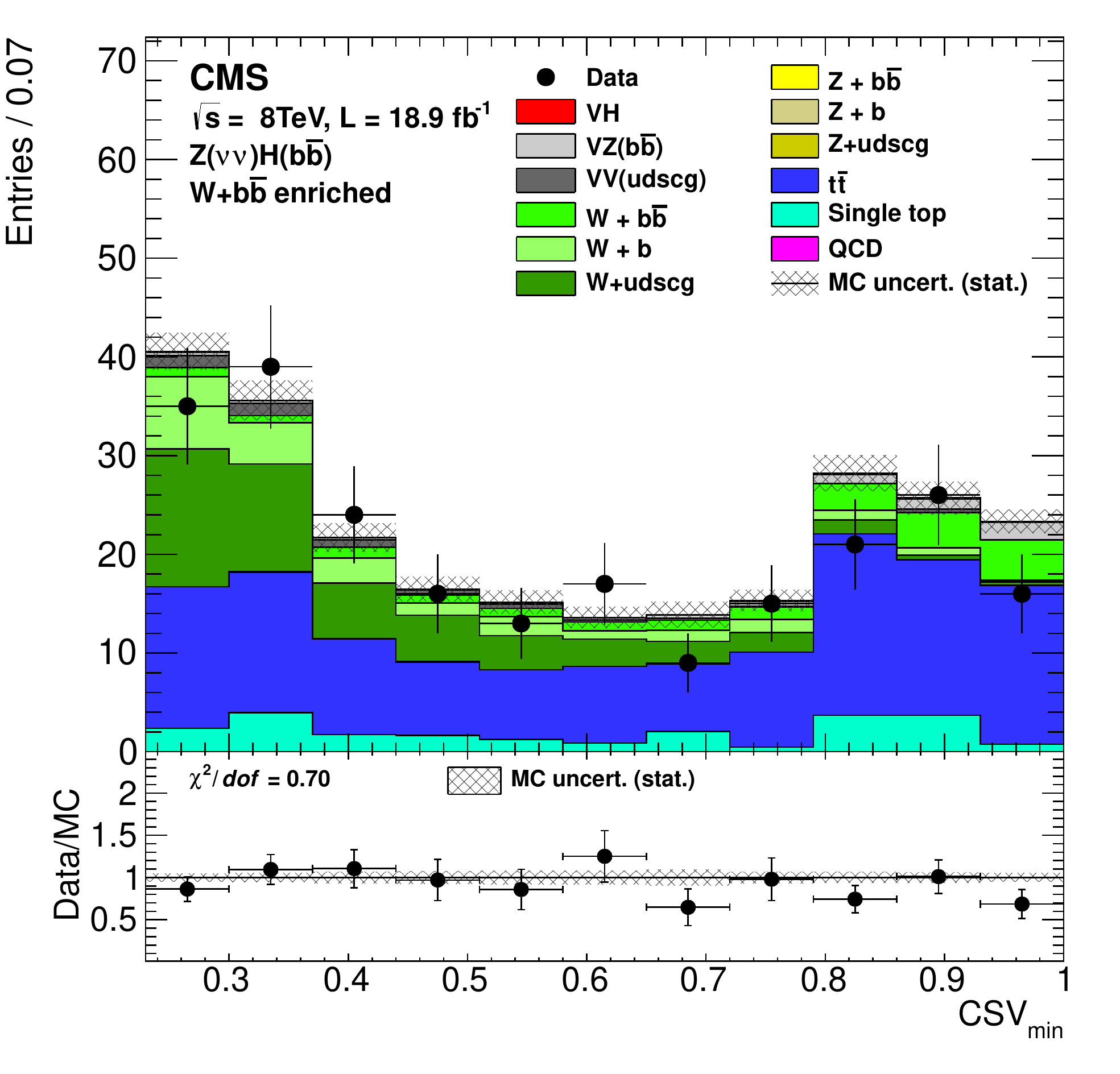}
    \includegraphics[width=0.45\textwidth]{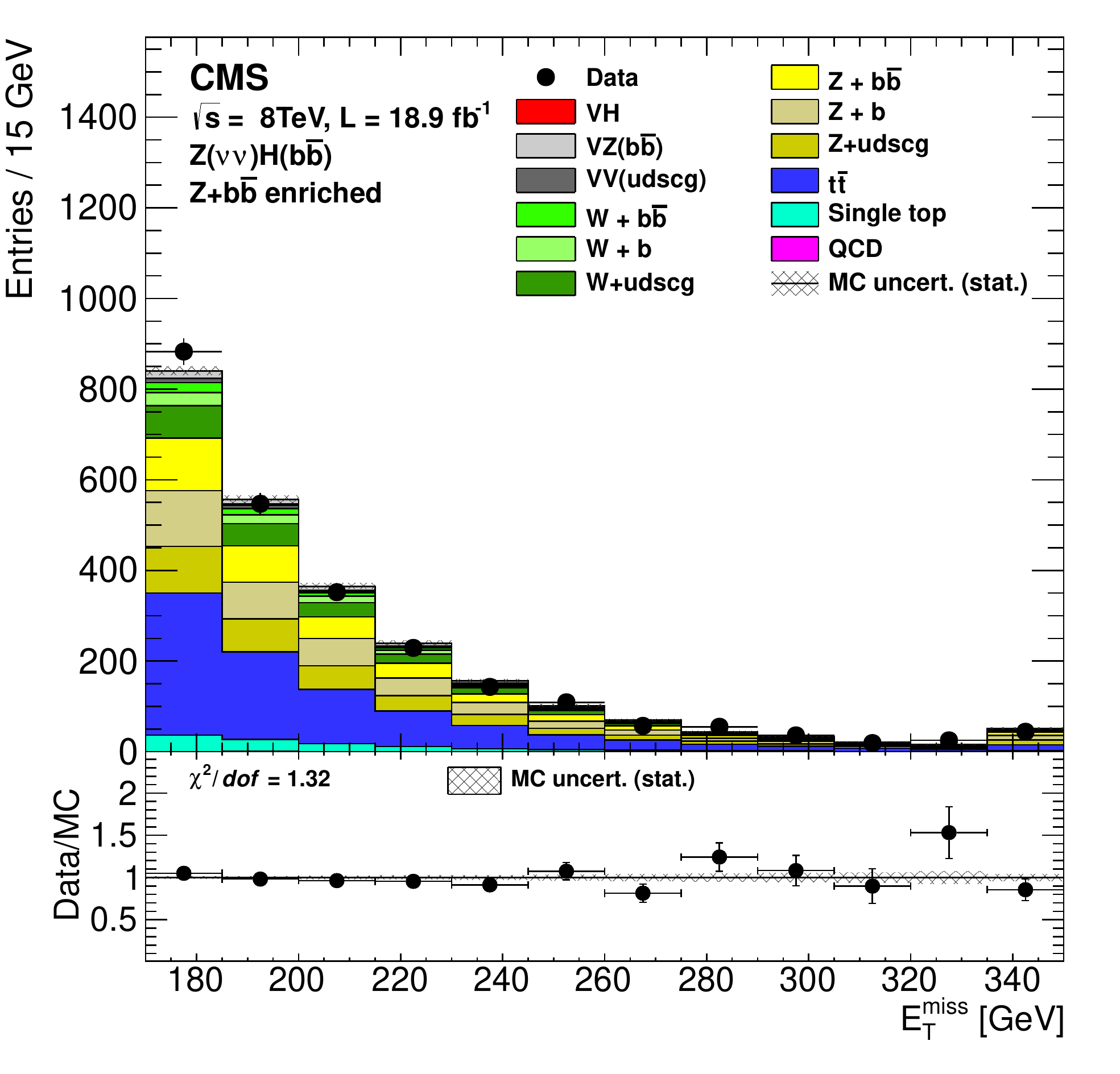}
    \caption{Examples of distributions for variables in the simulated
      samples and in data for different
      control regions and for different channels after applying the
      data/MC scale factors in Table~\ref{tab:SFs2012}. Top left:
      Dijet \pt\ distribution in the
      \cPZ+jets control region for the \ZeeH\ channel. Top
        right: \pt\ distribution in the  $\ttbar$ control region for
      the \WmnH\ channel. Bottom left:  CSV$_{\text{min}}$
      distribution for the \PW+HF high-boost control region for the \ZnnH\
      channel. Bottom right: \MET distribution for the \cPZ+HF
      high-boost control region for the \ZnnH\ channel. The bottom inset in each
      figure shows the ratio of the number of events observed in data to that
      of the Monte Carlo prediction for signal and backgrounds.
    }
    \label{fig:control_regions_ex}
  \end{center}
\end{figure*}

\begin{figure*}[tbh]
 \begin{center}
   \includegraphics[width=0.45\textwidth]{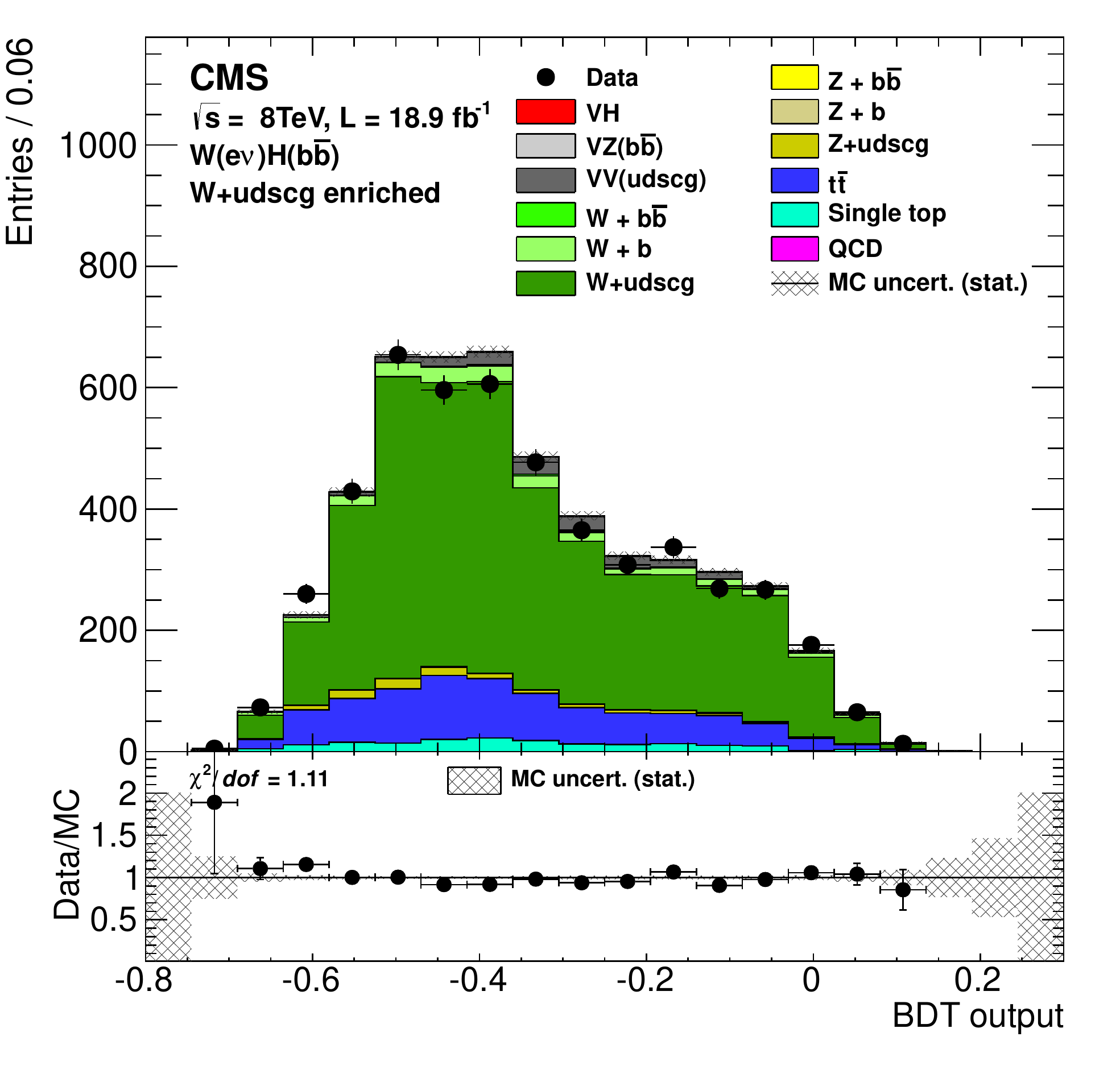}
    \includegraphics[width=0.45\textwidth]{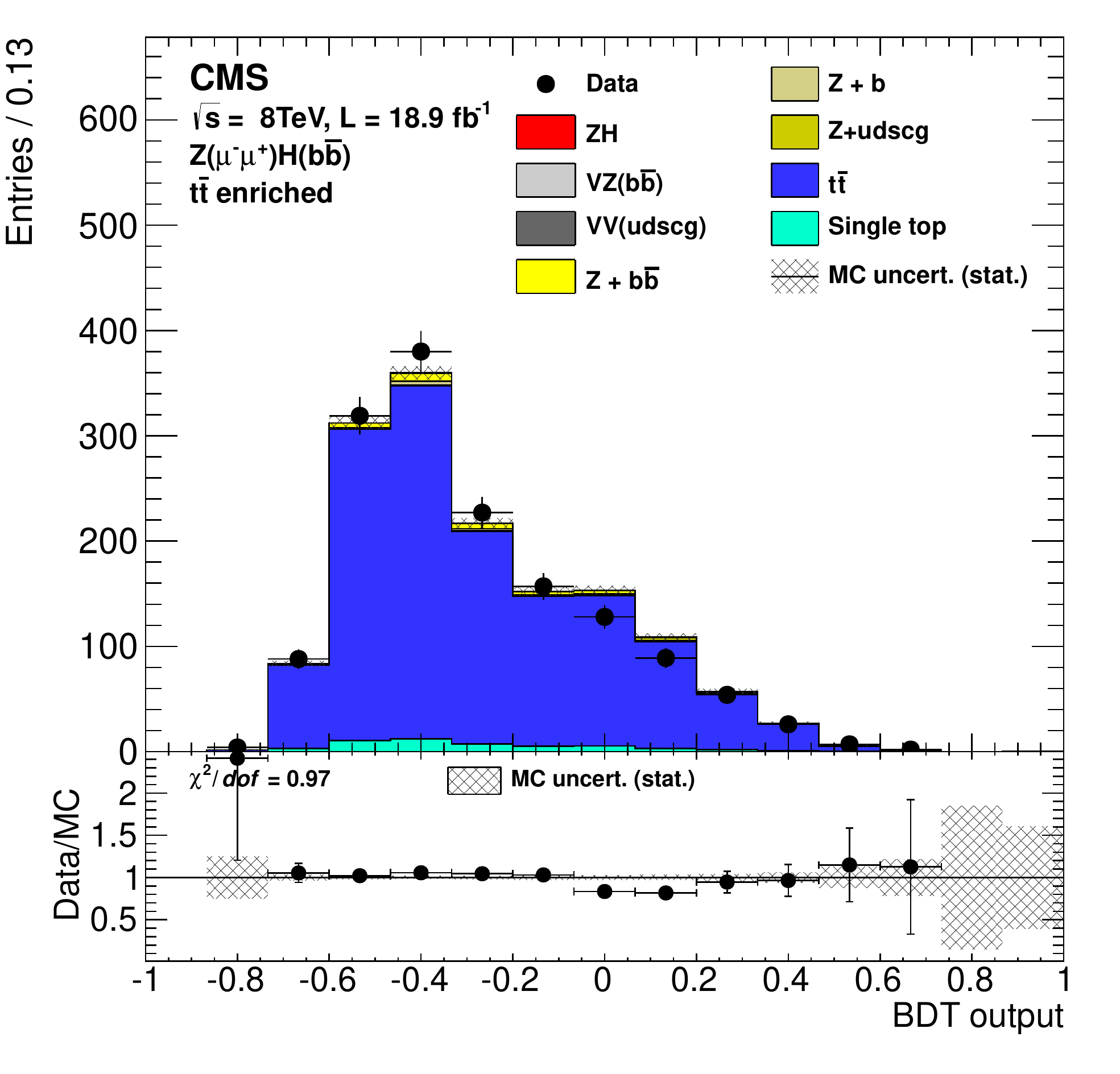}
    \includegraphics[width=0.45\textwidth]{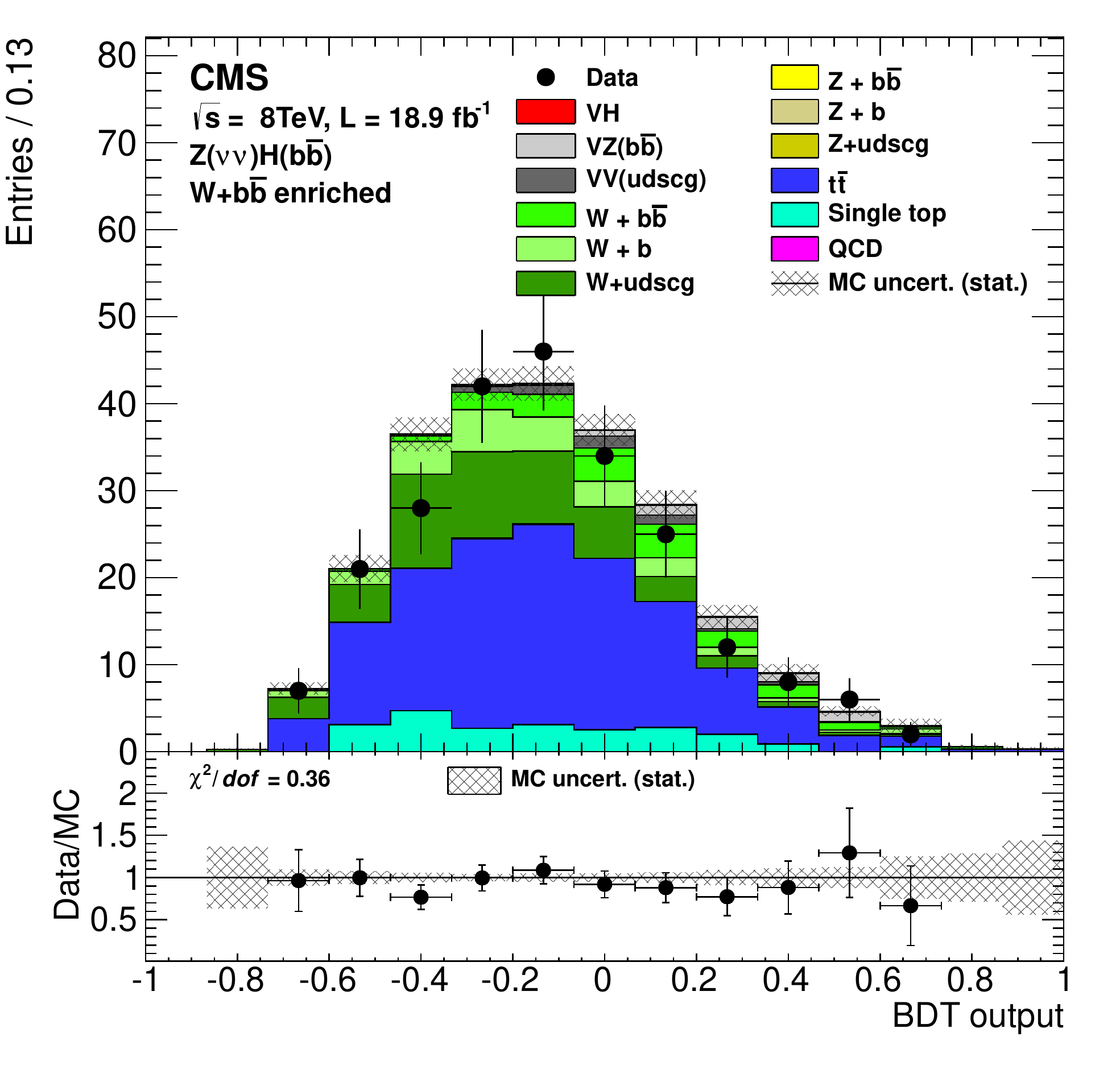}
    \includegraphics[width=0.45\textwidth]{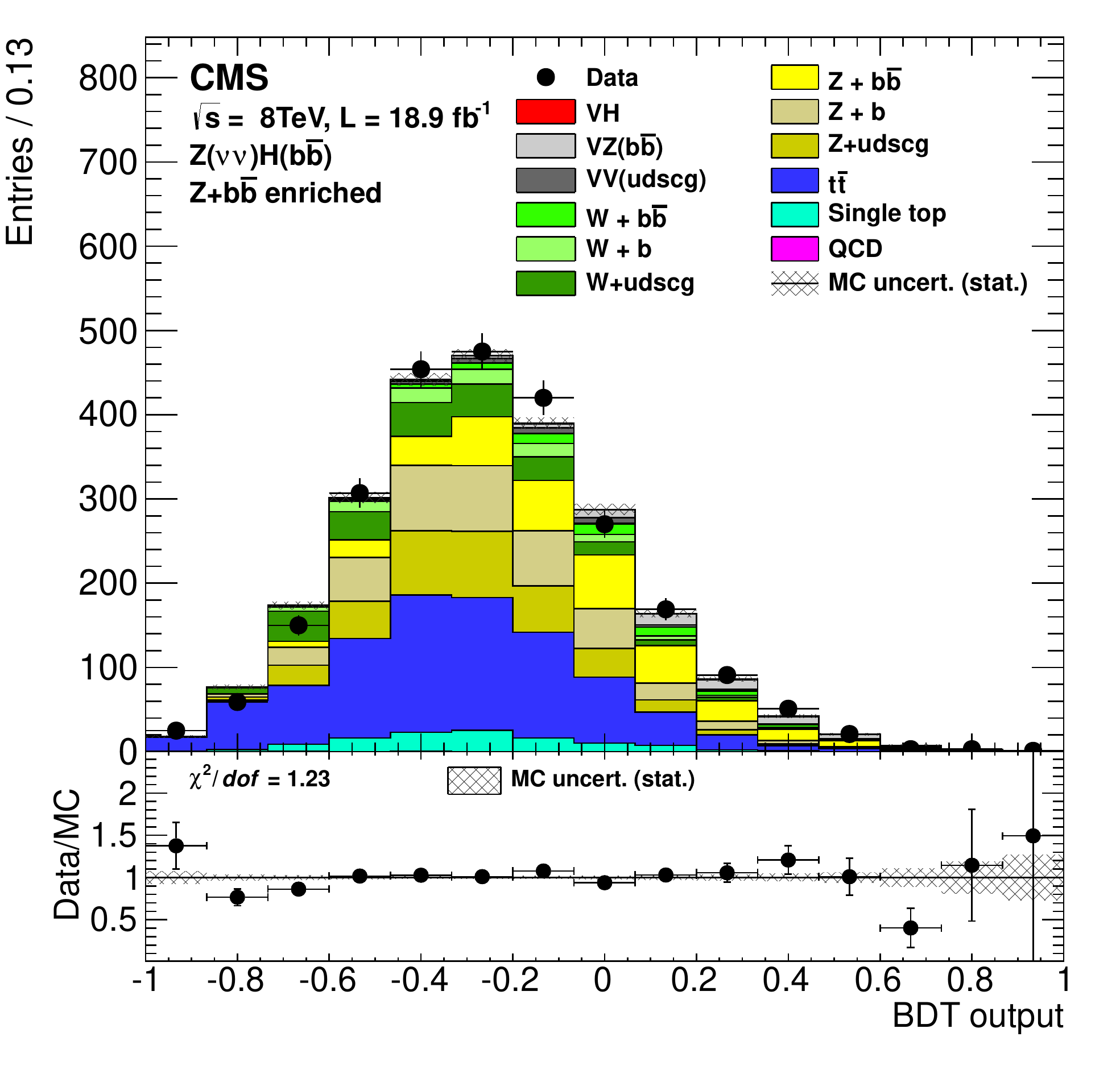}
    \caption{Examples of distributions of the event BDT discriminant output in the simulated
      samples and in data for different control regions and for different channels after applying the
      data/MC scale factors in Table~\ref{tab:SFs2012}. Top left:
      \PW+jets control region for the \WenH\ channel. Top
        right: $\ttbar$ control region for
      the \ZmmH\ channel. Bottom left: \PW+HF high-boost control region for the \ZnnH\
      channel. Bottom right: \cPZ+HF
      high-boost control region for the \ZnnH\ channel. The bottom inset in each
      figure shows the ratio of the number of events observed in data to that
      of the Monte Carlo prediction for signal and backgrounds.
    }
    \label{fig:control_regions_BDT}
  \end{center}
\end{figure*}

Figures~\ref{fig:control_regions_ex} and~\ref{fig:control_regions_BDT} show examples of distributions for variables in the simulated
      samples and in data for different
      control regions and for different channels. The scale factors
      described above have been applied to the corresponding simulated samples.

\begin{table}[tbp]
\topcaption{Definition of the control regions for the \WmnH\ and the
  \WenH\ channels. The same selection is used for all boost regions.
Here, LF and HF refer to light- and heavy-flavor
  jets. The values listed for kinematic variables are in units of
  \GeV. Because of the limited
  size of the simulated samples the scale factors derived in these
control regions are also applied to the \WtnH\ channel.}
\label{tab:WlnControl}
\centering
\begin{scotch}{cccc}
 Variable      & \PW+LF                  & \ttbar             & \PW+HF                  \\ \hline
$\pt(\mathrm{j}_1)$     & $>$30  	      & $>$30		    & $>$30		   \\
$\pt(\mathrm{j}_2)$     & $>$30  	      & $>$30		    & $>$30		   \\
\ptjj	       & $>$100 	       & $>$100 		    & $>$100		   \\
$\Mjj$         & $<$250 	       & $<$250 &$<$250, $\notin$[90-150] \\
CSV$_{\text{max}}$ 	       & $\in$[0.244--0.898]  & $>$0.898		    & $>$0.898	   \\
\Naj	       & $<$2		      & $>$1		    & $=$0		   \\
\Nal	       & $=$0		      & $=$0		    & $=$0		   \\
\MET	       & $>$45        & $>$45      & $>$45	    \\
\MET significance         & ${>}2.0 (\mu)\, {>}3.0(\Pe)$& --		     & --		    \\
\end{scotch}
\end{table}

\begin{table}[tbp]
\topcaption{Definition of the control regions for the \ZllH\ channel. The
same selection is used for both the low- and high-boost regions.
The values listed for kinematic variables are in units of \GeV.}
\label{tab:ZllControl}
\centering
\begin{scotch}{ccc}
Variable       & \cPZ+jets              & \ttbar                  \\ \hline
$m_{\ell\ell}$  & [75--105]       & $\notin$[75--105]             \\
$\pt(\mathrm{j}_1)$     & $>$20              & $>$20                   \\
$\pt(\mathrm{j}_2)$     & $>$20             & $>$20                   \\
\ptV            & $>$50       & [50--100]       \\
$\Mjj$       & $<$250, $\notin$[80--150]        & $<$250, $\notin$[80--150] \\
CSV$_{\text{max}}$            &  $>$0.244    & $>$0.244                     \\
CSV$_{\text{min}}$           &  $>$0.244      & $>$0.244                     \\
\end{scotch}
\end{table}

\begin{table*}[tbp]
\topcaption{Definition of the control regions for the \ZnnH\ channel.
If different, the entries in square brackets indicate the selection
for the different boost regions as defined by the \MET in the first row of the table.
 The values listed for kinematic variables are in units of \GeV, and
for angles in units of radians.}
\label{tab:ZnnControl}
\centering
\tiny
\resizebox{\linewidth}{!}{
  \begin{scotch}{cccccc}
    Variable             & \cPZ+LF                            & \cPZ+HF                               & \ttbar                             & \PW+LF                            & \PW+HF                               \\\hline
    \MET                 & [100--130] [130--170] [$>$170] &
    [100--130] [130--170] [$>$170] & [100--130] [130--170]
    [$>$170] & [100--130] [130--170] [$>$170] & [100--130]
    [130--170] $[>170]$  \\\hline
    $\pt(\mathrm{j}_1)$           & $>$60                              & $>$60                              & $>$60                              & $>$60                              & $>$60                               \\
    $\pt(\mathrm{j}_2)$           & $>$30                              & $>$30                              & $>$30                              & $>$30                              & $>$30                               \\
    \ptjj                & $[>100]$$[>130]$ $[>130]$           & $[>100]$$[>130]$ $[>130]$         & $[>100]$$[>130]$ $[>130]$   &$[>100]$$[>130]$ $[>130]$ & $[>100]$$[>130]$ $[>130]$           \\
    $\Mjj$               & $<$250                             &
    $<$250, $\notin$[100--140]           & $<$250, $\notin$[100--140]
    & $<$250                             & $<$250, $\notin$[100--140]            \\
    CSV$_{\text{max}}$ & $[0.244-0.898]$                    & $>$0.679                           & $>$0.898                           & $[0.244-0.898]$                    & $>$0.679                            \\
    CSV$_{\text{min}}$ & --                                 & $>$0.244                           & --                                 & --                                 & $>$0.244                            \\
    \Naj                 & [$<$2] [--] [--]                     &
    [$<$2] [--] [--]                     & $\geq 1$                           & $=$0                               & $=$0                                \\
    \Nal                 & $=$0                               & $=$0                               & $=$1                               & $=$1                               & $=$1                                \\
   \dphiVH              & --                                 & $>$2.0                             & --                                 & --                                 & $>$2.0                              \\
    \dphiMJ              & [$>$0.7] [$>$0.7] [$>$0.5]           &
    [$>$0.7] [$>$0.7] [$>$0.5] &  [$>$0.7] [$>$0.7] [$>$0.5]     &  [$>$0.7] [$>$0.7] [$>$0.5]   & [$>$0.7] [$>$0.7] [$>$0.5]       \\
    \dphiMtrk            & $<$0.5                             & $<$0.5                             & --                                 & --                                 & --                                  \\
    \MET significance   & [$>$3] [--] [--] &[$>$3] [--] [--]   & [$>$3] [--] [--]       &[$>$3] [--] [--]                 & [$>$3] [--] [--]                   \\
  \end{scotch}
}
\end{table*}

\begin{table*}[tbp]
\topcaption{Data/MC scale factors for 8\TeV data derived from the control
  regions, where the first quoted uncertainty
is statistical and the second is systematic. The muon and
electron channels in \ZllH\ and \WlnH\ are simultaneously fit to determine average scale
factors. For the \ZllH\ channel only four scale factors are derived, valid for
both the low and high \ptV\ boost regions. Because of the limited size of the simulated event
samples the scale factors obtained for the \WlnH\ channels are also applied to the \WtnH\ channel.}
\label{tab:SFs2012}
\centering
\begin{scotch}{cccccc}
 Process       & \WlnH                      & \ZllH                    & \ZnnH                     \\ \hline
   Low \ptV  \\ \hline
    \Wudscg\     & $1.03 \pm 0.01 \pm 0.05$  & --                        & $0.83 \pm 0.02 \pm 0.04$ \\
    \Woneb\     & $2.22 \pm 0.25 \pm 0.20$  & --                        & $2.30 \pm 0.21 \pm 0.11$ \\
    \Wtwob\     & $1.58 \pm 0.26 \pm 0.24$  & --                        & $0.85 \pm 0.24 \pm 0.14$ \\
    \Zudscg\     & --                        & $1.11 \pm 0.04 \pm 0.06$  & $1.24 \pm 0.03 \pm 0.09$ \\
    \Zoneb\     & --                        & $1.59 \pm 0.07 \pm 0.08$  & $2.06 \pm 0.06 \pm 0.09$ \\
    \Ztwob\     & --                        & $0.98 \pm 0.10 \pm 0.08$  & $1.25 \pm 0.05 \pm 0.11$ \\
   \ttbar   & $1.03 \pm 0.01 \pm 0.04$  & $1.10 \pm 0.05 \pm 0.06$  & $1.01 \pm 0.02 \pm 0.04$ \\
   \hline
   Intermediate \ptV \\ \hline
    \Wudscg\     & $1.02 \pm 0.01 \pm 0.07$  & --                        & $0.93 \pm 0.02 \pm 0.04$  \\
    \Woneb\     & $2.90 \pm 0.26 \pm 0.20$  & --                        & $2.08 \pm 0.20 \pm 0.12$  \\
    \Wtwob\     & $1.30 \pm 0.23 \pm 0.14$  & --                        & $0.75 \pm 0.26 \pm 0.11$  \\
    \Zudscg\     & --                        & --                        & $1.19 \pm 0.03 \pm 0.07$  \\
    \Zoneb\     & --                        & --                        & $2.30 \pm 0.07 \pm 0.08$  \\
    \Ztwob\     & --                        & --                        & $1.11 \pm 0.06 \pm 0.12$  \\
   \ttbar   & $1.02 \pm 0.01 \pm 0.15$  & --                        & $0.99 \pm 0.02 \pm 0.03$  \\
   \hline
   High \ptV \\ \hline
    \Wudscg\     & $1.04 \pm 0.01 \pm 0.07$  & --                        & $0.93 \pm 0.02 \pm 0.03$  \\
    \Woneb\     & $2.46 \pm 0.33 \pm 0.22$  & --                        & $2.12 \pm 0.22 \pm 0.10$  \\
    \Wtwob\     & $0.77 \pm 0.25 \pm 0.08$  & --                        & $0.71 \pm 0.25 \pm 0.15$  \\
    \Zudscg\     & --                        & $1.11 \pm 0.04 \pm 0.06$  & $1.17 \pm 0.02 \pm 0.08$  \\
    \Zoneb\     & --                        & $1.59 \pm 0.07 \pm 0.08$  & $2.13 \pm 0.05 \pm 0.07$  \\
    \Ztwob\     & --                        & $0.98 \pm 0.10 \pm 0.08$  & $1.12 \pm 0.04 \pm 0.10$  \\
   \ttbar   & $1.00 \pm 0.01 \pm 0.11$  & $1.10 \pm 0.05 \pm 0.06$  & $0.99 \pm 0.02 \pm 0.03$  \\
 \end{scotch}
\end{table*}

\section{Uncertainties}\label{sec:hbb_Uncertainties}

The systematic uncertainties that affect the results presented in this
article are listed in
Table~\ref{tab:syst} and are described in more detail below.

The uncertainty in the CMS luminosity measurement is estimated to be 2.2\%
for the 2011 data~\cite{CMS-PAS-SMP-12-008} and 2.6\% for the 2012 data~\cite{CMS-PAS-LUM-12-001}.
Muon and electron trigger,
reconstruction, and identification efficiencies are
determined in data from samples of
leptonic \cPZ-boson decays. The uncertainty on the event yields resulting
from the trigger efficiency estimate is
2\% per lepton and the uncertainty on the
identification efficiency is also 2\% per lepton. The parameters describing the
\ZnnH\ trigger efficiency turn-on curve have been
varied within their statistical uncertainties and also estimated for different
assumptions on the methods used to derive the efficiency.  This
results in an event yield
uncertainty of about 3\%.

The jet energy scale is
varied within its uncertainty as a function of jet \pt\ and
$\eta$. The efficiency of the analysis selection is
recomputed to assess the variation in event yields.  Depending on the
process, a 2--3\%
yield variation is found. The effect of the uncertainty
on the jet energy resolution is evaluated by
smearing the jet energies according to the measured
uncertainty. Depending on the process, a 3--6\% variation in event yields is
obtained. The uncertainties in the jet energy scale and
resolution also have an effect on the shape of the BDT
output distribution. The impact of the jet energy scale uncertainty is determined by recomputing
the BDT output distribution after shifting the energy scale up and down by its
uncertainty.   Similarly, the impact of the jet energy resolution is
determined by recomputing the BDT output distribution after increasing or decreasing the jet energy resolution. An uncertainty of 3\% is assigned to
the event yields of all processes in the \WlnH\ and \ZnnH\ channels due to the
uncertainty related to the missing transverse energy estimate.

Data/MC \cPqb-tagging scale factors are measured in heavy-flavor
enhanced samples of jets that contain muons and
are applied  consistently to jets in signal and background
events. The measured uncertainties for the \cPqb-tagging scale factors are:
3\% per \cPqb-quark tag, 6\% per charm-quark tag, and 15\% per
mistagged jet (originating from gluons and light \cPqu, \cPqd, or \cPqs\ quarks)~\cite{Chatrchyan:2012jua}. These translate into yield uncertainties in the 3--15\%
range, depending on the channel and the specific process. The
shape of the BDT output distribution is also affected by the shape of
the CSV distributions and an uncertainty is assigned according to a range of variations of the CSV distributions.

The total \VH signal cross section has
been calculated to NNLO accuracy,
and the total theoretical uncertainty is $\approx$4\%~\cite{Dittmaier:2012vm}, including
the effect of scale variations and PDF uncertainties~\cite{Alekhin:2011sk,Botje:2011sn,Lai:2010vv,Martin:2009iq,Ball:2011mu}. This analysis is performed
       in the boosted regime, and differences in
       the \pt\ spectrum of the \Vvar and $\PH$ bosons between data and
       MC introduce systematic effects in the
       signal acceptance and efficiency estimates.  Two
       calculations are available that evaluate the NLO
       electroweak (EW)~\cite{HAWK1,HAWK2,HAWK3} and NNLO QCD~\cite{Ferrera:2011bk}
       corrections to \VH production in the boosted regime. Both the electroweak and
       QCD corrections are applied to the signal samples. The estimated uncertainties of the NLO electroweak corrections
       are 2\% for both the \ZH and \WH production processes.
       The estimate for the NNLO QCD correction results in an uncertainty of $5\%$ for both the \ZH and \WH production
       processes.

The uncertainty in the background event yields
estimated from data is approximately 10\%. For {\Vvar}+jets, the
difference between the shape of the BDT output distribution for events generated
with the {\MADGRAPH} and the {\HERWIG++} Monte Carlo generators is
considered as a shape systematic uncertainty.
For \ttbar the differences in the shape of the BDT output distribution
between the one obtained from the nominal {\MADGRAPH}
samples and those obtained from the {\POWHEG} and {\MCATNLO}~\cite{Frixione:2002ik} generators are  considered as shape
systematic uncertainties.

An uncertainty of 15\% is assigned to the event yields obtained
from simulation for single-top-quark
production. For the diboson backgrounds, a 15\% cross section
uncertainty is assumed.
These uncertainties are consistent with the CMS measurements
of these processes~\cite{Chatrchyan:2012ep,Chatrchyan:2013oev}.
The limited number of MC simulated events
is also taken into account as a source of uncertainty.

The combined effect of the systematic uncertainties results in an
increase of about 15\% on the expected upper limit on the Higgs boson production
cross section and in a reduction of 15\%  on
the expected significance of an observation when the Higgs boson is
present in the data at the predicted standard model rate.

\begin{table*}[tbp]
\topcaption{Information about each source of systematic uncertainty,
including whether it affects the shape or normalization of the
BDT output, the uncertainty in signal or background
event yields, and the relative contribution to the expected
uncertainty
in the signal strength, $\mu$ (defined as the ratio of the best-fit value for the production cross section for a 125\GeV
Higgs boson, relative to the standard model cross section). Due to correlations, the total systematic
uncertainty is less than the sum in quadrature of the individual
uncertainties.  The last column shows the percentage decrease in
the total signal strength uncertainty, including statistical,
when removing that specific source of uncertainty.  The ranges quoted are
due to the difference between $7$ and $8\TeV$ data, different channels,
specific background processes, and the different Higgs boson mass
hypotheses. See text for details.}
\label{tab:syst}
\centering
\resizebox{\linewidth}{!}{
\begin{scotch}{lcccc}
         &           & Event yield uncertainty &  Individual contribution
         & Effect of removal\\
Source &    Type   & range (\%)  &  to $\mu$ uncertainty (\%)  &
on $\mu$ uncertainty (\%)\\  \hline
Luminosity & norm.   & 2.2--2.6      & $<$2    & $<$0.1 \\
Lepton efficiency and trigger (per lepton)    & norm.   &3 		 & $<$2    & $<$0.1 \\
\ZnnH\ triggers                               & shape           &3 		 & $<$2    & $<$0.1 \\
Jet energy scale                              & shape		&2--3 	 & 5.0     & $0.5$  \\
Jet energy resolution                         & shape 		&3--6          & 5.9     & $0.7$  \\
Missing transverse energy                     & shape           &3 		 & 3.2     & $0.2$  \\
b-tagging                                     & shape 		&3--15         & 10.2    & $2.1$  \\
Signal cross section (scale and PDF)          & norm.   &4 		 & 3.9    & $0.3$  \\
Signal cross section (\pt boost, EW/QCD)     & norm.   &2/5         & 3.9    & $0.3$  \\
Monte Carlo statistics                 & shape           &1--5 		 & 13.3    & $3.6$  \\
Backgrounds (data estimate)                   & norm. 	&10  & 15.9    & $5.2$  \\
Single-top-quark (simulation estimate)              & norm. 	&15        	 & 5.0    & $0.5$  \\
Dibosons (simulation estimate)                & norm.   &15 		 & 5.0     & $0.5$  \\
MC modeling ({\Vvar}+jets and \ttbar )              & shape           &10 		 & 7.4     & $1.1$  \\
\end{scotch}
}
\end{table*}

\section{Results}\label{sec:hbb_Results}

Results are obtained from combined signal and
background binned likelihood fits to the shape of the output distribution of the BDT
discriminants. These are trained separately for each channel and for each Higgs
boson mass hypothesis in the 110--135\GeV range.
In the simultaneous fit to all channels, in all boost regions, the BDT shape and
normalization for signal and for each background component
are allowed to vary within the systematic and statistical uncertainties described in
Section~\ref{sec:hbb_Uncertainties}. These uncertainties are treated as independent nuisance
parameters in the fit. All nuisance parameters,
including the scale factors described in
Section~\ref{sec:hbb_Background_Control_Regions}, are adjusted
by the fit.

\begin{figure}[htbp]
  \begin{center}
    \includegraphics[width=0.49\textwidth]{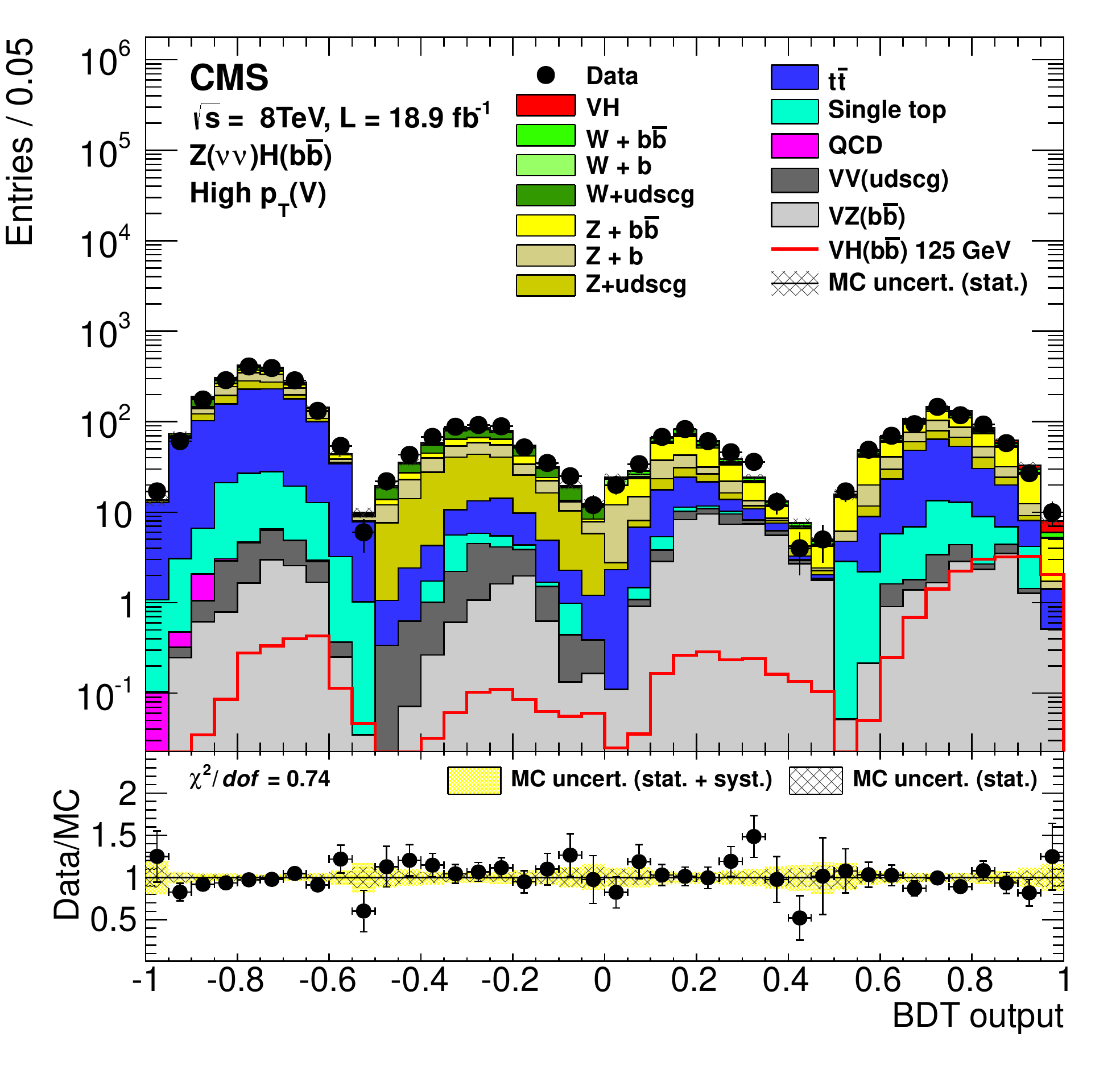}
    \includegraphics[width=0.49\textwidth]{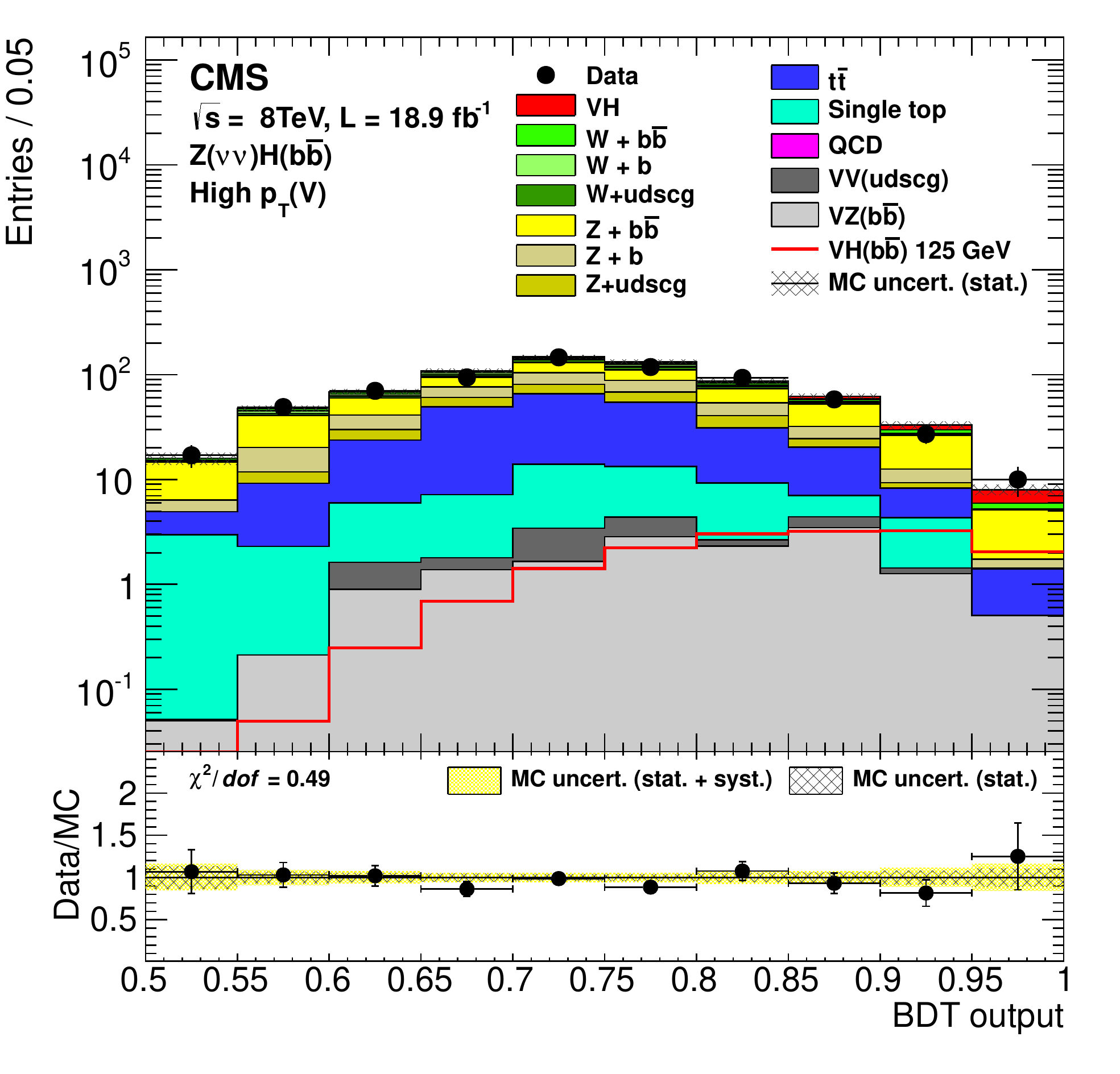}
\caption{Post-fit BDT output distributions for \ZnnH\ in the
      high-boost region for 8\TeV data (points with error bars), all
    backgrounds, and signal, after all selection criteria have been
    applied. The event BDT discriminant values for events in the four
    different subsets are rescaled and offset to assemble a single BDT output variable.
This leads to the four equally-sized partitions shown in the \cmsLeft
panel. The partitions correspond, starting from the left, to the event
subsets enriched in \ttbar, {\Vvar}+jets, diboson, and \VH production.
The \cmsRight panel shows the right-most, \VH-enriched, partition in more detail.
The bottom inset in each figure shows the ratio of the number of events observed in data to that of the Monte Carlo prediction for signal and backgrounds.}
    \label{fig:mBDT_Znn_example}
  \end{center}
\end{figure}

\begin{figure}[htb]
\begin{center}
\includegraphics[width=0.55\textwidth]{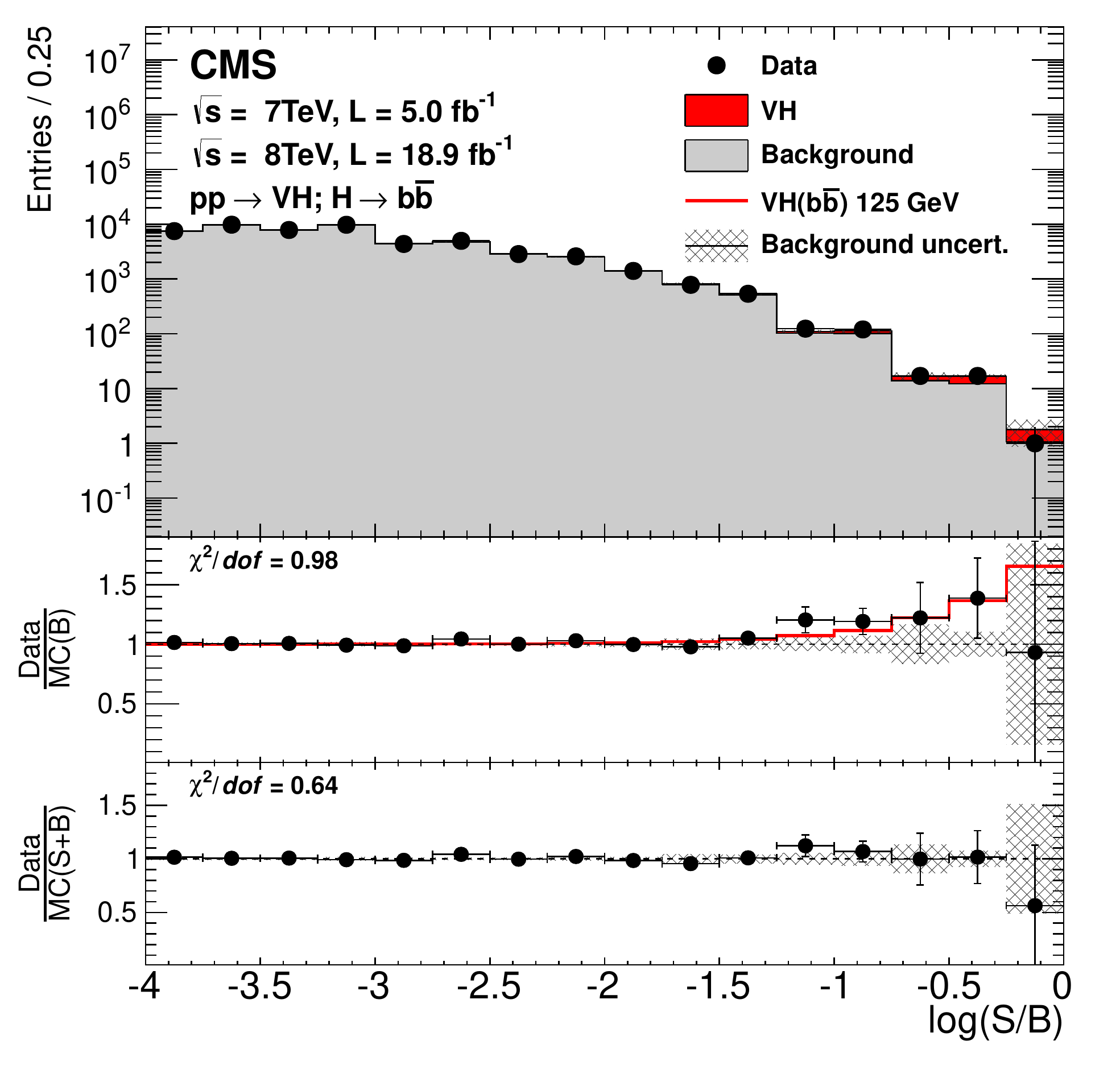}
\caption{Combination of all channels into a single distribution. Events are sorted in bins of similar expected signal-to-background
ratio, as given by the value of the output of their corresponding BDT
discriminant (trained with a Higgs boson mass hypothesis of 125\GeV).
The two bottom insets show the ratio of the data to the background-only
prediction (above) and to the predicted sum of background and SM Higgs
boson signal with a mass of 125\GeV
(below).}
    \label{fig:BDT_S_over_B_all}
  \end{center}
\end{figure}

In total 14 BDT distributions are considered. Figure~\ref{fig:mBDT_Znn_example} shows an example of
these distributions after the fit for the high-boost region of the
\ZnnH\ channel, for the  $\mH=125$\GeV  mass
hypothesis. The four partitions in the left panel
correspond to the subsets enriched in \ttbar, {\Vvar}+jets,
diboson, and \VH production, as described in
Section~\ref{sec:hbb_Event_Selection}. The right panel shows
the right-most, \VH-enriched, partition in more detail. For
completeness, all 14 BDT distributions used in the fit are shown in Figs.~\ref{fig:BDTWln8TeV_mu}--\ref{fig:BDTZnn8TeV} in
Appendix~\ref{sec:hbb_Appendix}. Table~\ref{table:3bin_yields} lists, for partial combinations of channels,
the total number of events in the four highest bins of their
corresponding BDT for the expected backgrounds, for the
125\GeV SM Higgs boson signal, and for data. An excess compatible with
the presence of the SM Higgs boson is observed. Figure~\ref{fig:BDT_S_over_B_all}
combines the BDT outputs of all channels where the events
are gathered in bins of similar expected signal-to-background
ratio, as given by the value of the output of their corresponding BDT
discriminant (trained with a Higgs boson mass hypothesis of 125\GeV). The
observed excess of events in the bins with the largest
signal-to-background ratio is consistent with what is expected from
the production of the standard model Higgs boson.

\begin{table*}[htbp]
\topcaption{The total number of events for partial combinations of channels in the four highest bins of their
corresponding BDT for the expected backgrounds
(B), for the 125\GeV SM Higgs boson \VH signal (S), and for data. Also
shown is the signal-to-background ratio (S/B).}
\begin{center}
\resizebox{\textwidth}{!}{
\begin{tabular}{lccccccccc} \hline\hline
         			&	\multicolumn{3}{c}{\WlnH}
                                &
                                \multicolumn{1}{c}{\WtnH}&\multicolumn{2}{c}{\ZllH}&\multicolumn{3}{c}{\ZnnH}
                                \\ \hline
Process  		   	&	Low \ptV
&	Int. \ptV		&High \ptV			&
&Low \ptV&High \ptV&Low \ptV&Int. \ptV&High \ptV \\ \hline
\Vtwob      	&	25.2          	&	22.4       	      &15.9        	&4.3  &158.6             &36.2               &177.3             &98.3              &68.2               \\
\Voneb     	&	3.1            	&	2.9     	     &9.6        	&1.2  &95.8              &14.6               &84.7              &58.3              &27.6               \\
\Vudscg     	&	4.5            	&	8.5      	    &10.0         	&2.5  &62.3              &8.7                &57.6              &31.0                &21.6               \\
\ttbar   	       &	113.2               &	106.5        &50.3               &22.6 &107.0               &6.9                &153.8             &87.4              &39.2               \\
Single-top-quark   &	24.1               &	20.3         &14.7     		&7.4  &2.9               &0.4                &54.5              &20.1              &11.7               \\
VV(udscg)     	&	0.3      		&	1.3          &1.2     		&0.2  &2.4               &0.4                &2.3               &1.5               &1.4                \\
VZ(\bbbar)     &	1.1        	&	1.4      &2.3                	&1.1  &11.0                &2.7                &9.5               &6.9               &7.7                \\  \hline
Total backgrounds       	&	171.7  	&	163.4     	      &104.1  	&39.4 &439.8             &69.8               &539.7             &303.5        &177.4              \\
\VH       		   	&	3.0    		&	6.0            &8.3              	&1.4  &5.5               &6.3                &8.5               &8.5               &11.5               \\
Data     		   	&	185         	&	182      	      &128        	&35   &425               &77                 &529               &322               &188                \\ \hline
S/B  (\%)  		   	&	1.7          		&	3.7     		      &8.0	&3.4  &1.3         & 9.0               &1.6             &2.8             &6.5              \\
\hline\hline
\end{tabular}
}
\label{table:3bin_yields}
\end{center}
\end{table*}

\begin{figure*}[tbh]
  \begin{center}
   \includegraphics[width=0.48\textwidth]{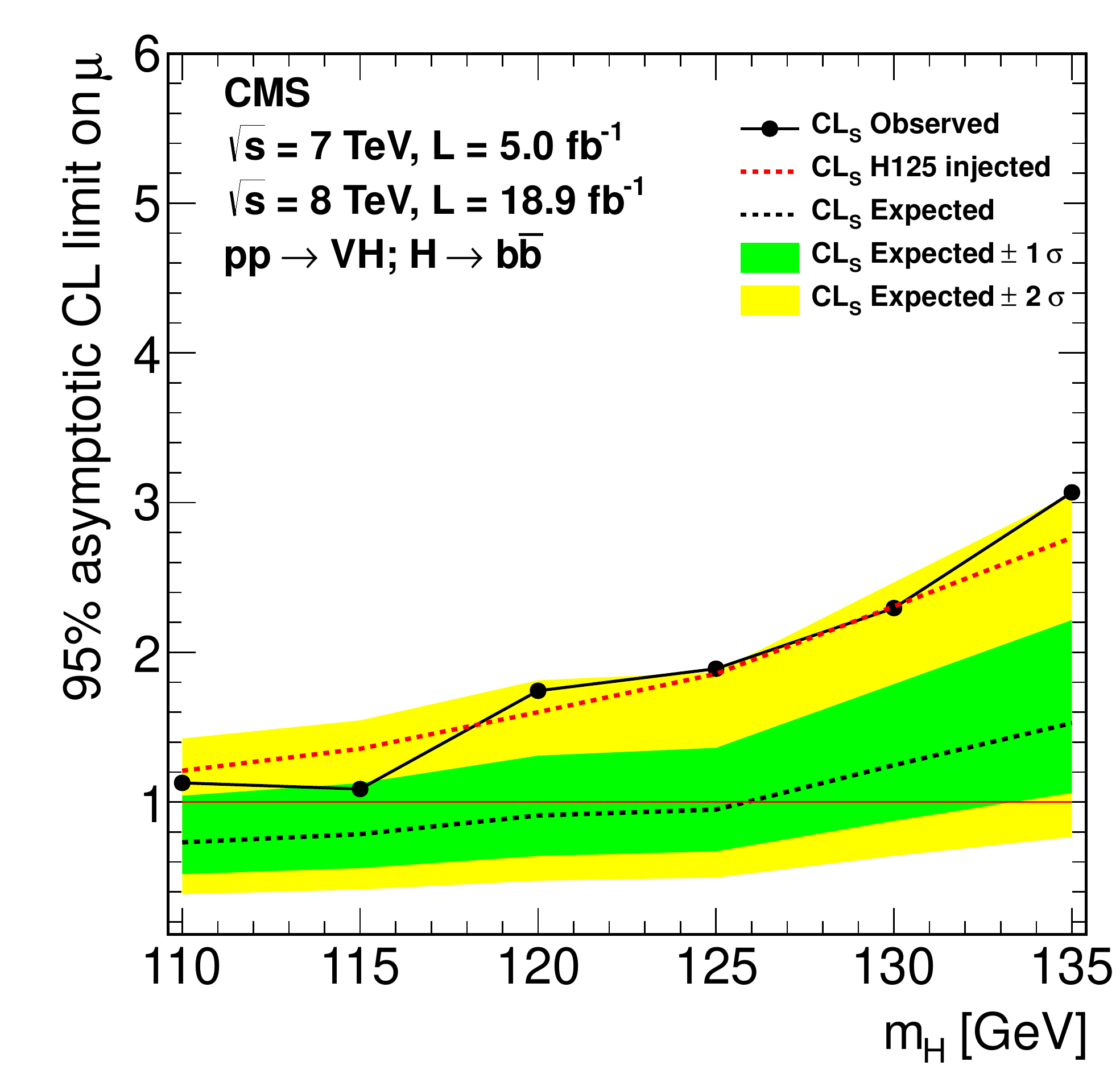}
   \includegraphics[width=0.48\textwidth]{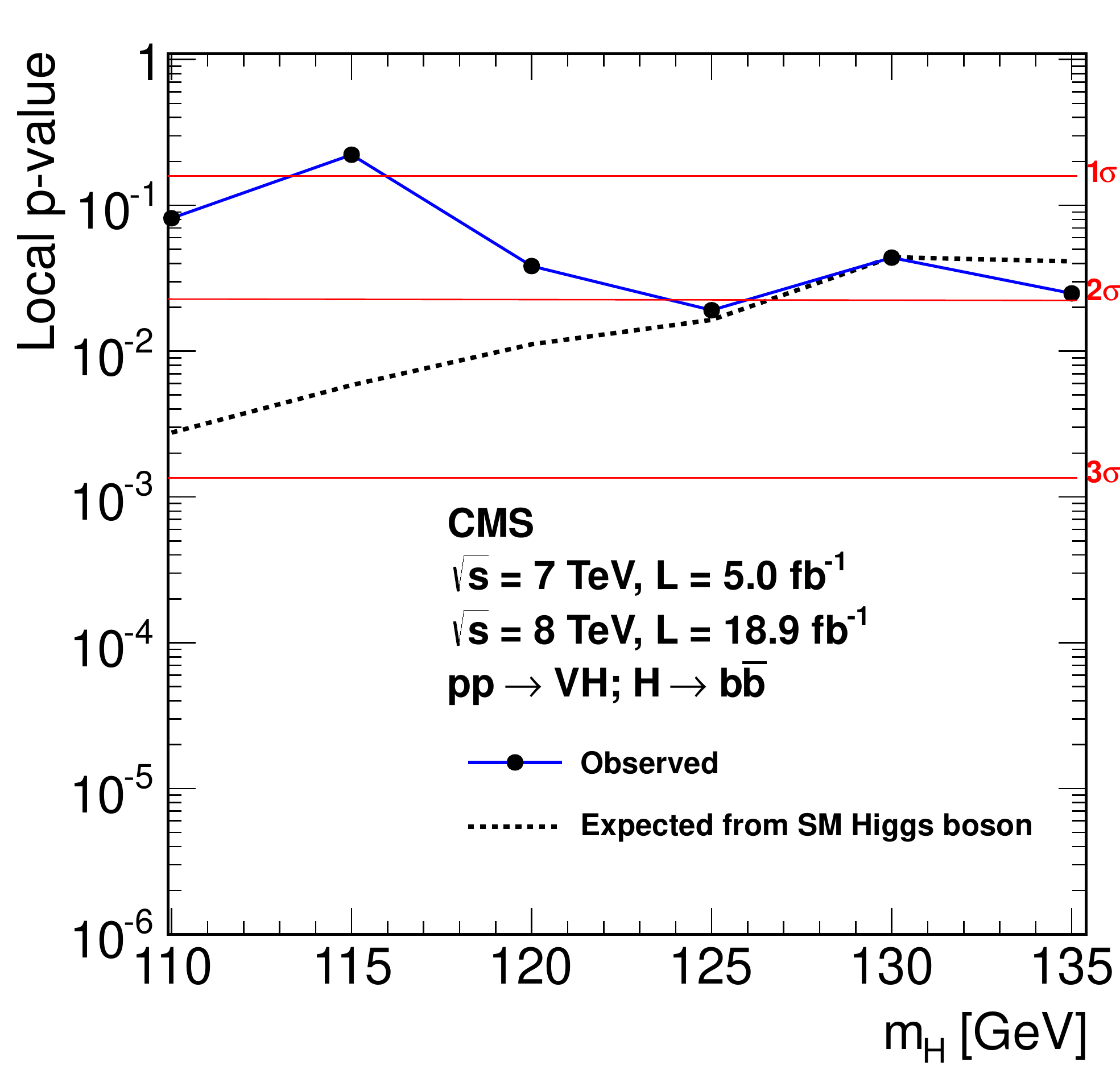}
   \caption{Left: The expected and observed $95\%$ CL upper limits on the
product of the \VH\ production cross section times the \HBB\
branching fraction, with respect to the expectations for the standard model
Higgs boson. The limits are obtained combining the results of the
searches using the 2011 (7\TeV) and 2012
(8\TeV) data. The red dashed line represents the
expected limit obtained from the sum of expected backgrounds and the SM Higgs boson signal
with a mass of 125\GeV. Right: local p-values and corresponding significance
(measured in standard deviations) for the background-only hypothesis to account for
the observed excess of events in the data.}
    \label{fig:Limits}
  \end{center}
\end{figure*}

The results of all
channels, for all boost regions and for the 7 and
8\TeV data, are combined to obtain 95\% confidence level (CL) upper limits
on the product of the \VH production cross section times the \HBB\
branching fraction, with respect to the expectations for a standard model
Higgs boson ($\sigma/\sigma_{\mathrm{SM}}$). At
each mass point the observed limit,  the median expected limit, and the 1 and 2 standard
deviation bands are
calculated using the modified frequentist method
CL$_\mathrm{s}$~\cite{Read:2002hq,junkcls,LHC-HCG}. Figure~\ref{fig:Limits}
displays the results.

For a Higgs
boson mass of 125\GeV the expected limit is 0.95 and the observed
limit is 1.89. Given that the resolution for the reconstructed Higgs
boson mass is $\approx$10\%, these results are compatible with a
Higgs mass of 125\GeV. This is demonstrated by the red dashed line in
the left panel of Fig.~\ref{fig:Limits}, which is the expected
limit obtained from the sum of expected background and the signal
of a SM Higgs boson with a mass of 125\GeV.

For all channels an excess of events over the expected background
contributions is indicated by the fits of the BDT output distributions.
The probability  (p-value)  to observe data as discrepant as observed
under the background-only hypothesis is shown in
the right panel of Fig.~\ref{fig:Limits} as a function of the assumed
Higgs boson mass.
For $\mH =125$\GeV, the excess of observed events corresponds to a
local significance of 2.1 standard deviations away from the background-only hypothesis.
This is consistent with the 2.1 standard deviations expected when assuming the standard model prediction for Higgs boson
production.

The relative sensitivity of the channels that are topologically
distinct is demonstrated in Table~\ref{tab:limits_by_mode} for
$\mH=125$\GeV. The table lists
the expected and observed limits and local significance for the \WlnH\ and \WtnH\ channels combined, for the \ZllH\
channels combined, and for the \ZnnH\ channel.

\begin{table*}[htbp]
\topcaption{The expected and observed $95\%$ CL upper limits on the
product of the \VH\ production cross section times the \HBB\
branching fraction, with respect to the expectations for the standard model
Higgs boson, for partial combinations of channels and for all channels
combined, for
$\mH=125$\GeV. Also shown are the expected and observed local significances.}
\label{tab:limits_by_mode}
\centering
\begin{scotch}{ccccc}
   $\mH=125$\GeV          & $\sigma /\sigma_{\mathrm{SM}}$ (95\% CL) &  $\sigma/\sigma_{\mathrm{SM}}$ (95\% CL)  &  Significance &
         Significance  \\
      & median expected & observed & expected & observed \\\hline
$\PW(\ell\cPgn,\Pgt\cPgn)\PH$ &  1.6  &   2.3    &  1.3  &  1.4 \\
\ZllH\ &  1.9  &   2.8    &  1.1  &  0.8 \\
 \ZnnH\ &  1.6  &   2.6    &  1.3  &  1.3   \\\hline
All channels & 0.95 & 1.89 & 2.1 & 2.1 \\
\end{scotch}
\end{table*}

\begin{figure*}[htbp]
  \begin{center}
\includegraphics[width=0.48\textwidth]{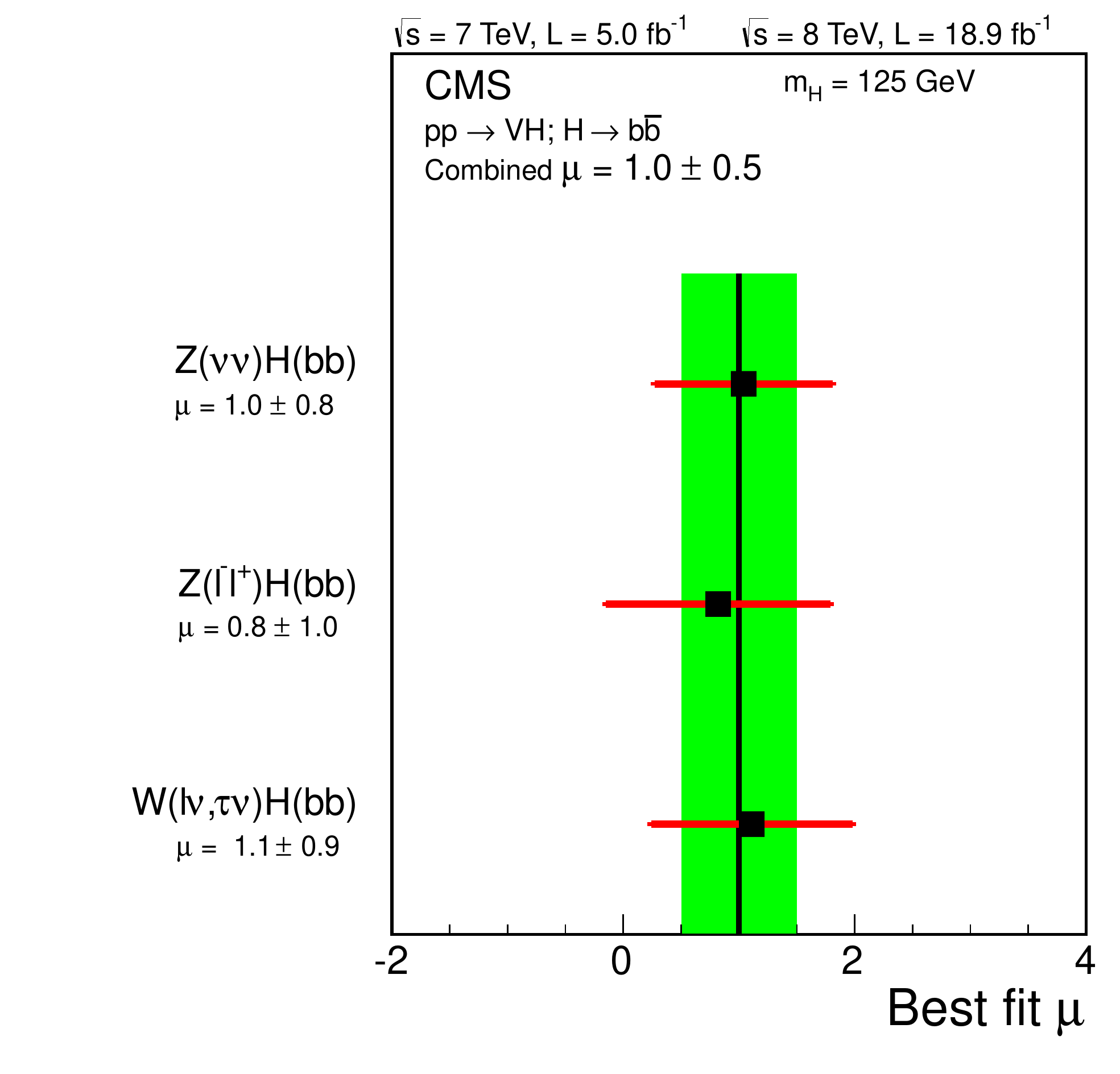}
\includegraphics[width=0.48\textwidth]{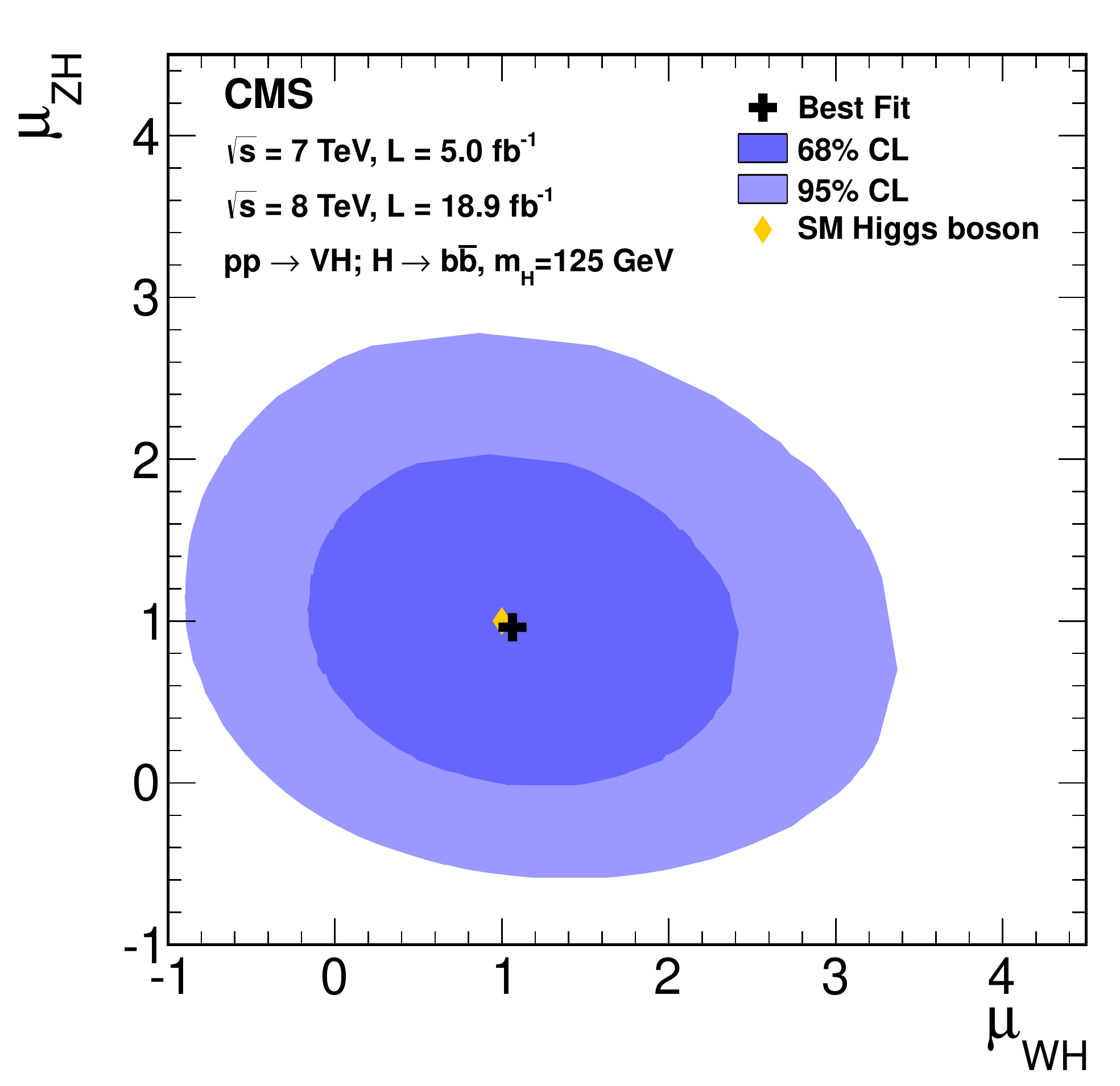}
\caption{Left: The best-fit value of the production cross section for a 125\GeV
Higgs boson relative to the standard model cross section, \ie, signal
strength $\mu$, for partial
combinations of channels and for all channels combined (band). Right:
The best-fit values and the 68\% and 95\% CL
contour regions for the $\mu_{\ZH}$, $\mu_{\WH}$ signal strength parameters for a 125\GeV Higgs boson.}
    \label{fig:mu-values}
  \end{center}
\end{figure*}

\begin{figure*}[htbp]
\centering
\includegraphics[width=0.48\textwidth]{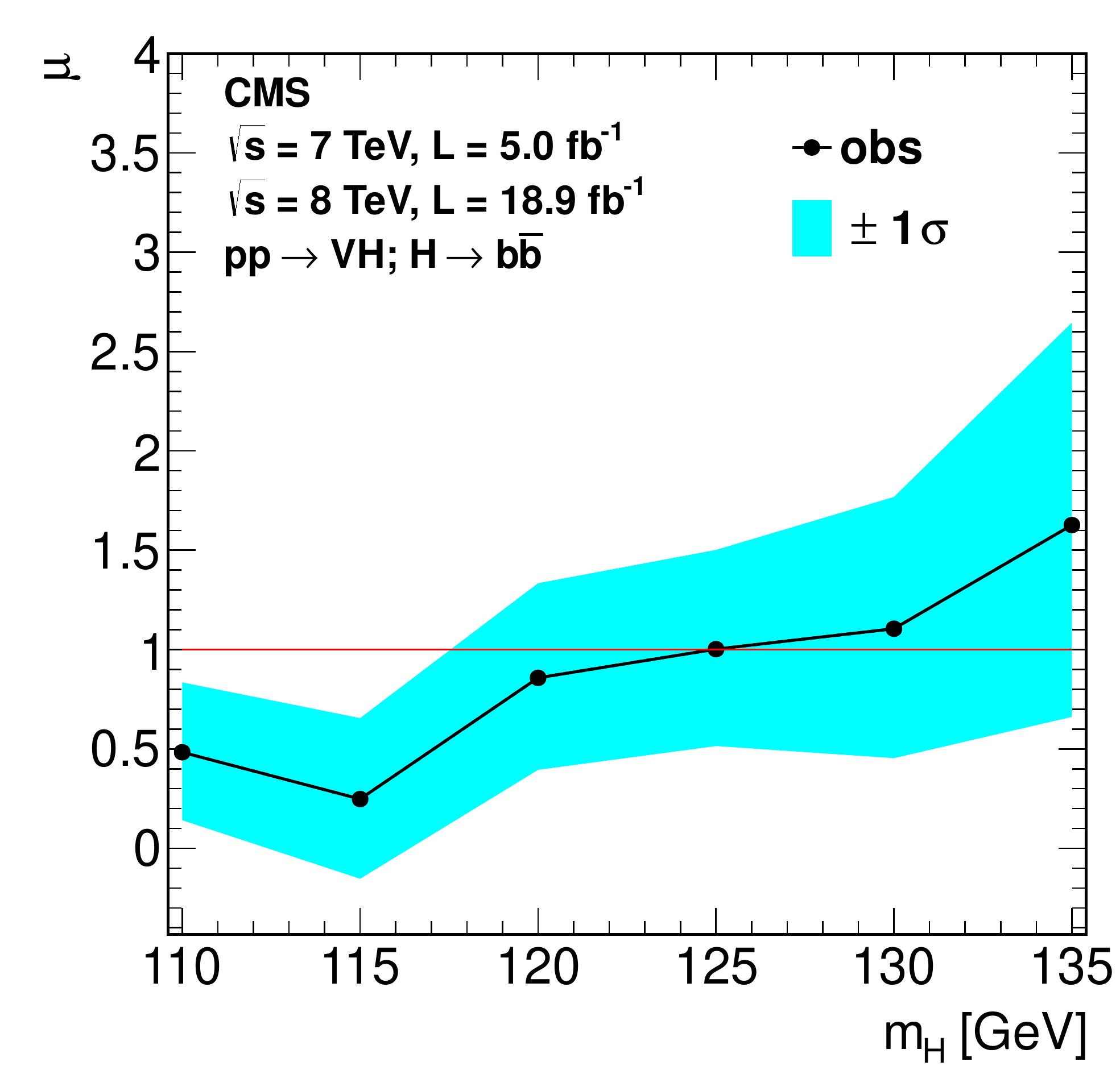}
\includegraphics[width=0.48\textwidth]{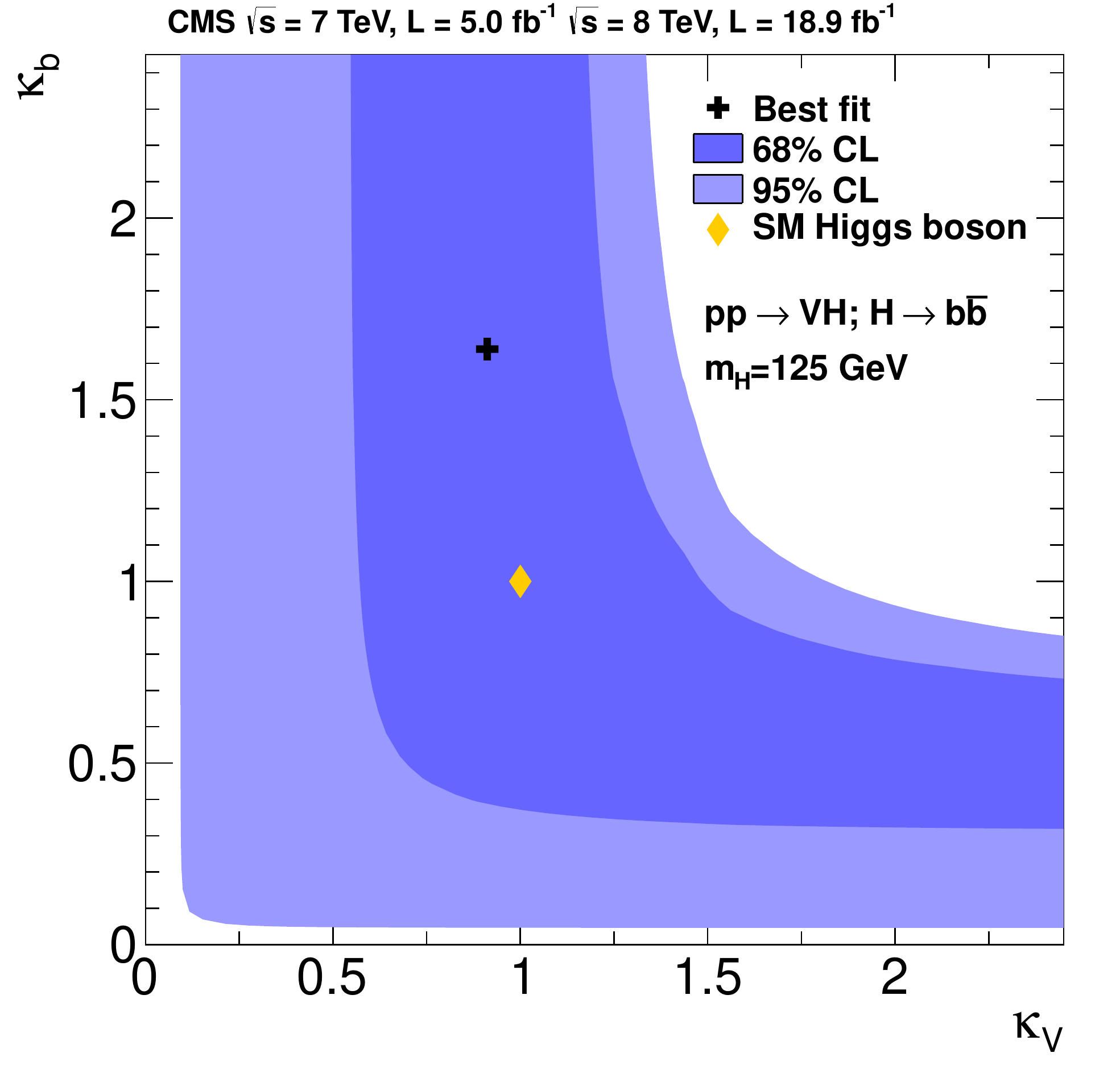}
\caption{Left: Signal strength for all channels combined as a
   function of the value assumed for the Higgs boson mass. Right: The
   best-fit values and the 68\% and 95\% CL contour
regions for the  $\kappa_\Vvar$ and $\kappa_\cPqb$
   parameters. The cross indicates the best-fit
values and the  yellow diamond shows the SM point  $(\kappa_\Vvar,
\kappa_\cPqb) = (1, 1)$. The likelihood fit is performed in the
 positive quadrant only.
}
    \label{fig:mu-values1}
\end{figure*}

The best-fit values of the production cross section for a 125\GeV
Higgs boson, relative to the standard model cross section (signal
strength, $\mu$),  are shown in the left panel of Fig.~\ref{fig:mu-values}
for the \WlnH\ and \WtnH\ channels combined, for the \ZllH\ channels
combined, and for the \ZnnH\ channel.
The observed signal strengths are consistent
with each other, and the value for the signal strength for the
combination of all channels is $1.0\pm 0.5$. In the right
panel of Fig.~\ref{fig:mu-values} the correlation between the signal
strengths for the separate  \WH\ and \ZH\ production processes is
shown. The two production modes are consistent with the SM
expectation, within
uncertainties. This figure contains slightly different information than the
one on the left panel as some final states contain signal events that
originate from both \WH\ and \ZH\ production processes. The \WH\
process contributes approximately 20\%  of the Higgs boson signal
event yields in the \ZnnH\ channel, resulting from events in which the
lepton is outside the detector acceptance, and the \ZllH\ process
contributes less than 5\% to the \WlnH\ channel when one of the
leptons is outside the detector acceptance. The dependency of the combined
signal strength on the value assumed for the Higgs boson mass is
shown in the left panel of Fig.~\ref{fig:mu-values1}.

In the right panel of Fig.~\ref{fig:mu-values1} the best-fit values
for the $\kappa_\Vvar$ and $\kappa_\cPqb$ parameters are shown. The parameter $\kappa_\Vvar$
quantifies the ratio of the measured Higgs boson couplings to vector
bosons relative to the SM value. The parameter $\kappa_\cPqb$ quantifies
the ratio of the measured Higgs
boson partial width into \bbbar relative to the SM value.
They are defined as:  ${\kappa_\Vvar}^2 =
\left.\sigma_{\VH}\middle/\sigma^{\mathrm{SM}}_{\VH}\right. $
and $ {\kappa_\cPqb}^2 = \left.\Gamma_{\bbbar}\middle/\Gamma^{\mathrm{SM}}_{\bbbar}\right. $,
with the SM scaling of the total
width~\cite{Heinemeyer:2013tqa}. The measured
couplings are consistent with the expectations from the standard
model, within uncertainties.

\subsection{Results for the dijet mass cross-check analysis}

\begin{figure*}[htbp]
  \begin{center}
   \includegraphics[width=0.48\textwidth]{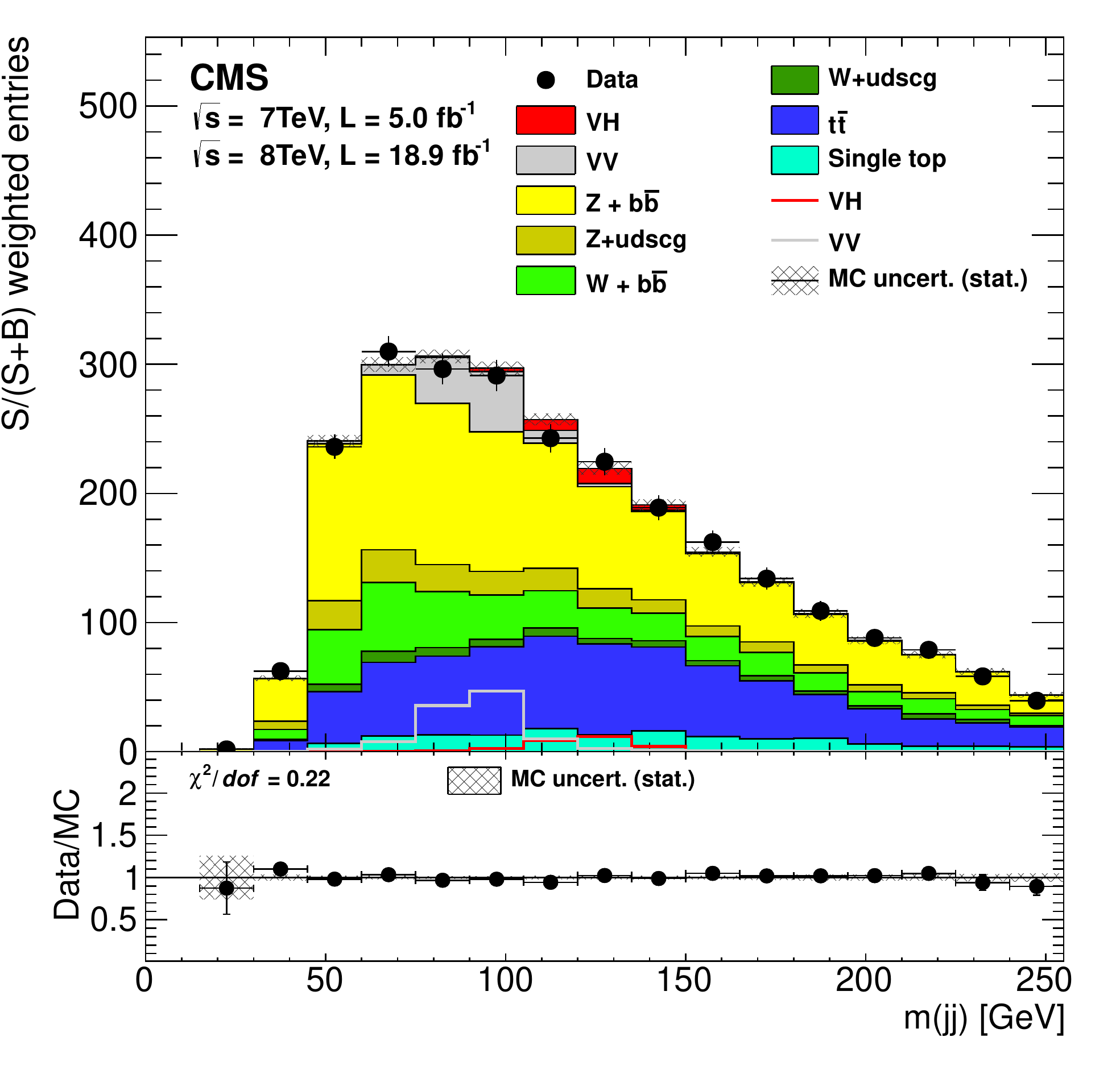}
   \includegraphics[width=0.48\textwidth]{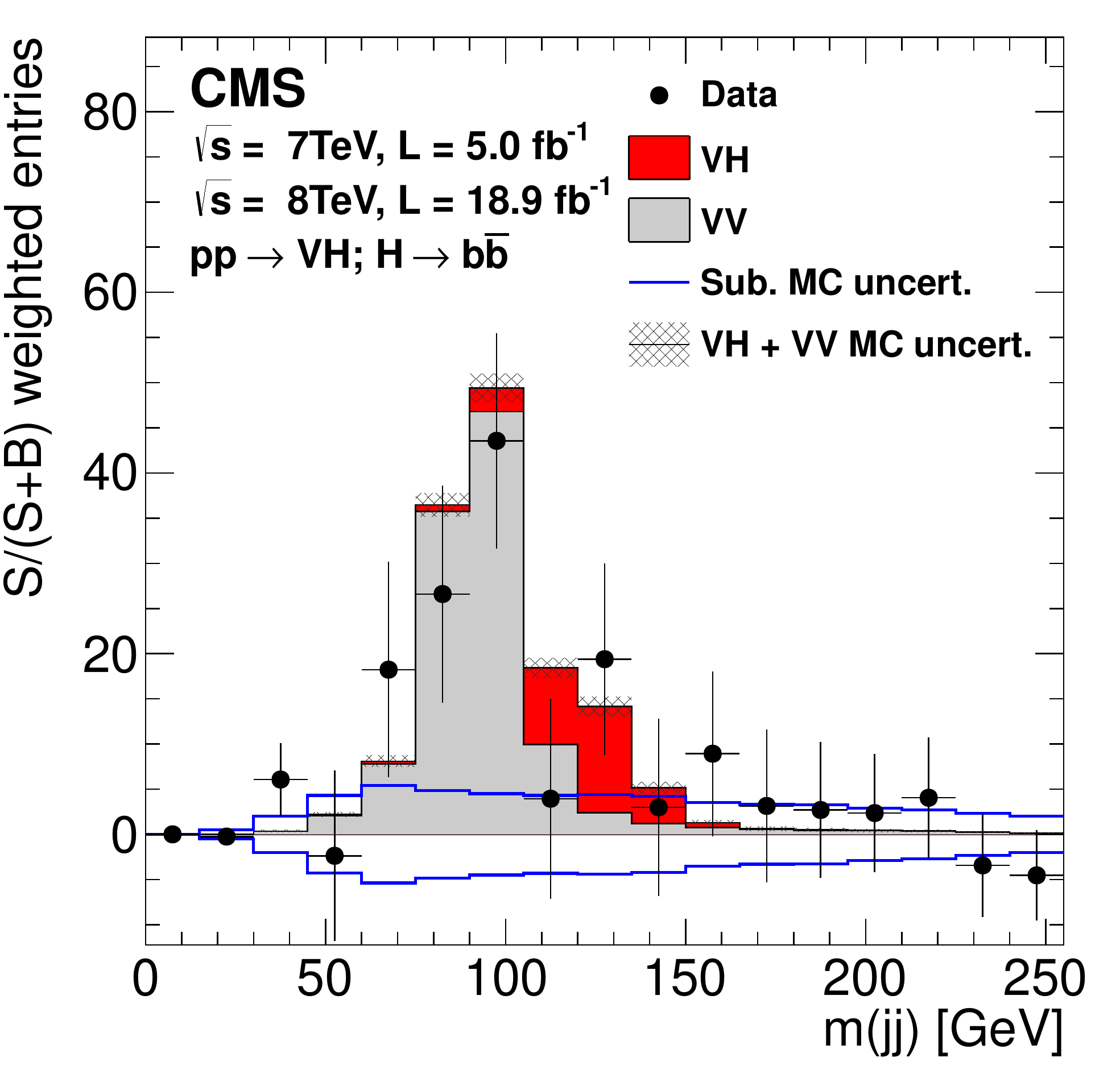}
\caption{Dijet mass cross-check analysis. Left: weighted dijet invariant mass distribution, combined for all
channels. For each channel, the relative dijet mass distribution weight
for each boost region is obtained from the ratio of the expected number of
signal events to the sum of expected signal and background events in a
window of \Mjj\ values between 105 and 150\GeV. The expected signal
used corresponds to the production of the SM Higgs boson with a mass of 125\GeV. The weight for the
highest-boost region is set to 1.0 and all other weights are adjusted
proportionally. The solid histograms for the
backgrounds and the signal are summed cumulatively. The line histogram for
signal and for {\Vvar}{\Vvar} backgrounds are also shown superimposed. The data is
represented by points. The bottom inset shows the ratio of the number of events observed in data to that
of the Monte Carlo prediction for signal and backgrounds. Right: same distribution with all backgrounds, except
dibosons, subtracted.}
    \label{fig:MJJ-combined}
  \end{center}
\end{figure*}

The left panel of Fig.~\ref{fig:MJJ-combined} shows a weighted dijet invariant mass distribution
for the combination of all channels, in all
boost regions, in the
combined 7 and 8\TeV data, using the event selection for the \Mjj\
cross-check analysis described in
Section~\ref{sec:hbb_Event_Selection}. For each channel, the relative
event weight in each boost region is obtained from the ratio of the expected number of
signal events to the sum of expected signal and background events in a
window of \Mjj\ values between 105 and 150\GeV. The expected signal
used corresponds to the production of the SM Higgs boson with a mass of
125\GeV. The weight for the
highest-boost region is set to 1.0 and all other weights are adjusted proportionally.
Figure~\ref{fig:MJJ-combined} also shows the same weighted
dijet invariant mass
distribution with all backgrounds, except diboson production, subtracted. The
data are consistent with the presence of a diboson signal from $\cPZ\cPZ$
and \WZ channels, with
$\cPZ\to \bbbar$), with a rate
consistent with the standard model prediction from the {\MADGRAPH}
generator, together with a
small excess consistent with the production
of the standard model Higgs boson with a mass of 125\GeV. For the \Mjj\ analysis, a
fit to the dijet invariant mass distribution results in a measured Higgs boson signal
strength, relative to that predicted by the standard model, of
$\mu = 0.8\pm 0.7$, with a local significance of 1.1 standard deviations
with respect to the background-only hypothesis.
For a Higgs boson of mass 125\GeV, the expected and observed 95\% CL upper limits
on the production cross section, relative to the standard model
prediction, are 1.4 and 2.0, respectively.

\subsection{Diboson signal extraction}

As a validation of the multivariate technique, BDT discriminants are trained
using the diboson sample as signal, and all other processes, including
\VH production (at the predicted standard model rate for a 125\GeV
Higgs mass), as background. This is done for the 8\TeV dataset only.
The observed excess of events for the combined $\PW\cPZ$ and $\cPZ\cPZ$ processes, with
$\cPZ\to\bbbar$, differs by over 7 standard deviations from the event yield expectation
from the background-only hypothesis. The corresponding signal
strength, relative to
the prediction from the diboson {\MADGRAPH} generator mentioned in
Section~\ref{sec:hbb_Simulations},
and rescaled to the cross section from the NLO {\MCFM} generator,
is measured to be  $\mu_{\Vvar\Vvar} = {1.19}_{-0.23}^{+0.28}$.

\section{Summary}\label{sec:hbb_Conclusions}

 A search for the standard model Higgs boson when produced in association with an electroweak
   vector boson and decaying to
\bbbar is reported for the $\PW(\mu\nu)\PH$,
   $\PW(\Pe\nu)\PH$,
$\PW(\tau\nu)\PH$, $\cPZ(\mu\mu)\PH$, $\cPZ(\Pe\Pe)\PH$, and
   $\cPZ(\nu\nu)\PH$ channels. The search is performed in data samples corresponding to integrated
luminosities of up to 5.1\fbinv at $\sqrt{s}=7\TeV$ and up to
18.9\fbinv at $\sqrt{s}=8\TeV$, recorded by the CMS experiment
at the LHC.

Upper limits, at the 95\% confidence level, on the
   \VH production cross section times the $\PH \to \bbbar$
   branching fraction, with respect to the expectations for a standard
   model Higgs boson, are derived for the Higgs boson in the mass range
   110--135\GeV. For a Higgs
   boson mass of 125\GeV the expected limit is 0.95 and the observed
   limit is 1.89.

An excess of events is observed above the expected
   background with a local significance of 2.1 standard deviations. The
   expected significance when taking into account the production of the
   standard model Higgs boson is also 2.1 standard deviations. The sensitivity of this
   search, as represented by the expected significance, is the highest
   for a single experiment thus far.  The signal strength corresponding to this excess, relative to that of
   the standard model Higgs boson, is  $\mu = 1.0\pm 0.5$. The
   measurements presented in this article represent the first indication
   of the $\PH \to \bbbar$ decay at the LHC.

\section*{Acknowledgments}

\hyphenation{Bundes-ministerium Forschungs-gemeinschaft Forschungs-zentren} We congratulate our colleagues in the CERN accelerator departments for the excellent performance of the LHC and thank the technical and administrative staffs at CERN and at other CMS institutes for their contributions to the success of the CMS effort. In addition, we gratefully acknowledge the computing centres and personnel of the Worldwide LHC Computing Grid for delivering so effectively the computing infrastructure essential to our analyses. Finally, we acknowledge the enduring support for the construction and operation of the LHC and the CMS detector provided by the following funding agencies: the Austrian Federal Ministry of Science and Research and the Austrian Science Fund; the Belgian Fonds de la Recherche Scientifique, and Fonds voor Wetenschappelijk Onderzoek; the Brazilian Funding Agencies (CNPq, CAPES, FAPERJ, and FAPESP); the Bulgarian Ministry of Education and Science; CERN; the Chinese Academy of Sciences, Ministry of Science and Technology, and National Natural Science Foundation of China; the Colombian Funding Agency (COLCIENCIAS); the Croatian Ministry of Science, Education and Sport; the Research Promotion Foundation, Cyprus; the Ministry of Education and Research, Recurrent financing contract SF0690030s09 and European Regional Development Fund, Estonia; the Academy of Finland, Finnish Ministry of Education and Culture, and Helsinki Institute of Physics; the Institut National de Physique Nucl\'eaire et de Physique des Particules~/~CNRS, and Commissariat \`a l'\'Energie Atomique et aux \'Energies Alternatives~/~CEA, France; the Bundesministerium f\"ur Bildung und Forschung, Deutsche Forschungsgemeinschaft, and Helmholtz-Gemeinschaft Deutscher Forschungszentren, Germany; the General Secretariat for Research and Technology, Greece; the National Scientific Research Foundation, and National Office for Research and Technology, Hungary; the Department of Atomic Energy and the Department of Science and Technology, India; the Institute for Studies in Theoretical Physics and Mathematics, Iran; the Science Foundation, Ireland; the Istituto Nazionale di Fisica Nucleare, Italy; the Korean Ministry of Education, Science and Technology and the World Class University program of NRF, Republic of Korea; the Lithuanian Academy of Sciences; the Mexican Funding Agencies (CINVESTAV, CONACYT, SEP, and UASLP-FAI); the Ministry of Business, Innovation and Employment, New Zealand; the Pakistan Atomic Energy Commission; the Ministry of Science and Higher Education and the National Science Centre, Poland; the Funda\c{c}\~ao para a Ci\^encia e a Tecnologia, Portugal; JINR, Dubna; the Ministry of Education and Science of the Russian Federation, the Federal Agency of Atomic Energy of the Russian Federation, Russian Academy of Sciences, and the Russian Foundation for Basic Research; the Ministry of Education, Science and Technological Development of Serbia; the Secretar\'{\i}a de Estado de Investigaci\'on, Desarrollo e Innovaci\'on and Programa Consolider-Ingenio 2010, Spain; the Swiss Funding Agencies (ETH Board, ETH Zurich, PSI, SNF, UniZH, Canton Zurich, and SER); the National Science Council, Taipei; the Thailand Center of Excellence in Physics, the Institute for the Promotion of Teaching Science and Technology of Thailand, Special Task Force for Activating Research and the National Science and Technology Development Agency of Thailand; the Scientific and Technical Research Council of Turkey, and Turkish Atomic Energy Authority; the Science and Technology Facilities Council, UK; the US Department of Energy, and the US National Science Foundation.

Individuals have received support from the Marie-Curie programme and the European Research Council and EPLANET (European Union); the Leventis Foundation; the A. P. Sloan Foundation; the Alexander von Humboldt Foundation; the Belgian Federal Science Policy Office; the Fonds pour la Formation \`a la Recherche dans l'Industrie et dans l'Agriculture (FRIA-Belgium); the Agentschap voor Innovatie door Wetenschap en Technologie (IWT-Belgium); the Ministry of Education, Youth and Sports (MEYS) of Czech Republic; the Council of Science and Industrial Research, India; the Compagnia di San Paolo (Torino); the HOMING PLUS programme of Foundation for Polish Science, cofinanced by EU, Regional Development Fund; and the Thalis and Aristeia programmes cofinanced by EU-ESF and the Greek NSRF.

\bibliography{auto_generated}   

\providecommand{\href}[2]{#2}\begingroup\raggedright\begin{thebibliography}{10}%
\makeatletter
\providecommand{\hrefCMSnoop }[0]{\@secondoftwo}%
\makeatother
\providecommand{\doi}{\texttt{doi:}\begingroup \urlstyle{tt}\Url}

\bibitem{Chatrchyan:2012ufa}
\hrefCMSnoop {} {{ CMS} Collaboration, ``{Observation of a new boson at a mass
  of 125 GeV with the CMS experiment at the LHC}'',} \textit{ Phys. Lett. B}
  \textbf{ 716} (2012) 30,
  \href{http://dx.doi.org/10.1016/j.physletb.2012.08.021}{\doi{10.1016/j.physletb.2012.08.021}},
\href{http://www.arXiv.org/abs/1207.7235}{\texttt{ arXiv:1207.7235}}.

\bibitem{Aad:2012tfa}
\hrefCMSnoop {} {{ ATLAS} Collaboration, ``{Observation of a new particle in
  the search for the Standard Model Higgs boson with the ATLAS detector at the
  LHC}'',} \textit{ Phys. Lett. B} \textbf{ 716} (2012) 1,
  \href{http://dx.doi.org/10.1016/j.physletb.2012.08.020}{\doi{10.1016/j.physletb.2012.08.020}},
\href{http://www.arXiv.org/abs/1207.7214}{\texttt{ arXiv:1207.7214}}.

\bibitem{Englert:1964et}
\hrefCMSnoop {} {F.~Englert and R.~Brout, ``Broken symmetry and the mass of
  gauge vector mesons'',} \textit{ Phys. Rev. Lett.} \textbf{ 13} (1964) 321,
  \href{http://dx.doi.org/10.1103/PhysRevLett.13.321}{\doi{10.1103/PhysRevLett.13.321}}.

\bibitem{Higgs:1964ia}
\hrefCMSnoop {} {P.~W. Higgs, ``Broken symmetries, massless particles and gauge
  fields'',} \textit{ Phys. Lett.} \textbf{ 12} (1964) 132,
  \href{http://dx.doi.org/10.1016/0031-9163(64)91136-9}{\doi{10.1016/0031-9163(64)91136-9}}.

\bibitem{Higgs:1964pj}
\hrefCMSnoop {} {P.~W. Higgs, ``Broken symmetries and the masses of gauge
  bosons'',} \textit{ Phys. Rev. Lett.} \textbf{ 13} (1964) 508,
  \href{http://dx.doi.org/10.1103/PhysRevLett.13.508}{\doi{10.1103/PhysRevLett.13.508}}.

\bibitem{Guralnik:1964eu}
\hrefCMSnoop {} {G.~S. Guralnik, C.~R. Hagen, and T.~W.~B. Kibble, ``Global
  conservation laws and massless particles'',} \textit{ Phys. Rev. Lett.}
  \textbf{ 13} (1964) 585,
  \href{http://dx.doi.org/10.1103/PhysRevLett.13.585}{\doi{10.1103/PhysRevLett.13.585}}.

\bibitem{Higgs:1966ev}
\hrefCMSnoop {} {P.~W. Higgs, ``Spontaneous symmetry breakdown without massless
  bosons'',} \textit{ Phys. Rev.} \textbf{ 145} (1966) 1156,
  \href{http://dx.doi.org/10.1103/PhysRev.145.1156}{\doi{10.1103/PhysRev.145.1156}}.

\bibitem{Kibble:1967sv}
\hrefCMSnoop {} {T.~W.~B. Kibble, ``Symmetry breaking in non-{A}belian gauge
  theories'',} \textit{ Phys. Rev.} \textbf{ 155} (1967) 1554,
  \href{http://dx.doi.org/10.1103/PhysRev.155.1554}{\doi{10.1103/PhysRev.155.1554}}.

\bibitem{Dittmaier:2011ti}
\hrefCMSnoop {} {{ LHC Higgs Cross Section Working Group} Collaboration,
  ``{Handbook of LHC Higgs Cross Sections: 1. Inclusive Observables}'',}
  (2011).
\href{http://www.arXiv.org/abs/1101.0593}{\texttt{ arXiv:1101.0593}}.

\bibitem{PhysRevD.88.052014}
\hrefCMSnoop {} {{CDF and D0 Collaborations}, ``Higgs boson studies at the
  Tevatron'',} \textit{ Phys. Rev. D} \textbf{ 88} (2013) 052014,
  \href{http://dx.doi.org/10.1103/PhysRevD.88.052014}{\doi{10.1103/PhysRevD.88.052014}}.

\bibitem{PhysRevLett.109.071804}
\hrefCMSnoop {} {{CDF and D0 Collaborations}, ``{Evidence for a particle
  produced in association with weak bosons and decaying to a bottom-antibottom
  quark pair in Higgs boson searches at the Tevatron}'',} \textit{ Phys. Rev.
  Lett.} \textbf{ 109} (2012) 071804,
  \href{http://dx.doi.org/10.1103/PhysRevLett.109.071804}{\doi{10.1103/PhysRevLett.109.071804}},
\href{http://www.arXiv.org/abs/1207.6436}{\texttt{ arXiv:1207.6436}}.

\bibitem{Aad:2012gxa}
\hrefCMSnoop {} {{ ATLAS} Collaboration, ``{Search for the Standard Model Higgs
  boson produced in association with a vector boson and decaying to a $b$-quark
  pair with the ATLAS detector}'',} \textit{ Phys. Lett. B} \textbf{ 718}
  (2012) 369,
  \href{http://dx.doi.org/10.1016/j.physletb.2012.10.061}{\doi{10.1016/j.physletb.2012.10.061}},
\href{http://www.arXiv.org/abs/1207.0210}{\texttt{ arXiv:1207.0210}}.

\bibitem{Chatrchyan:2013lba}
\hrefCMSnoop {} {{ CMS} Collaboration, ``{Observation of a new boson with mass
  near 125 GeV in pp collisions at $\sqrt{s}=7\text{\,}\text{\,}\mathrm{TeV}$
  and $8\text{\,}\text{\,}\mathrm{TeV}$}'',} \textit{ JHEP} \textbf{ 06} (2013)
  081,
  \href{http://dx.doi.org/10.1007/JHEP06(2013)081}{\doi{10.1007/JHEP06(2013)081}},
\href{http://www.arXiv.org/abs/1303.4571}{\texttt{ arXiv:1303.4571}}.

\bibitem{Roe:2004na}
B.~P. Roe\hrefCMSnoop {} { {et~al.}, ``{Boosted decision trees, an alternative
  to artificial neural networks}'',} \textit{ Nucl. Instrum. Meth. A} \textbf{
  543} (2005) 577,
  \href{http://dx.doi.org/10.1016/j.nima.2004.12.018}{\doi{10.1016/j.nima.2004.12.018}},
\href{http://www.arXiv.org/abs/physics/0408124}{\texttt{
  arXiv:physics/0408124}}.

\bibitem{Hocker:2007ht}
A.~Hocker\hrefCMSnoop {} { {et~al.}, ``{TMVA---Toolkit for Multivariate Data
  Analysis}'',} \textit{ PoS} \textbf{ ACAT} (2007) 040,
\href{http://www.arXiv.org/abs/physics/0703039}{\texttt{
  arXiv:physics/0703039}}.

\bibitem{Chatrchyan:2008aa}
\hrefCMSnoop {} {{ CMS} Collaboration, ``{The CMS experiment at the CERN
  LHC}'',} \textit{ JINST} \textbf{ 3} (2008) S08004,
\href{http://dx.doi.org/10.1088/1748-0221/3/08/S08004}{\doi{10.1088/1748-0221/3/08/S08004}}.

\bibitem{GEANT4}
\hrefCMSnoop {} {{ GEANT4} Collaboration, ``{GEANT4}---a simulation toolkit'',}
  \textit{ Nucl. Instrum. Meth. A} \textbf{ 506} (2003) 250,
\href{http://dx.doi.org/10.1016/S0168-9002(03)01368-8}{\doi{10.1016/S0168-9002(03)01368-8}}.

\bibitem{POWHEG}
\hrefCMSnoop {} {{S. Frixione, P. Nason, and C. Oleari}, ``Matching NLO QCD
  computations with parton shower simulations: the POWHEG method'',} \textit{
  JHEP} \textbf{ 11} (2007) 070,
  \href{http://dx.doi.org/10.1088/1126-6708/2007/11/070}{\doi{10.1088/1126-6708/2007/11/070}},
  \href{http://www.arXiv.org/abs/0709.2092}{\texttt{ arXiv:0709.2092}}.

\bibitem{Alwall:2011uj}
J.~Alwall\hrefCMSnoop {} { {et~al.}, ``{MadGraph 5: going beyond}'',} \textit{
  JHEP} \textbf{ 06} (2011) 128,
  \href{http://dx.doi.org/10.1007/JHEP06(2011)128}{\doi{10.1007/JHEP06(2011)128}},
\href{http://www.arXiv.org/abs/1106.0522}{\texttt{ arXiv:1106.0522}}.

\bibitem{Sjostrand:2006za}
\hrefCMSnoop {} {T.~Sj{\"o}strand, S.~Mrenna, and P.~Z. Skands, ``{PYTHIA} 6.4
  physics and manual'',} \textit{ JHEP} \textbf{ 05} (2006) 026,
  \href{http://dx.doi.org/10.1088/1126-6708/2006/05/026}{\doi{10.1088/1126-6708/2006/05/026}},
\href{http://www.arXiv.org/abs/hep-ph/0603175}{\texttt{ arXiv:hep-ph/0603175}}.

\bibitem{Campbell:2010ff}
\hrefCMSnoop {} {J.~M. Campbell and R.~K. Ellis, ``{MCFM for the Tevatron and
  the LHC}'',} \textit{ Nucl. Phys. Proc. Suppl.} \textbf{ 205-206} (2010) 10,
  \href{http://dx.doi.org/10.1016/j.nuclphysbps.2010.08.011}{\doi{10.1016/j.nuclphysbps.2010.08.011}},
\href{http://www.arXiv.org/abs/1007.3492}{\texttt{ arXiv:1007.3492}}.

\bibitem{Gavin:2010az}
\hrefCMSnoop {} {R.~Gavin, Y.~Li, F.~Petriello, and S.~Quackenbush, ``{FEWZ
  2.0: A code for hadronic Z production at next-to-next-to-leading order}'',}
  \textit{ Comput. Phys. Commun.} \textbf{ 182} (2011) 2388,
  \href{http://dx.doi.org/10.1016/j.cpc.2011.06.008}{\doi{10.1016/j.cpc.2011.06.008}},
\href{http://www.arXiv.org/abs/1011.3540}{\texttt{ arXiv:1011.3540}}.

\bibitem{Li:2012wna}
\hrefCMSnoop {} {Y.~Li and F.~Petriello, ``{Combining QCD and electroweak
  corrections to dilepton production in FEWZ}'',} \textit{ Phys. Rev. D}
  \textbf{ 86} (2012) 094034,
  \href{http://dx.doi.org/10.1103/PhysRevD.86.094034}{\doi{10.1103/PhysRevD.86.094034}},
\href{http://www.arXiv.org/abs/1208.5967}{\texttt{ arXiv:1208.5967}}.

\bibitem{Gavin:2012sy}
\hrefCMSnoop {} {R.~Gavin, Y.~Li, F.~Petriello, and S.~Quackenbush, ``{W
  Physics at the LHC with FEWZ 2.1}'',} \textit{ Comput. Phys. Commun.}
  \textbf{ 184} (2013) 208,
  \href{http://dx.doi.org/10.1016/j.cpc.2012.09.005}{\doi{10.1016/j.cpc.2012.09.005}},
\href{http://www.arXiv.org/abs/1201.5896}{\texttt{ arXiv:1201.5896}}.

\bibitem{Martin:2009iq}
\hrefCMSnoop {} {A.~D. Martin, W.~J. Stirling, R.~S. Thorne, and G.~Watt,
  ``{Parton distributions for the LHC}'',} \textit{ Eur. Phys. J. C} \textbf{
  63} (2009) 189,
  \href{http://dx.doi.org/10.1140/epjc/s10052-009-1072-5}{\doi{10.1140/epjc/s10052-009-1072-5}},
  \href{http://www.arXiv.org/abs/0901.0002}{\texttt{ arXiv:0901.0002}}.

\bibitem{Pumplin:2002vw}
J.~Pumplin\hrefCMSnoop {} { {et~al.}, ``{New generation of parton distributions
  with uncertainties from global QCD analysis}'',} \textit{ JHEP} \textbf{ 07}
  (2002) 012,
  \href{http://dx.doi.org/10.1088/1126-6708/2002/07/012}{\doi{10.1088/1126-6708/2002/07/012}},
\href{http://www.arXiv.org/abs/hep-ph/0201195}{\texttt{ arXiv:hep-ph/0201195}}.

\bibitem{Bahr:2008pv}
M.~B{\"a}hr\hrefCMSnoop {} { {et~al.}, ``{Herwig++} physics and manual'',}
  \textit{ Eur. Phys. J. C} \textbf{ 58} (2008) 639,
  \href{http://dx.doi.org/10.1140/epjc/s10052-008-0798-9}{\doi{10.1140/epjc/s10052-008-0798-9}},
\href{http://www.arXiv.org/abs/0803.0883}{\texttt{ arXiv:0803.0883}}.

\bibitem{Chatrchyan:2011id}
\hrefCMSnoop {} {{ {CMS}} Collaboration, ``{Measurement of the underlying event
  activity at the LHC with $\sqrt{s}= 7$ TeV and comparison with $\sqrt{s} =
  0.9$ TeV}'',} \textit{ JHEP} \textbf{ 09} (2011) 109,
  \href{http://dx.doi.org/10.1007/JHEP09(2011)109}{\doi{10.1007/JHEP09(2011)109}},
\href{http://www.arXiv.org/abs/1107.0330}{\texttt{ arXiv:1107.0330}}.

\bibitem{Jadach1991275}
\hrefCMSnoop {} {S.~Jadach, J.~H. K{\"u}hn, and Z.~W{\c{a}}s, ``TAUOLA---a
  library of Monte Carlo programs to simulate decays of polarized tau
  leptons'',} \textit{ Comput. Phys. Commun.} \textbf{ 64} (1991) 275,
  \href{http://dx.doi.org/10.1016/0010-4655(91)90038-M}{\doi{10.1016/0010-4655(91)90038-M}}.

\bibitem{CMS-PAS-PFT-09-001}
\href {http://cdsweb.cern.ch/record/1194487} {{ CMS} Collaboration,
  ``Particle--Flow Event Reconstruction in {CMS} and Performance for Jets,
  Taus, and {\MET}'',} CMS Physics Analysis Summary CMS-PAS-PFT-09-001, (2009).

\bibitem{CMS-PAS-PFT-10-002}
\href {http://cdsweb.cern.ch/record/1279341} {{ CMS} Collaboration,
  ``Commissioning of the Particle-flow Event Reconstruction in Minimum-Bias and
  Jet Events from pp Collisions at 7 TeV'',} CMS Physics Analysis Summary
  CMS-PAS-PFT-10-002, (2010).

\bibitem{Cacciari:subtraction}
\hrefCMSnoop {} {{M. Cacciari and G. P. Salam}, ``{Pileup subtraction using jet
  areas}'',} \textit{ Phys. Lett. B} \textbf{ 659} (2008) 119,
  \href{http://dx.doi.org/10.1016/j.physletb.2007.09.077}{\doi{10.1016/j.physletb.2007.09.077}},
  \href{http://www.arXiv.org/abs/0707.1378}{\texttt{ arXiv:0707.1378}}.

\bibitem{antikt}
\hrefCMSnoop {} {{M. Cacciari, G. P. Salam and G. Soyez}, ``{The anti-$k_t$ jet
  clustering algorithm}'',} \textit{ JHEP} \textbf{ 04} (2008) 063,
  \href{http://dx.doi.org/10.1088/1126-6708/2008/04/063}{\doi{10.1088/1126-6708/2008/04/063}},
  \href{http://www.arXiv.org/abs/0802.1189}{\texttt{ arXiv:0802.1189}}.

\bibitem{Cacciari:fastjet1}
\hrefCMSnoop {} {M.~Cacciari, G.~P. Salam, and G.~Soyez, ``{FastJet User
  Manual}'',} \textit{ Eur. Phys. J. C} \textbf{ 72} (2012) 1896,
  \href{http://dx.doi.org/10.1140/epjc/s10052-012-1896-2}{\doi{10.1140/epjc/s10052-012-1896-2}},
\href{http://www.arXiv.org/abs/1111.6097}{\texttt{ arXiv:1111.6097}}.

\bibitem{Cacciari:fastjet2}
\hrefCMSnoop {} {{M. Cacciari and G. P. Salam}, ``{Dispelling the $N^{3}$ myth
  for the $k_t$ jet-finder}'',} \textit{ Phys. Lett. B} \textbf{ 641} (2006)
  57,
  \href{http://dx.doi.org/10.1016/j.physletb.2006.08.037}{\doi{10.1016/j.physletb.2006.08.037}},
\href{http://www.arXiv.org/abs/hep-ph/0512210}{\texttt{ arXiv:hep-ph/0512210}}.

\bibitem{Chatrchyan:2011ds}
\hrefCMSnoop {} {{ CMS} Collaboration, ``Determination of jet energy
  calibration and transverse momentum resolution in {CMS}'',} \textit{ JINST}
  \textbf{ 6} (2011) P11002,
  \href{http://dx.doi.org/10.1088/1748-0221/6/11/P11002}{\doi{10.1088/1748-0221/6/11/P11002}},
\href{http://www.arXiv.org/abs/1107.4277}{\texttt{ arXiv:1107.4277}}.

\bibitem{Chatrchyan:2012xi}
\hrefCMSnoop {} {{ {CMS}} Collaboration, ``{Performance of CMS muon
  reconstruction in pp collision events at $\sqrt{7}$~\TeV}'',} \textit{ JINST}
  \textbf{ 7} (2012) P10002,
  \href{http://dx.doi.org/10.1088/1748-0221/7/10/P10002}{\doi{10.1088/1748-0221/7/10/P10002}},
\href{http://www.arXiv.org/abs/1206.4071}{\texttt{ arXiv:1206.4071}}.

\bibitem{CMS-PAS-EGM-10-004}
\href {http://cdsweb.cern.ch/record/1299116} {{ {CMS}} Collaboration,
  ``Electron reconstruction and identification at
  $\sqrt{s}=7\text{\,}\text{\,}\mathrm{TeV}$'',} CMS Physics Analysis Summary
  CMS-PAS-EGM-10-004, (2010).

\bibitem{Chatrchyan:2012zz}
\hrefCMSnoop {} {{ CMS} Collaboration, ``{Performance of tau-lepton
  reconstruction and identification in CMS}'',} \textit{ JINST} \textbf{ 7}
  (2012) P01001,
  \href{http://dx.doi.org/10.1088/1748-0221/7/01/P01001}{\doi{10.1088/1748-0221/7/01/P01001}},
\href{http://www.arXiv.org/abs/1109.6034}{\texttt{ arXiv:1109.6034}}.

\bibitem{Chatrchyan:2012jua}
\hrefCMSnoop {} {{ CMS} Collaboration, ``{Identification of b-quark jets with
  the CMS experiment}'',} \textit{ JINST} \textbf{ 8} (2013) P04013,
  \href{http://dx.doi.org/10.1088/1748-0221/8/04/P04013}{\doi{10.1088/1748-0221/8/04/P04013}},
\href{http://www.arXiv.org/abs/1211.4462}{\texttt{ arXiv:1211.4462}}.

\bibitem{PhysRevLett.100.242001}
\hrefCMSnoop {} {J.~M. Butterworth, A.~R. Davison, M.~Rubin, and G.~P. Salam,
  ``Jet Substructure as a New Higgs-Search Channel at the Large Hadron
  Collider'',} \textit{ Phys. Rev. Lett.} \textbf{ 100} (2008) 242001,
  \href{http://dx.doi.org/10.1103/PhysRevLett.100.242001}{\doi{10.1103/PhysRevLett.100.242001}}.

\bibitem{1107.3026}
T.~Aaltonen\hrefCMSnoop {} { {et~al.}, ``{Improved $b$-jet Energy Correction
  for $H \to b\bar{b}$ Searches at CDF}'',} (2011).
  \href{http://www.arXiv.org/abs/1107.3026}{\texttt{ arXiv:1107.3026}}.

\bibitem{Verkerke:2003ir}
\hrefCMSnoop {} {W.~Verkerke and D.~P. Kirkby, ``{The RooFit toolkit for data
  modeling}'',} in \textit{ Computing in High Energy and Nuclear Physics,
  CHEP03}.
\newblock 2003.
\newblock \href{http://www.arXiv.org/abs/physics/0306116}{\texttt{
  arXiv:physics/0306116}}.
\newblock
eConf/C0303241/MOLT007.

\bibitem{Aaltonen:2012id}
\hrefCMSnoop {} {{ CDF} Collaboration, ``{Search for the standard model Higgs
  boson decaying to a bb pair in events with two oppositely-charged leptons
  using the full CDF data set}'',} \textit{ Phys. Rev. Lett.} \textbf{ 109}
  (2012) 111803,
  \href{http://dx.doi.org/10.1103/PhysRevLett.109.111803}{\doi{10.1103/PhysRevLett.109.111803}},
\href{http://www.arXiv.org/abs/1207.1704}{\texttt{ arXiv:1207.1704}}.

\bibitem{Gallicchio:2010sw}
\hrefCMSnoop {} {J.~Gallicchio and M.~D. Schwartz, ``{Seeing in Color: Jet
  Superstructure}'',} \textit{ Phys. Rev. Lett.} \textbf{ 105} (2010) 022001,
  \href{http://dx.doi.org/10.1103/PhysRevLett.105.022001}{\doi{10.1103/PhysRevLett.105.022001}},
\href{http://www.arXiv.org/abs/1001.5027}{\texttt{ arXiv:1001.5027}}.

\bibitem{Aad:2013vka}
\hrefCMSnoop {} {{ ATLAS} Collaboration, ``{Measurement of the cross-section
  for W boson production in association with b-jets in pp collisions at
  $\sqrt{s}=7\text{\,}\text{\,}\mathrm{TeV}$ with the ATLAS detector}'',}
  \textit{ JHEP} \textbf{ 06} (2013) 084,
  \href{http://dx.doi.org/10.1007/JHEP06(2013)084}{\doi{10.1007/JHEP06(2013)084}},
\href{http://www.arXiv.org/abs/1302.2929}{\texttt{ arXiv:1302.2929}}.

\bibitem{Chatrchyan:2013zja}
\hrefCMSnoop {} {{CMS Collaboration}, ``{Measurement of the cross section and
  angular correlations for associated production of a Z boson with b hadrons in
  pp collisions at $\sqrt{s}$ = 7 TeV}'',} (2013).
  \href{http://www.arXiv.org/abs/1310.1349}{\texttt{ arXiv:1310.1349}}.
Submitted to JHEP.

\bibitem{Chatrchyan:2012vr}
\hrefCMSnoop {} {{ CMS} Collaboration, ``{Measurement of the
  Z/$\gamma^*$*+b-jet cross section in pp collisions at 7 TeV}'',} \textit{
  JHEP} \textbf{ 06} (2012) 126,
  \href{http://dx.doi.org/10.1007/JHEP06(2012)126}{\doi{10.1007/JHEP06(2012)126}},
\href{http://www.arXiv.org/abs/1204.1643}{\texttt{ arXiv:1204.1643}}.

\bibitem{CMS-PAS-SMP-12-008}
\href {http://cdsweb.cern.ch/record/1434360} {{ {CMS}} Collaboration,
  ``Absolute Calibration of the Luminosity Measurement at {CMS}: {W}inter 2012
  Update'',} CMS Physics Analysis Summary CMS-PAS-SMP-12-008, (2012).

\bibitem{CMS-PAS-LUM-12-001}
\href {http://cdsweb.cern.ch/record/1482193} {{ {CMS}} Collaboration, ``{CMS}
  Luminosity Measurement Based on Pixel Cluster Counting: {S}ummer 2012
  Update'',} CMS Physics Analysis Summary CMS-PAS-LUM-12-001, (2012).

\bibitem{Dittmaier:2012vm}
\hrefCMSnoop {} {S.~Dittmaier {et~al.}, ``{Handbook of LHC Higgs Cross
  Sections: 2. Differential Distributions}'',} CERN Report CERN-2012-002,
  (2012).

\bibitem{Alekhin:2011sk}
\hrefCMSnoop {} {S.~Alekhin {et~al.}, ``The {PDF4LHC Working Group} Interim
  Report'',} (2011). \href{http://www.arXiv.org/abs/1101.0536}{\texttt{
  arXiv:1101.0536}}.

\bibitem{Botje:2011sn}
\hrefCMSnoop {} {M.~Botje {et~al.}, ``The {PDF4LHC Working Group} Interim
  Recommendations'',} (2011).
  \href{http://www.arXiv.org/abs/1101.0538}{\texttt{ arXiv:1101.0538}}.

\bibitem{Lai:2010vv}
H.-L. Lai\hrefCMSnoop {} { {et~al.}, ``New parton distributions for collider
  physics'',} \textit{ Phys. Rev. D} \textbf{ 82} (2010) 074024,
  \href{http://dx.doi.org/10.1103/PhysRevD.82.074024}{\doi{10.1103/PhysRevD.82.074024}},
  \href{http://www.arXiv.org/abs/1007.2241}{\texttt{ arXiv:1007.2241}}.

\bibitem{Ball:2011mu}
R.~D. Ball\hrefCMSnoop {} { {et~al.}, ``{Impact of Heavy Quark Masses on Parton
  Distributions and LHC Phenomenology}'',} \textit{ Nucl. Phys. B} \textbf{
  849} (2011) 296,
  \href{http://dx.doi.org/10.1016/j.nuclphysb.2011.03.021}{\doi{10.1016/j.nuclphysb.2011.03.021}},
\href{http://www.arXiv.org/abs/1101.1300}{\texttt{ arXiv:1101.1300}}.

\bibitem{HAWK1}
\hrefCMSnoop {} {M.~Ciccolini, A.~Denner, and S.~Dittmaier, ``{Strong and
  electroweak corrections to the production of Higgs+2jets via weak
  interactions at the LHC}'',} \textit{ Phys. Rev. Lett.} \textbf{ 99} (2007)
  161803,
  \href{http://dx.doi.org/10.1103/PhysRevLett.99.161803}{\doi{10.1103/PhysRevLett.99.161803}},
  \href{http://www.arXiv.org/abs/0707.0381}{\texttt{ arXiv:0707.0381}}.

\bibitem{HAWK2}
\hrefCMSnoop {} {M.~Ciccolini, A.~Denner, and S.~Dittmaier, ``{Electroweak and
  QCD corrections to Higgs production via vector-boson fusion at the LHC}'',}
  \textit{ Phys. Rev. D} \textbf{ 77} (2008) 013002,
  \href{http://dx.doi.org/10.1103/PhysRevD.77.013002}{\doi{10.1103/PhysRevD.77.013002}},
  \href{http://www.arXiv.org/abs/0710.4749}{\texttt{ arXiv:0710.4749}}.

\bibitem{HAWK3}
\hrefCMSnoop {} {A.~Denner, S.~Dittmaier, S.~Kallweit, and A.~Muck,
  ``{Electroweak corrections to Higgs-strahlung off W/Z bosons at the Tevatron
  and the LHC with HAWK}'',} \textit{ JHEP} \textbf{ 03} (2012) 075,
  \href{http://dx.doi.org/10.1007/JHEP03(2012)075}{\doi{10.1007/JHEP03(2012)075}},
\href{http://www.arXiv.org/abs/1112.5142}{\texttt{ arXiv:1112.5142}}.

\bibitem{Ferrera:2011bk}
\hrefCMSnoop {} {G.~Ferrera, M.~Grazzini, and F.~Tramontano, ``{Associated WH
  production at hadron colliders: a fully exclusive QCD calculation at
  NNLO}'',} \textit{ Phys. Rev. Lett.} \textbf{ 107} (2011) 152003,
  \href{http://dx.doi.org/10.1103/PhysRevLett.107.152003}{\doi{10.1103/PhysRevLett.107.152003}},
\href{http://www.arXiv.org/abs/1107.1164}{\texttt{ arXiv:1107.1164}}.

\bibitem{Frixione:2002ik}
\hrefCMSnoop {} {S.~Frixione and B.~R. Webber, ``{Matching NLO QCD computations
  and parton shower simulations}'',} \textit{ JHEP} \textbf{ 06} (2002) 029,
  \href{http://dx.doi.org/10.1088/1126-6708/2002/06/029}{\doi{10.1088/1126-6708/2002/06/029}},
\href{http://www.arXiv.org/abs/hep-ph/0204244}{\texttt{ arXiv:hep-ph/0204244}}.

\bibitem{Chatrchyan:2012ep}
\hrefCMSnoop {} {{ CMS} Collaboration, ``{Measurement of the single-top-quark
  $t$-channel cross section in $pp$ collisions at $\sqrt{s}=7$ TeV}'',}
  \textit{ JHEP} \textbf{ 12} (2012) 035,
  \href{http://dx.doi.org/10.1007/JHEP12(2012)035}{\doi{10.1007/JHEP12(2012)035}},
\href{http://www.arXiv.org/abs/1209.4533}{\texttt{ arXiv:1209.4533}}.

\bibitem{Chatrchyan:2013oev}
\hrefCMSnoop {} {{ CMS} Collaboration, ``Measurement of the {$\PWp\PWm$} and
  {$\cPZ\cPZ$} production cross sections in pp collisions at {$\sqrt{s} =
  8\TeV$}'',} \textit{ Phys. Lett. B} \textbf{ 721} (2013) 190,
  \href{http://dx.doi.org/10.1016/j.physletb.2013.03.027}{\doi{10.1016/j.physletb.2013.03.027}}.

\bibitem{Read:2002hq}
\hrefCMSnoop {} {A.~L. Read, ``{Presentation of search results: The $CL_s$
  technique}'',} \textit{ J. Phys. G} \textbf{ 28} (2002) 2693,
\href{http://dx.doi.org/10.1088/0954-3899/28/10/313}{\doi{10.1088/0954-3899/28/10/313}}.

\bibitem{junkcls}
\hrefCMSnoop {} {T.~Junk, ``{Confidence level computation for combining
  searches with small statistics}'',} \textit{ Nucl. Instrum. Meth. A} \textbf{
  434} (1999) 435,
  \href{http://dx.doi.org/10.1016/S0168-9002(99)00498-2}{\doi{10.1016/S0168-9002(99)00498-2}}.

\bibitem{LHC-HCG}
\href {http://cdsweb.cern.ch/record/1379837} {{ATLAS and CMS Collaborations,
  LHC Higgs Combination Group}, ``Procedure for the {LHC} Higgs boson search
  combination in {S}ummer 2011'',} {ATL-PHYS-PUB 011-11/CMS NOTE 2011-005},
  (2011).

\bibitem{Heinemeyer:2013tqa}
\hrefCMSnoop {} {S.~Heinmeyer {et~al.}, ``{Handbook of LHC Higgs Cross
  Sections: 3. Higgs Properties}'',} CERN Report CERN-2013-004, (2013).

\end{thebibliography}\endgroup
\appendix

\section{Post-fit BDT distributions}\label{sec:hbb_Appendix}

\begin{figure*}[htbp]
\centering
\includegraphics[width=0.49\textwidth]{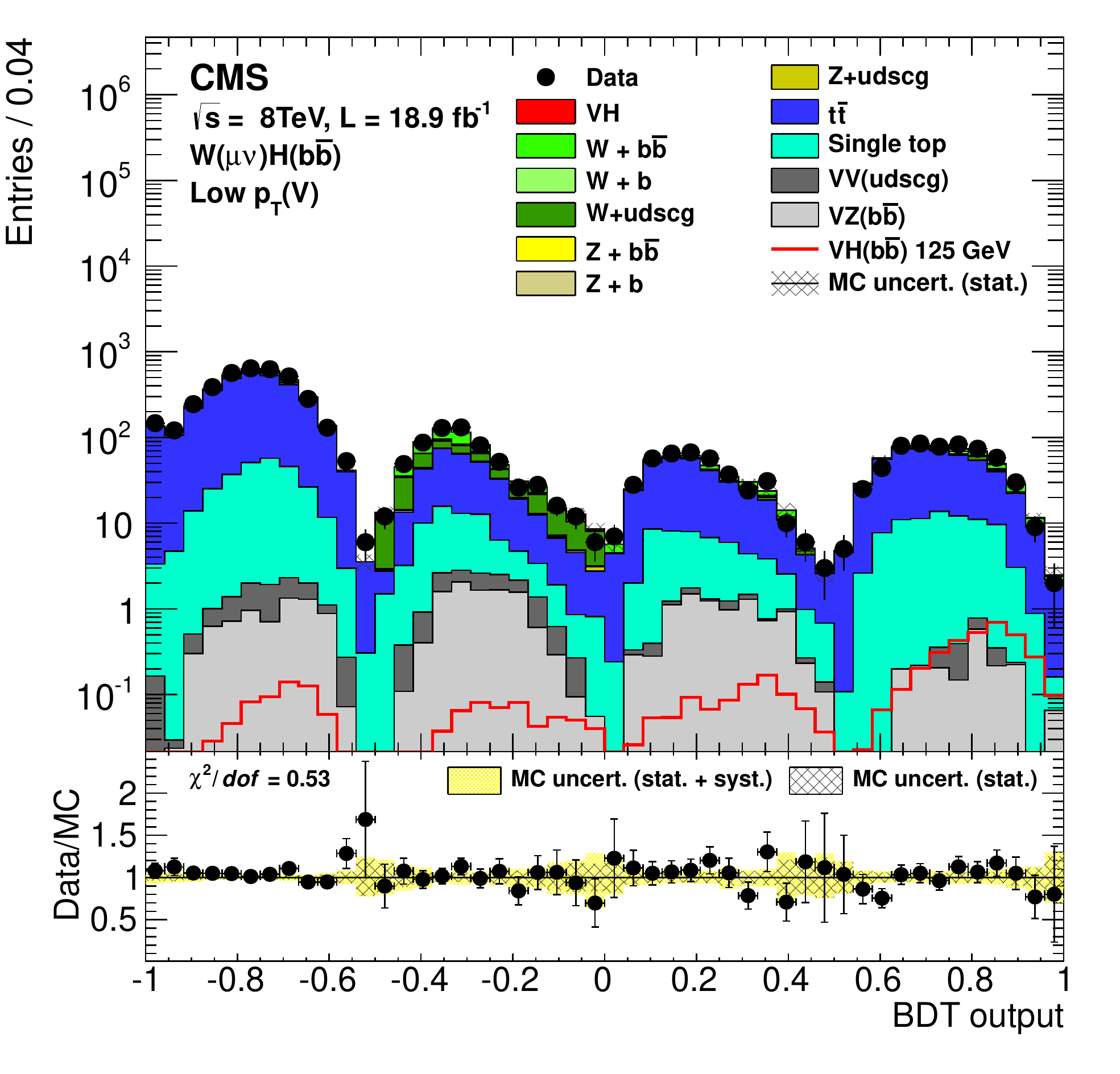}
\includegraphics[width=0.49\textwidth]{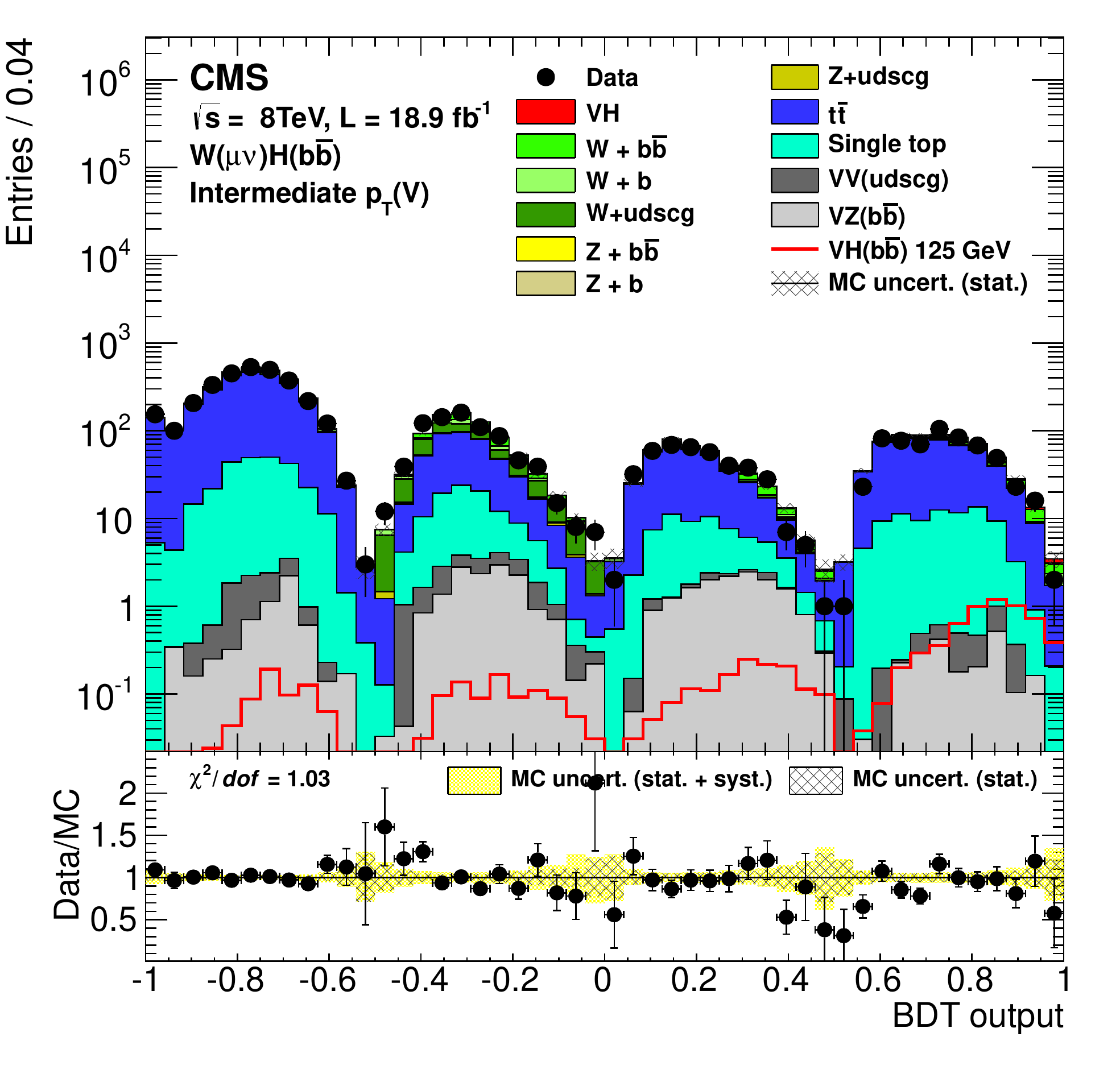}
\includegraphics[width=0.49\textwidth]{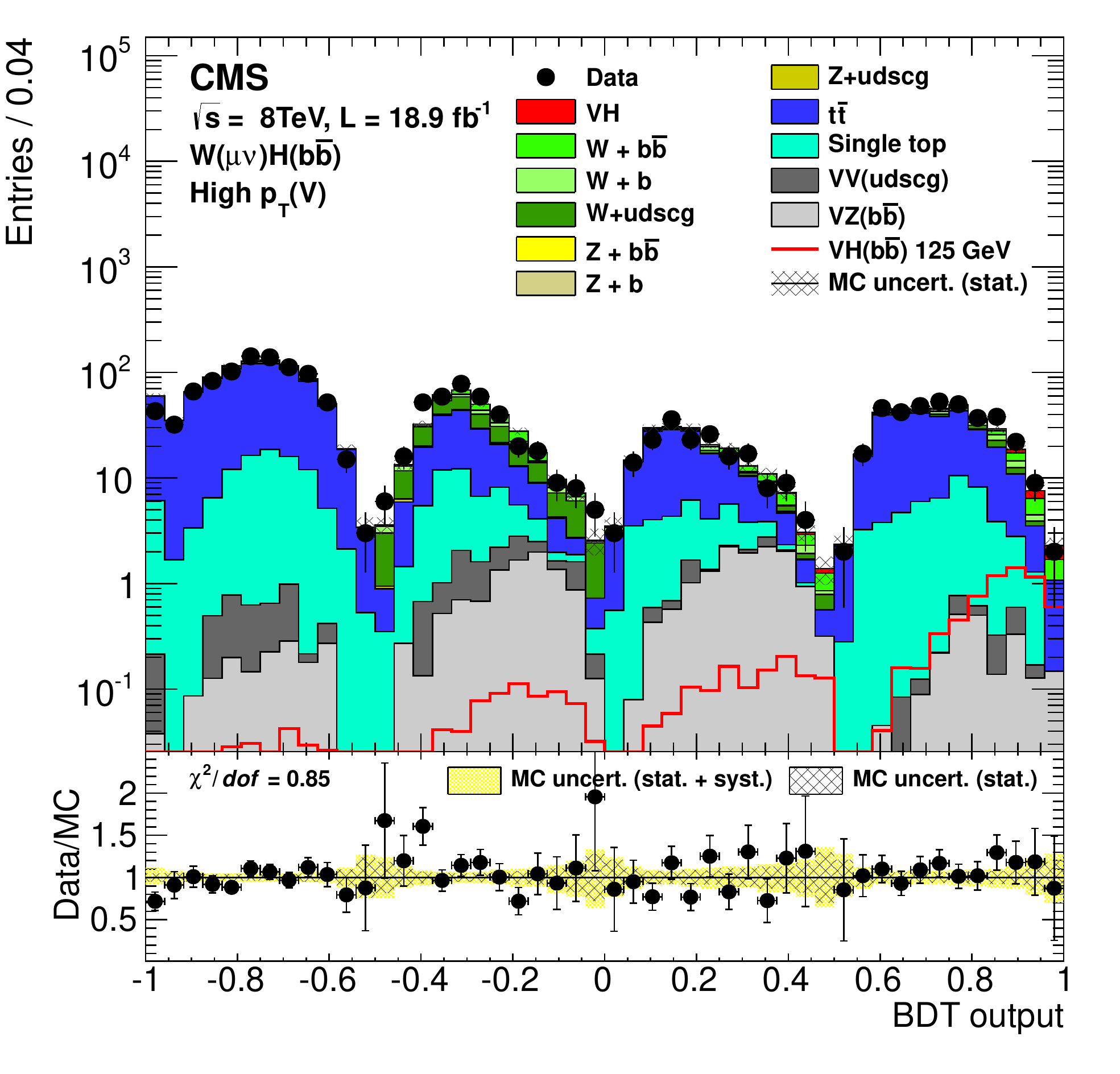}
\includegraphics[width=0.49\textwidth]{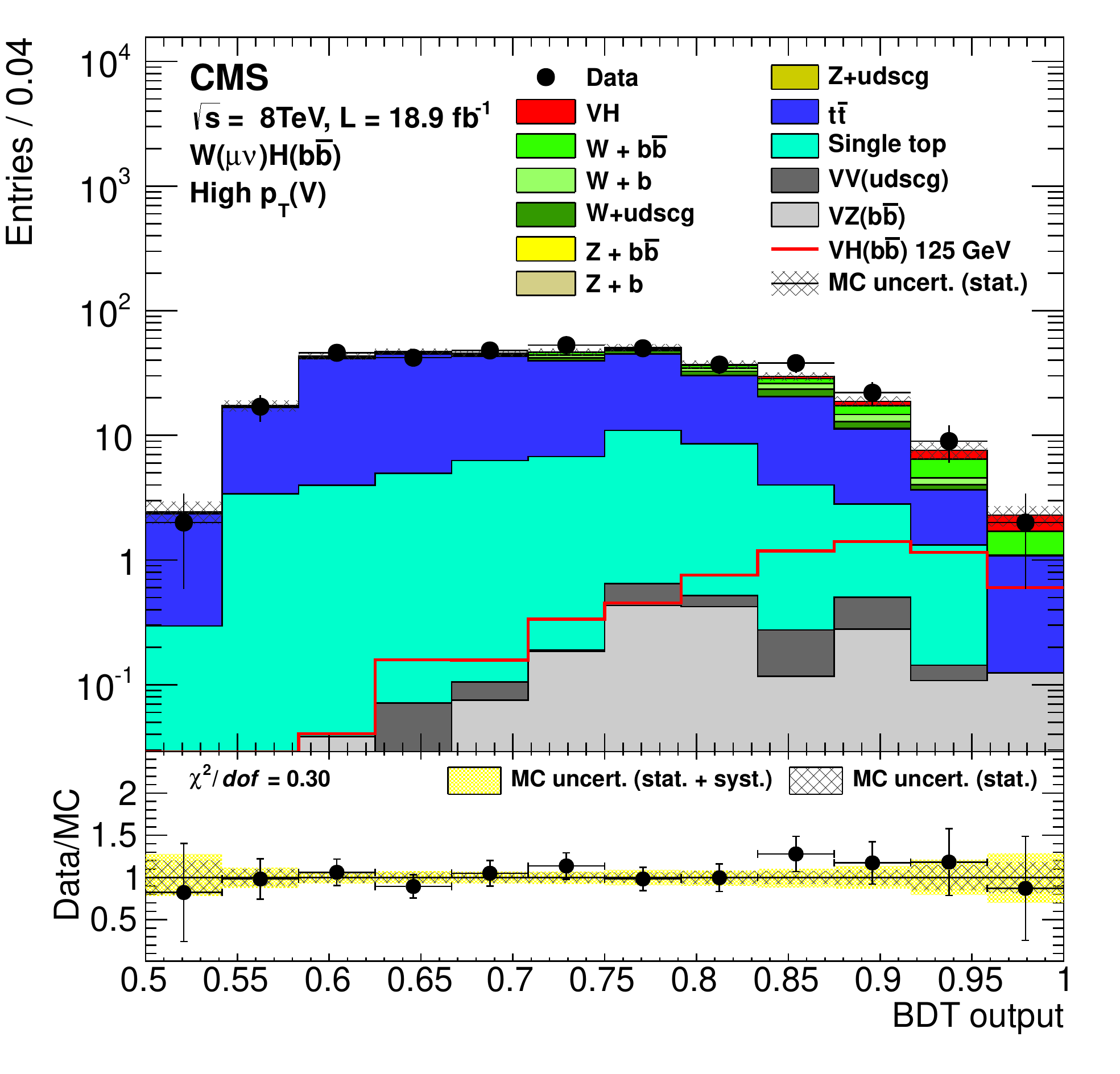}
   \caption{Post-fit BDT output distributions for \WmnH\ in the
    low-boost region (left),  the intermediate-boost (right), and the
    high-boost  (bottom), for 8\TeV data (points with error bars),
    all backgrounds, and signal, after all selection criteria have been
    applied. Bottom right: the \VH-enriched partition of the high-boost region is
    shown in more detail. The bottom inset in each
      figure shows the ratio of the number of events observed in data to that
      of the Monte Carlo prediction for signal and backgrounds.}
    \label{fig:BDTWln8TeV_mu}
\end{figure*}

\begin{figure*}[htbp]
\centering
\includegraphics[width=0.49\textwidth]{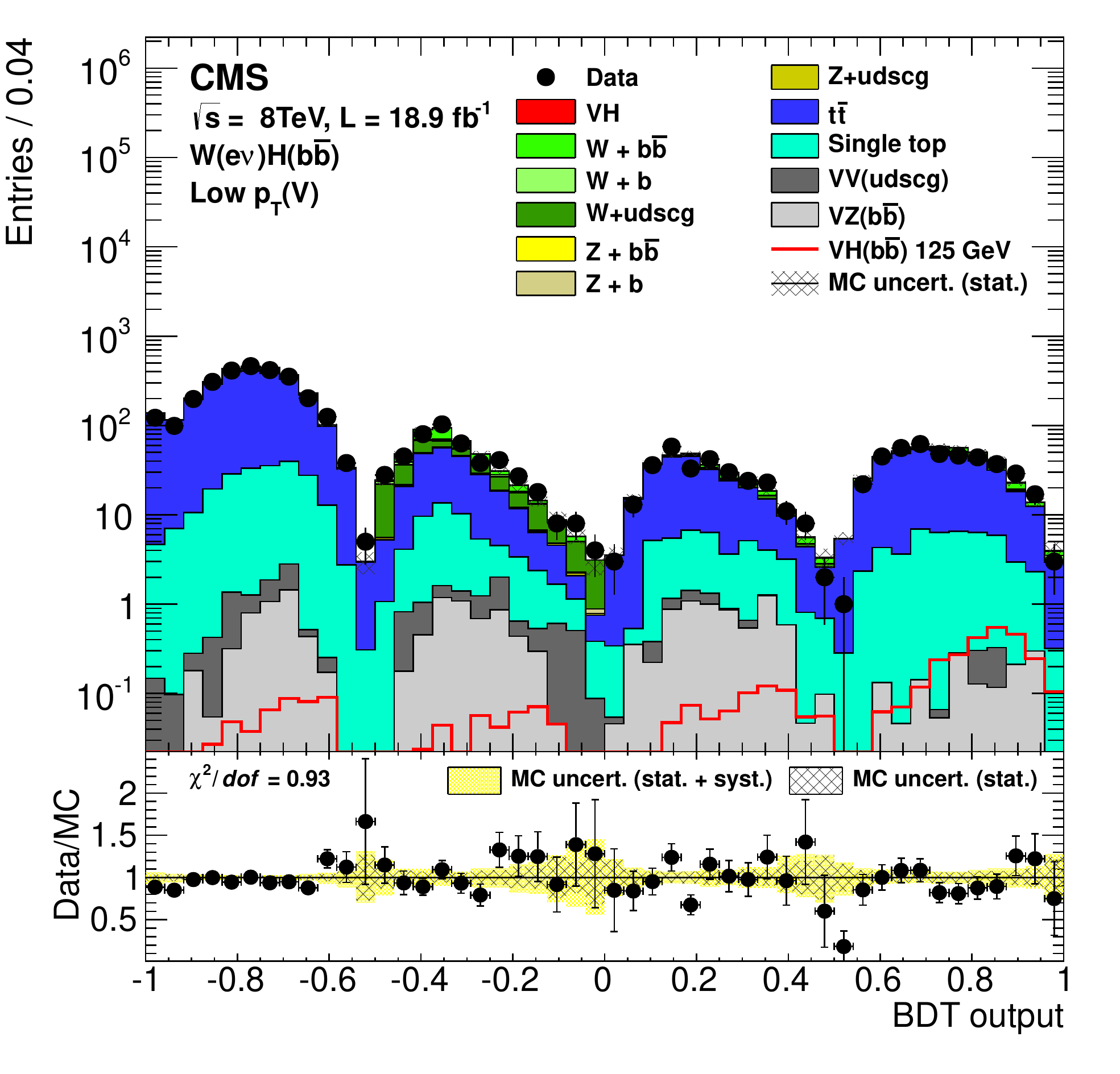}
\includegraphics[width=0.49\textwidth]{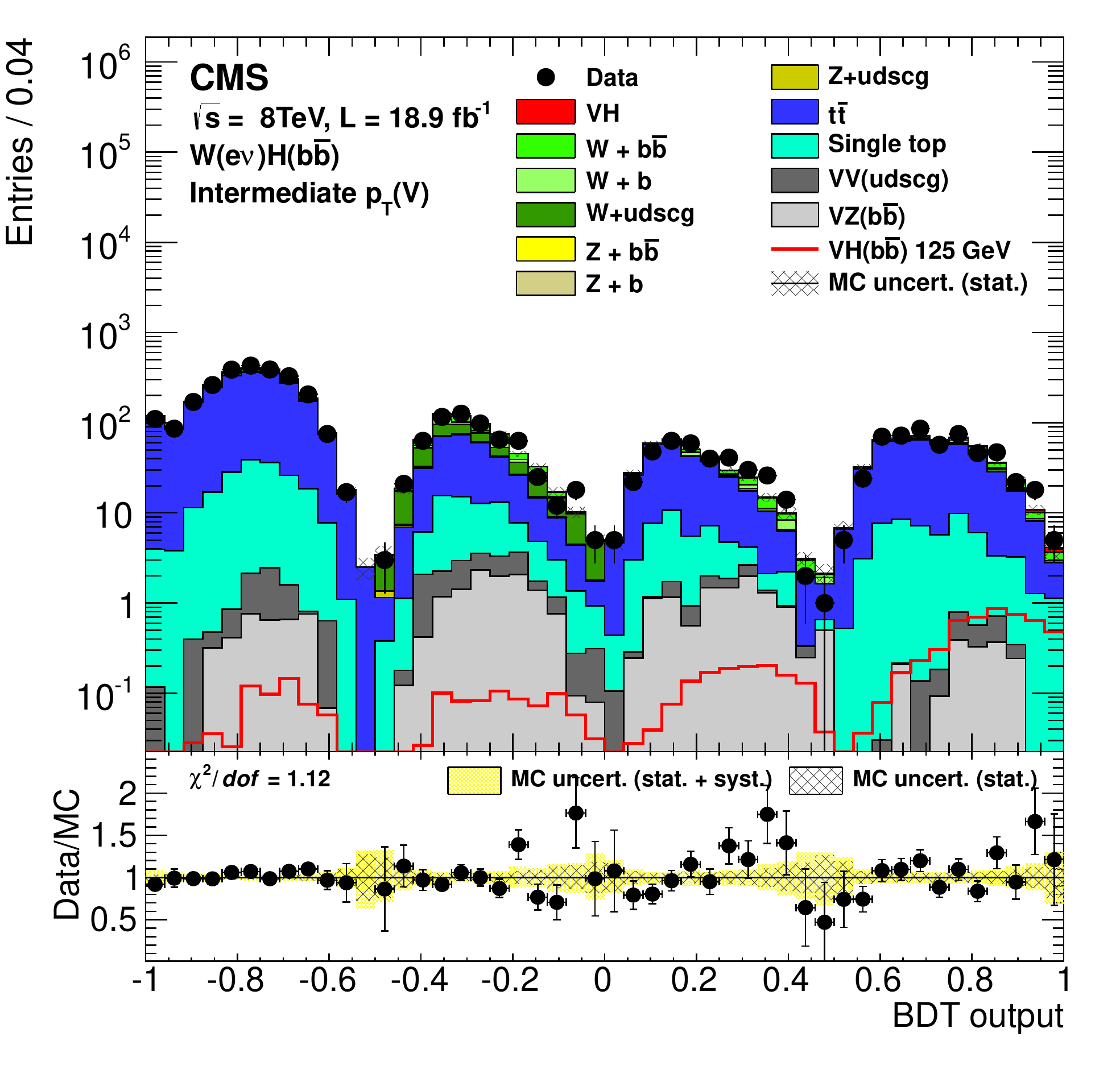}
\includegraphics[width=0.49\textwidth]{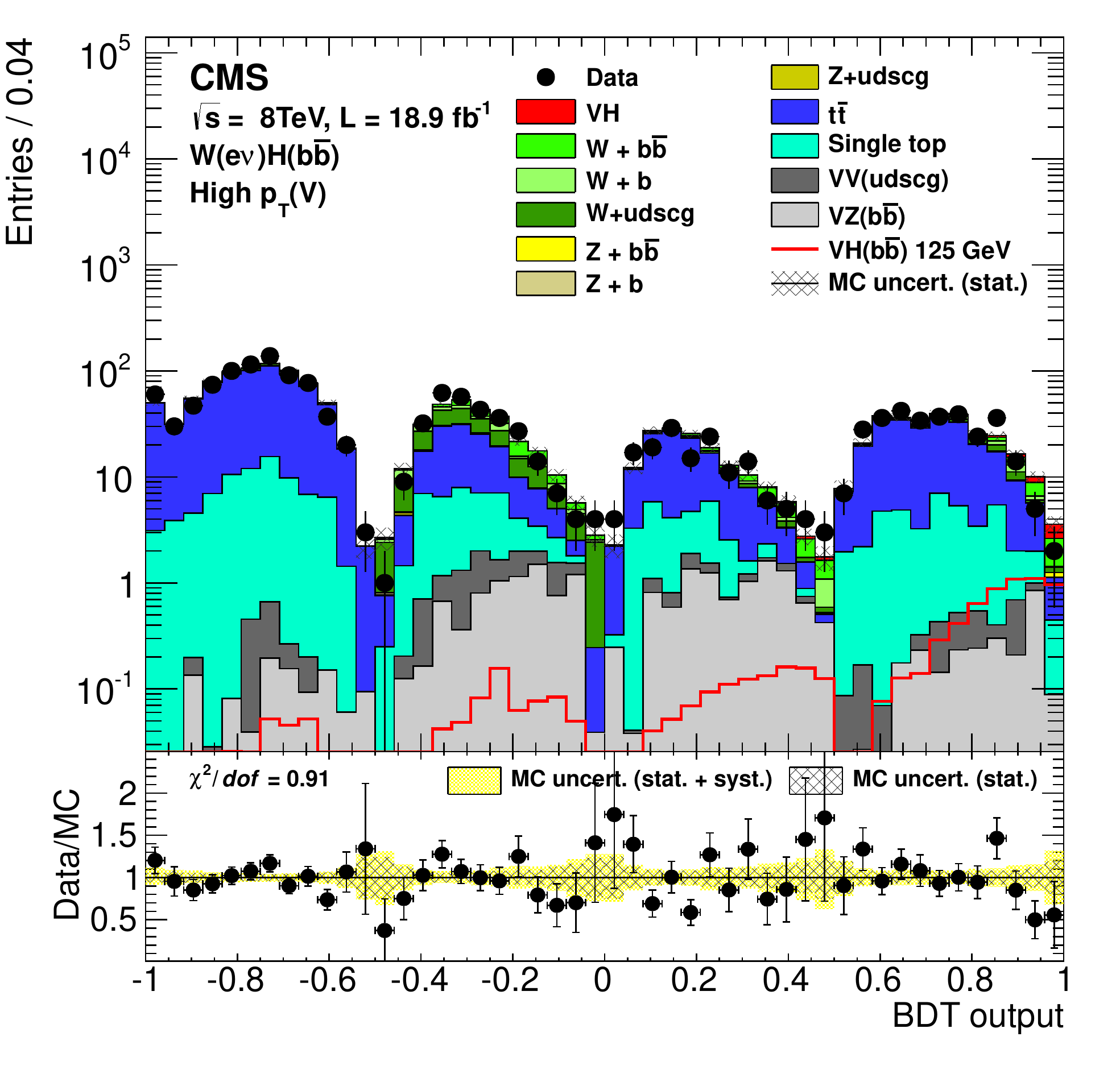}
\includegraphics[width=0.49\textwidth]{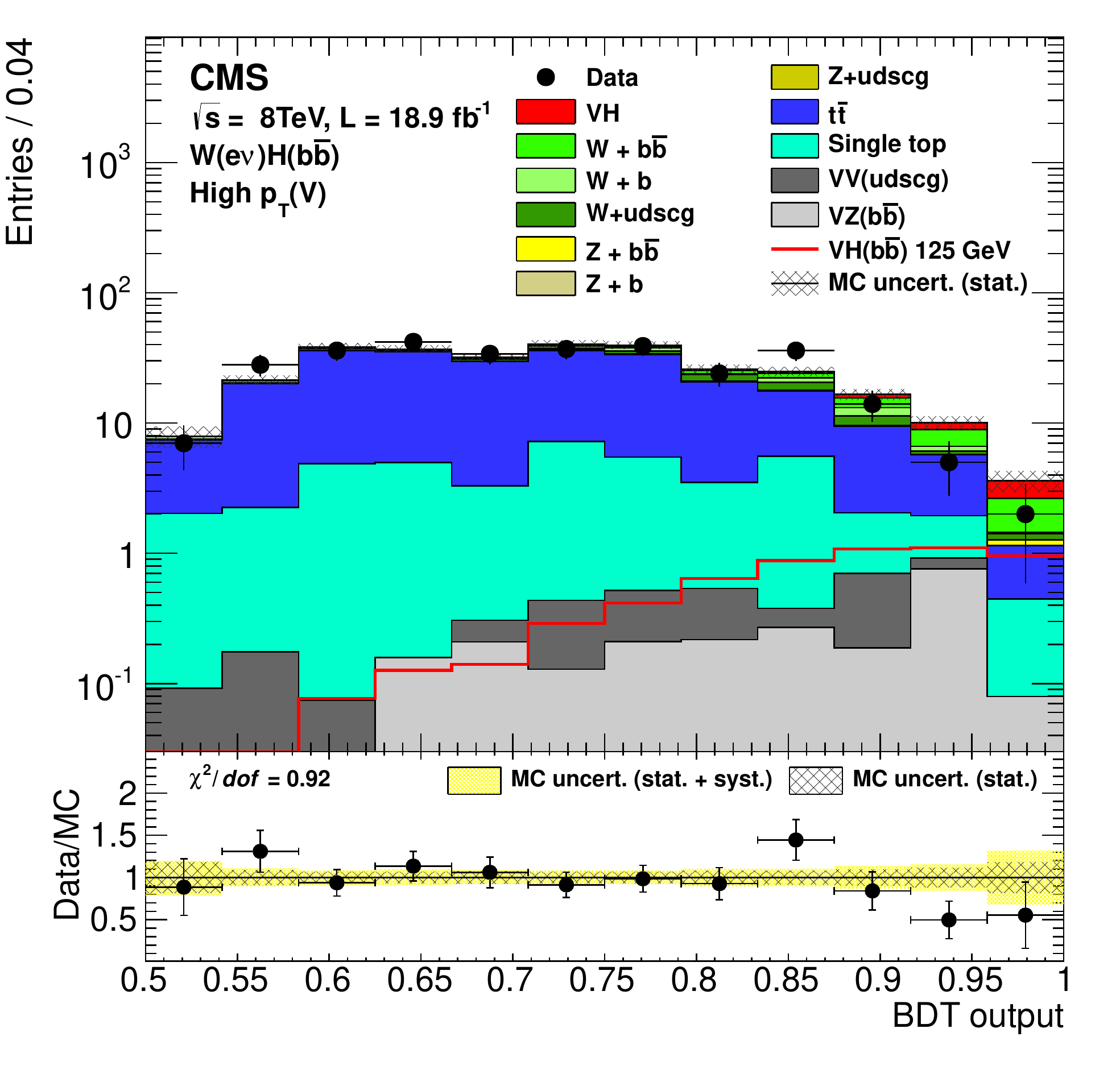}
   \caption{Post-fit BDT output distributions for \WenH\ in the
    low-boost region (left),  the intermediate-boost (right), and the
    high-boost  (bottom), for 8\TeV data (points with error bars),
    all backgrounds, and signal, after all selection criteria have been
    applied.  Bottom right: the \VH-enriched partition of the high-boost region
    is shown in more detail. The bottom inset in each
      figure shows the ratio of the number of events observed in data to that
      of the Monte Carlo prediction for signal and backgrounds.}
    \label{fig:BDTWln8TeV_e}
\end{figure*}

\begin{figure}[htbp]
\centering
\includegraphics[width=0.49\textwidth]{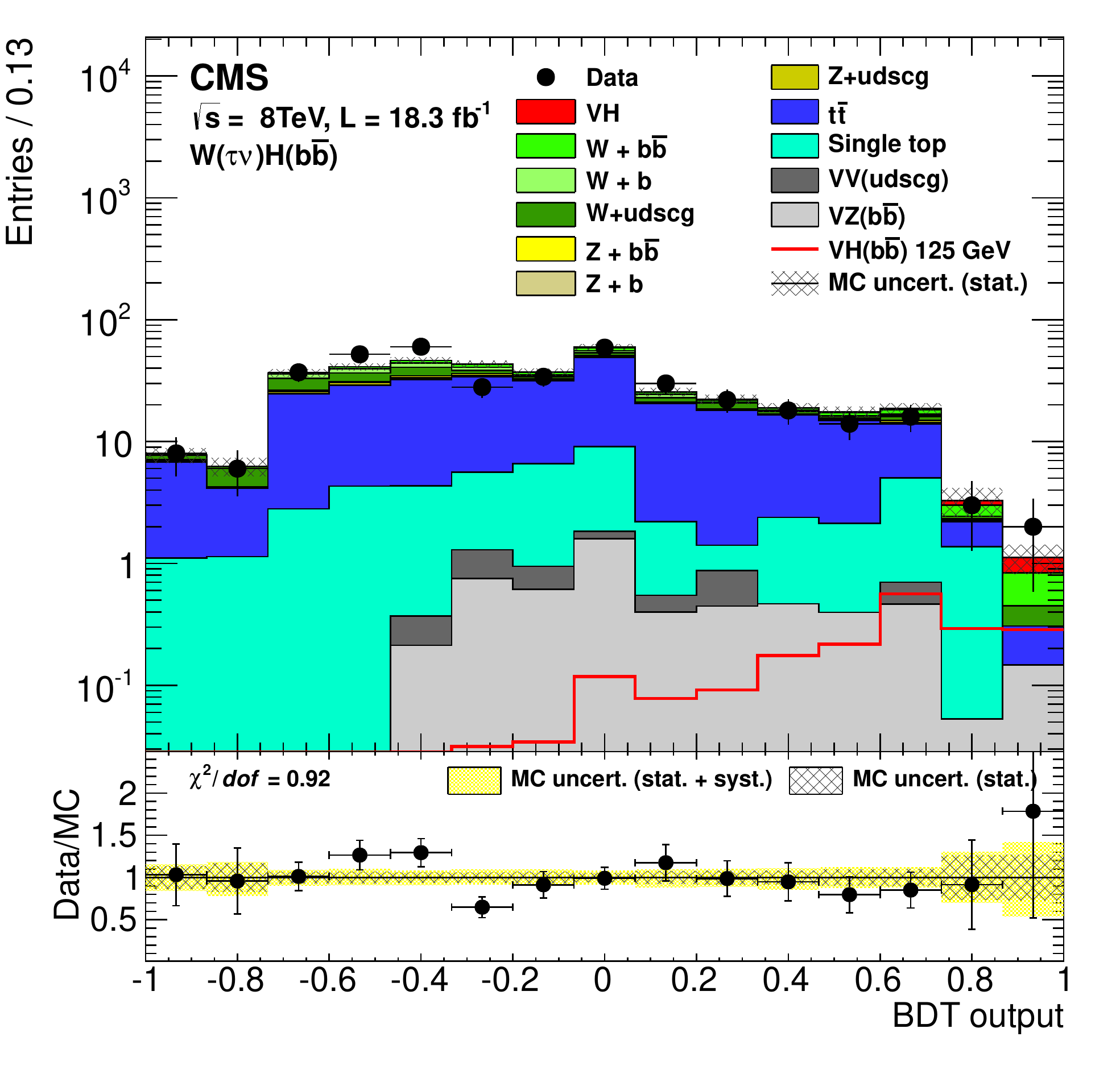}
  \caption{Post-fit BDT output distributions for \WtnH\ for 8\TeV data
    (points with error bars),
    all backgrounds, and signal, after all selection criteria have been
    applied. The bottom inset
      shows the ratio of the number of events observed in data to that
      of the Monte Carlo prediction for signal and backgrounds.}
    \label{fig:BDTWln8TeV_t}
\end{figure}

\begin{figure*}[htbp]
\centering
    \includegraphics[width=0.49\textwidth]{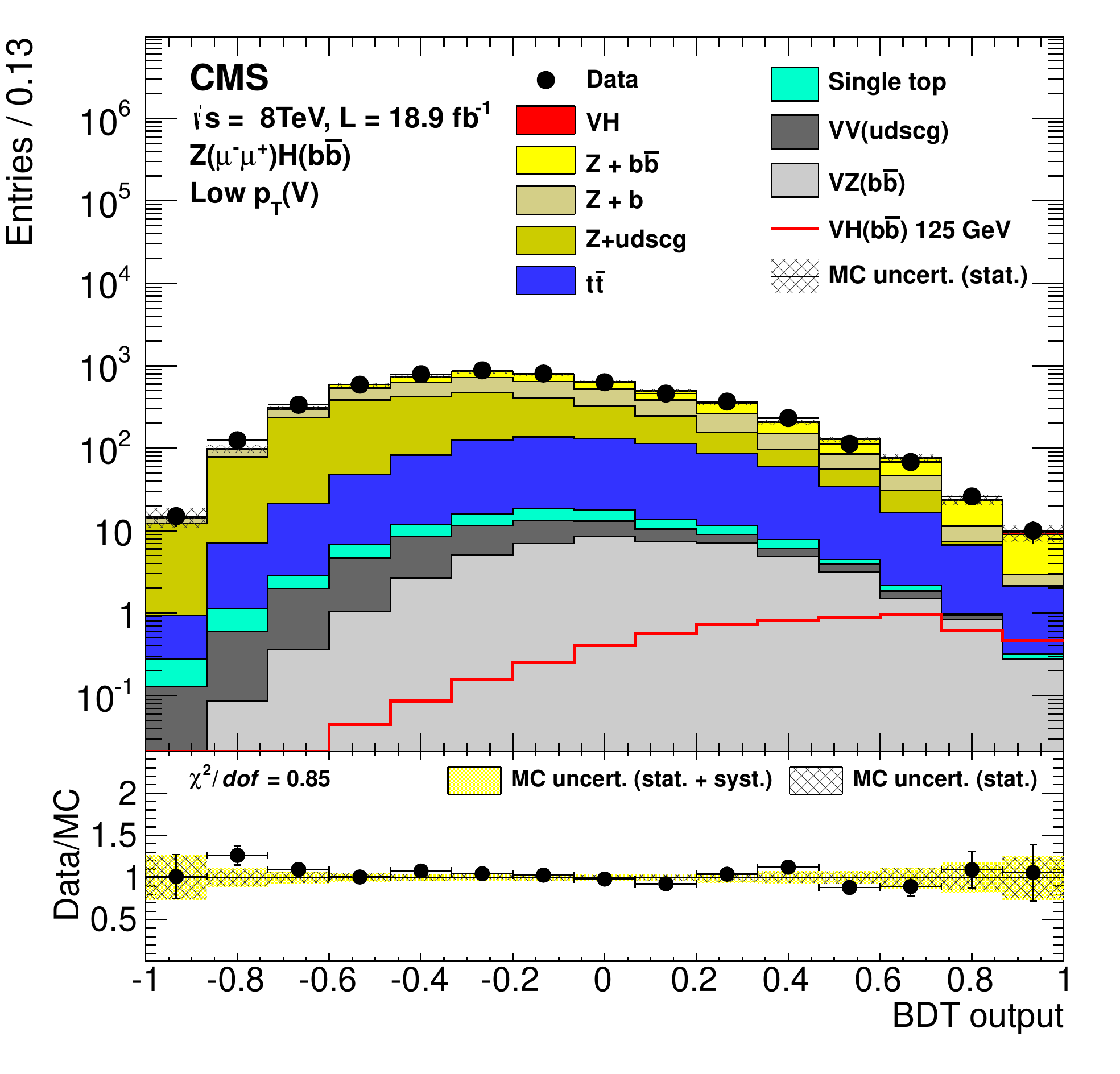}
    \includegraphics[width=0.49\textwidth]{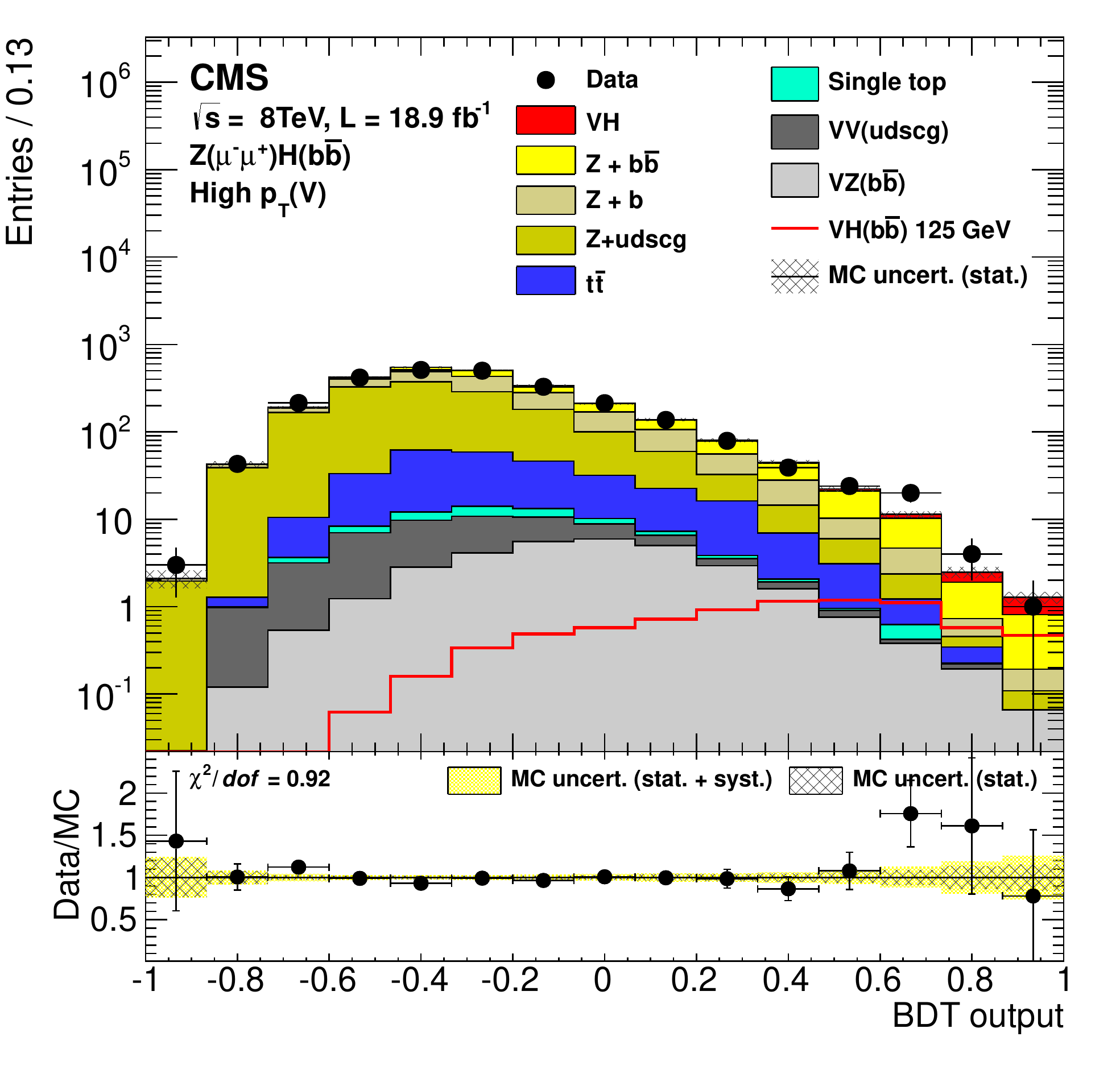}
   \includegraphics[width=0.49\textwidth]{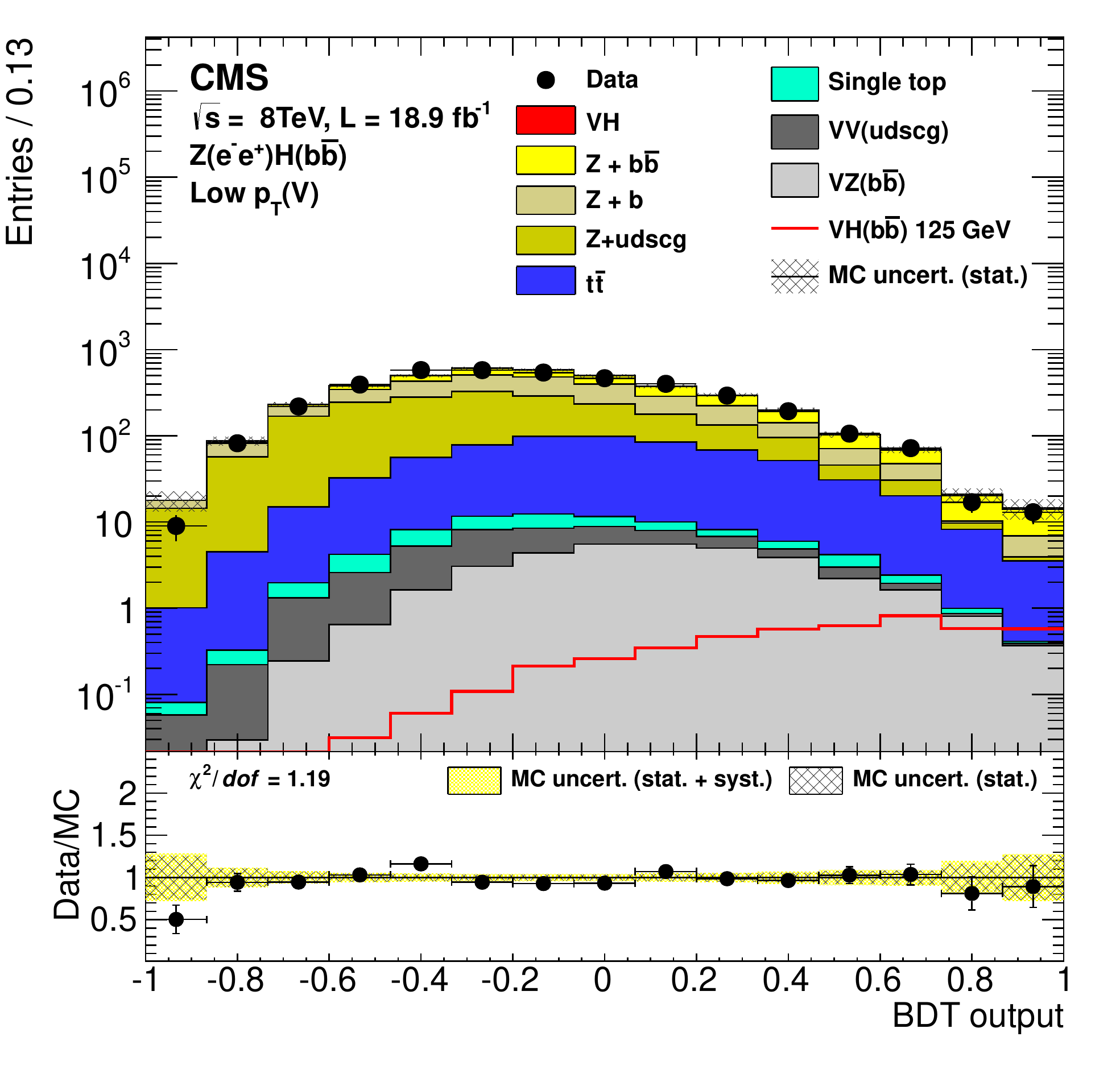}
    \includegraphics[width=0.49\textwidth]{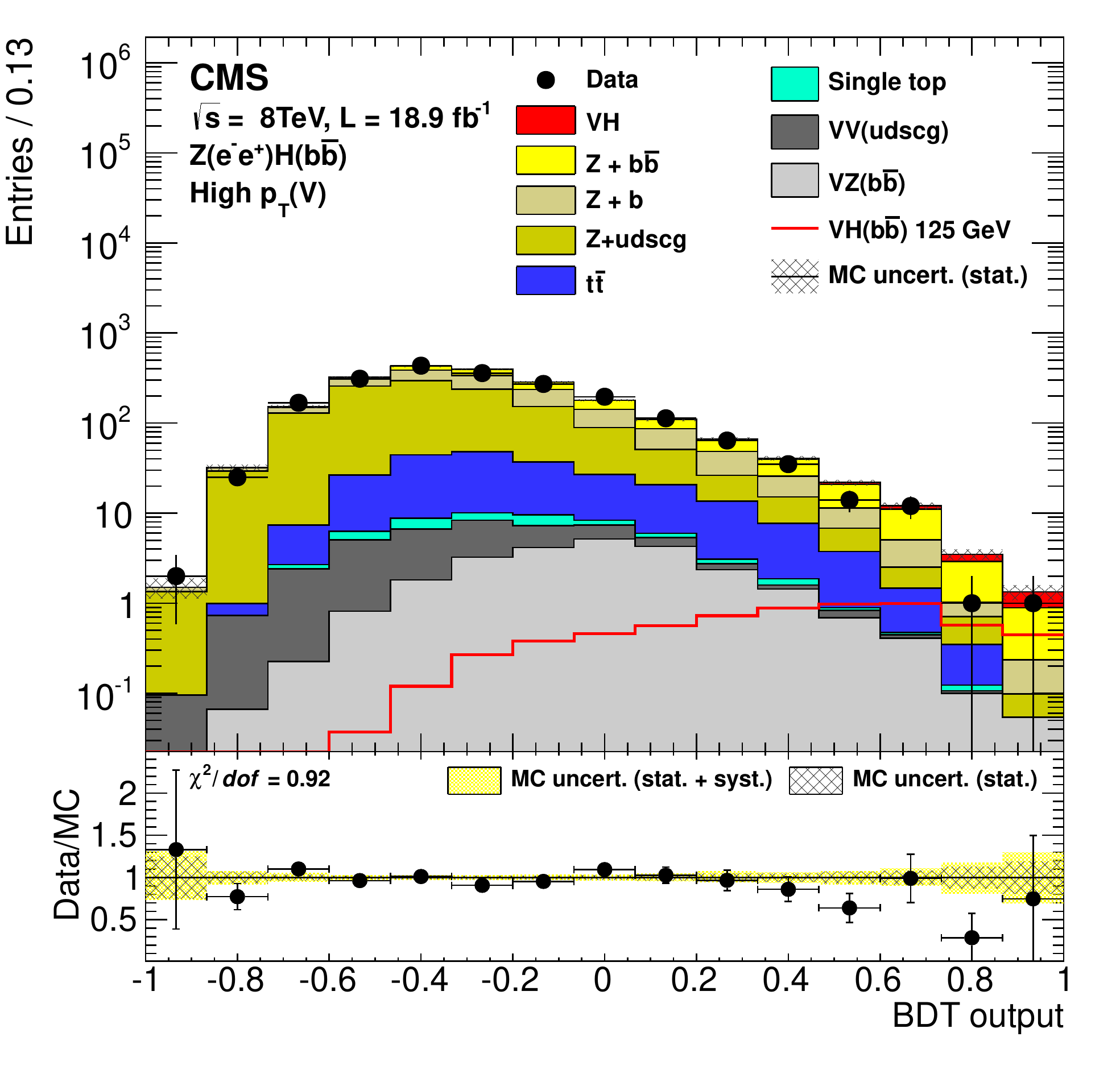}
    \caption{Post-fit BDT output distributions for \ZllH\ in the
    low-boost region (left) and high-boost region (right), for 8\TeV
    data (points with error bars),
    all backgrounds, and signal, after all selection criteria have been
    applied. Top: \ZmmH, bottom: \ZeeH. The bottom inset in each
      figure shows the ratio of the number of events observed in data to that
      of the Monte Carlo prediction for signal and backgrounds.}
    \label{fig:BDTZmm8TeV}
\end{figure*}

\begin{figure*}[htbp]
\centering
    \includegraphics[width=0.49\textwidth]{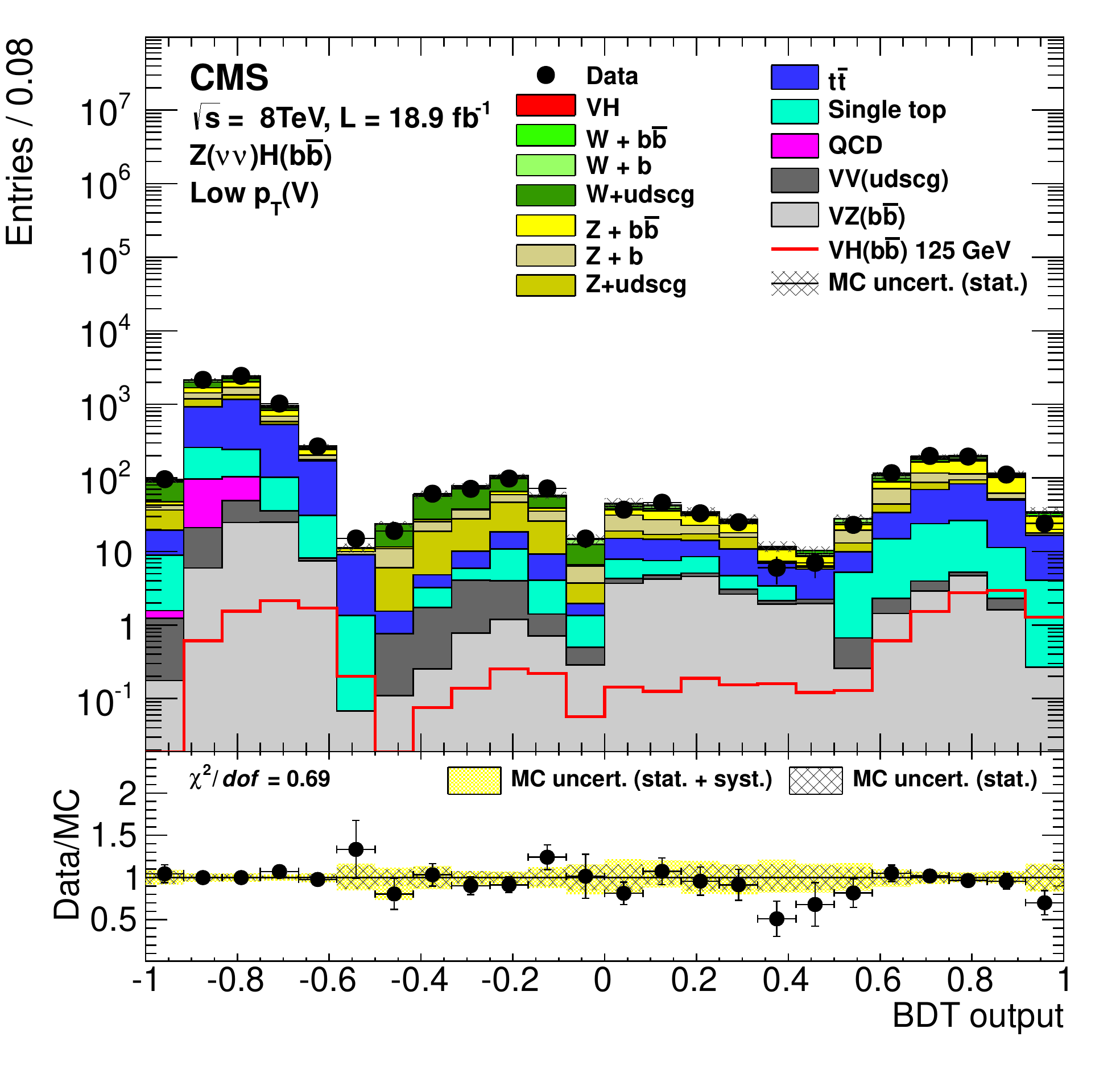}
    \includegraphics[width=0.49\textwidth]{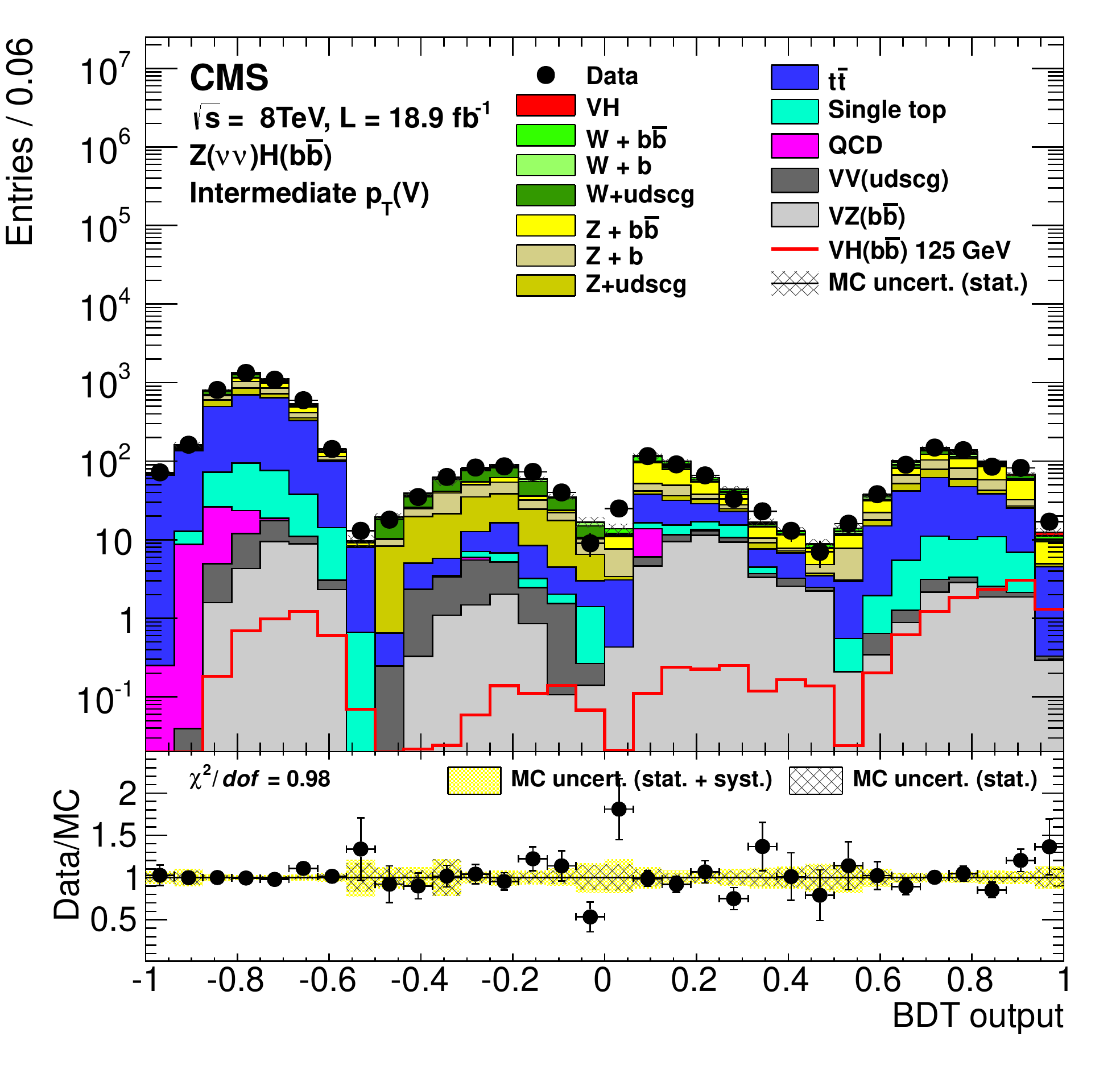}
    \includegraphics[width=0.49\textwidth]{BDT_Znn_HighPt_ZnunuHighPt_8TeV_PostFit_s}
    \includegraphics[width=0.49\textwidth]{BDT_Znn_HighPt_Last_ZnunuHighPt_8TeV_PostFit_s}
    \caption{Post-fit BDT output distributions for \ZnnH\ in the low-boost region (left),
    the intermediate-boost (right), and the high-boost  (bottom), for 8\TeV data
    (points with error bars), all
    backgrounds, and signal, after all selection criteria have been
    applied.  Bottom right: the \VH-enriched partition of the high-boost region
    is shown in more detail. The bottom inset in each
      figure shows the ratio of the number of events observed in data to that
      of the Monte Carlo prediction for signal and backgrounds.}
    \label{fig:BDTZnn8TeV}
\end{figure*}

\begin{figure*}[htbp]
\centering
    \includegraphics[width=0.4\textwidth]{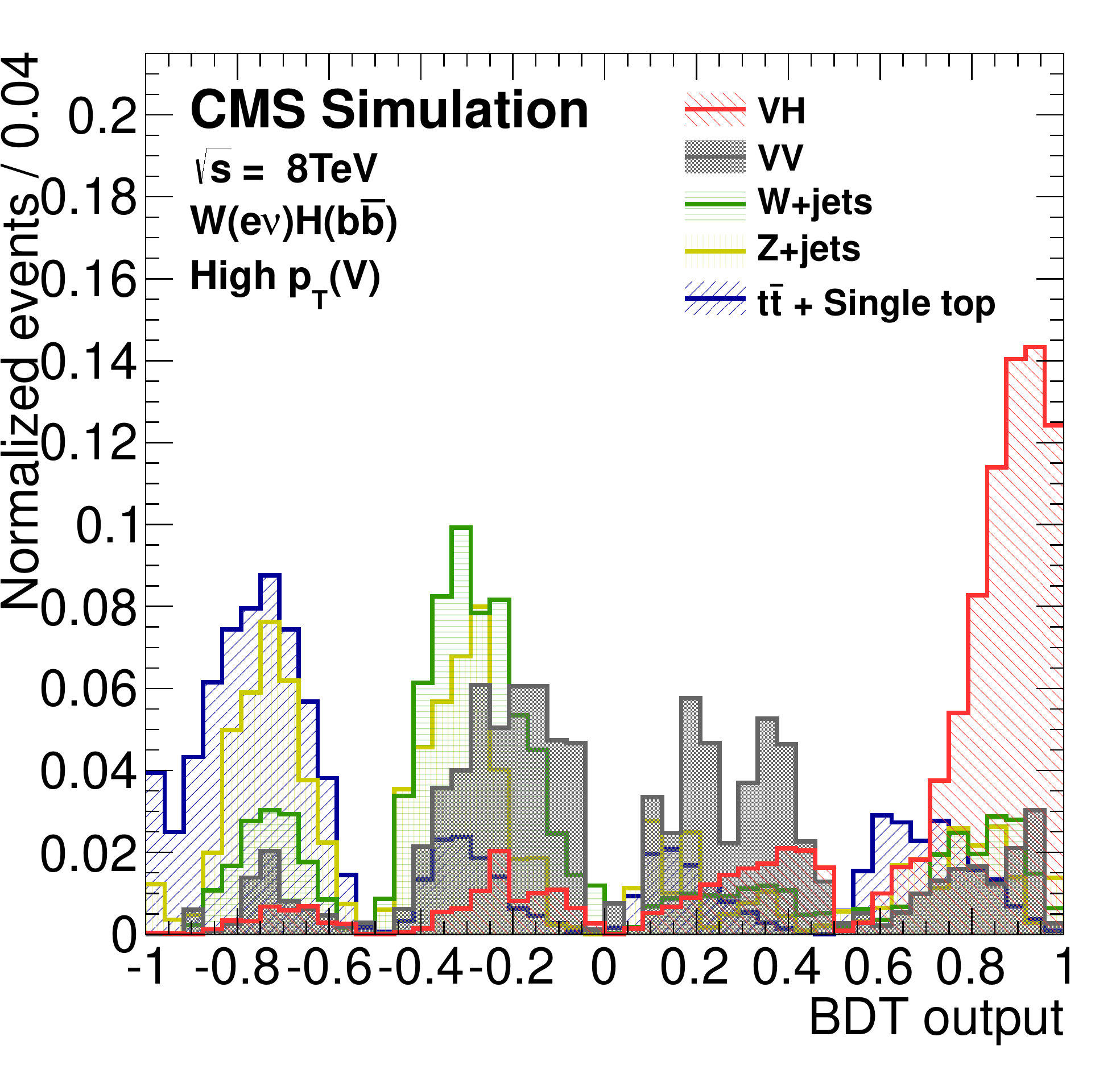}
    \includegraphics[width=0.4\textwidth]{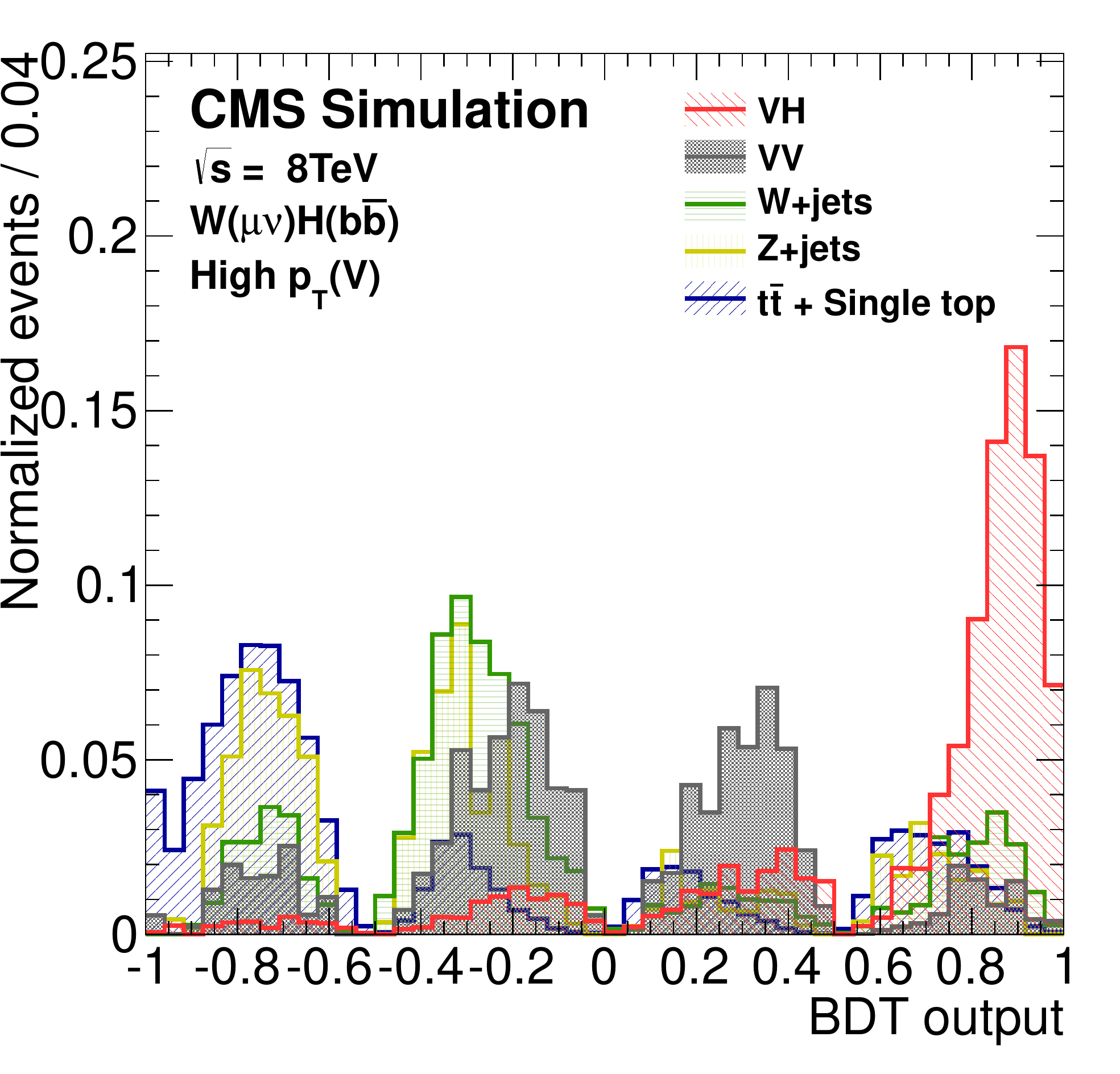}
    \includegraphics[width=0.4\textwidth]{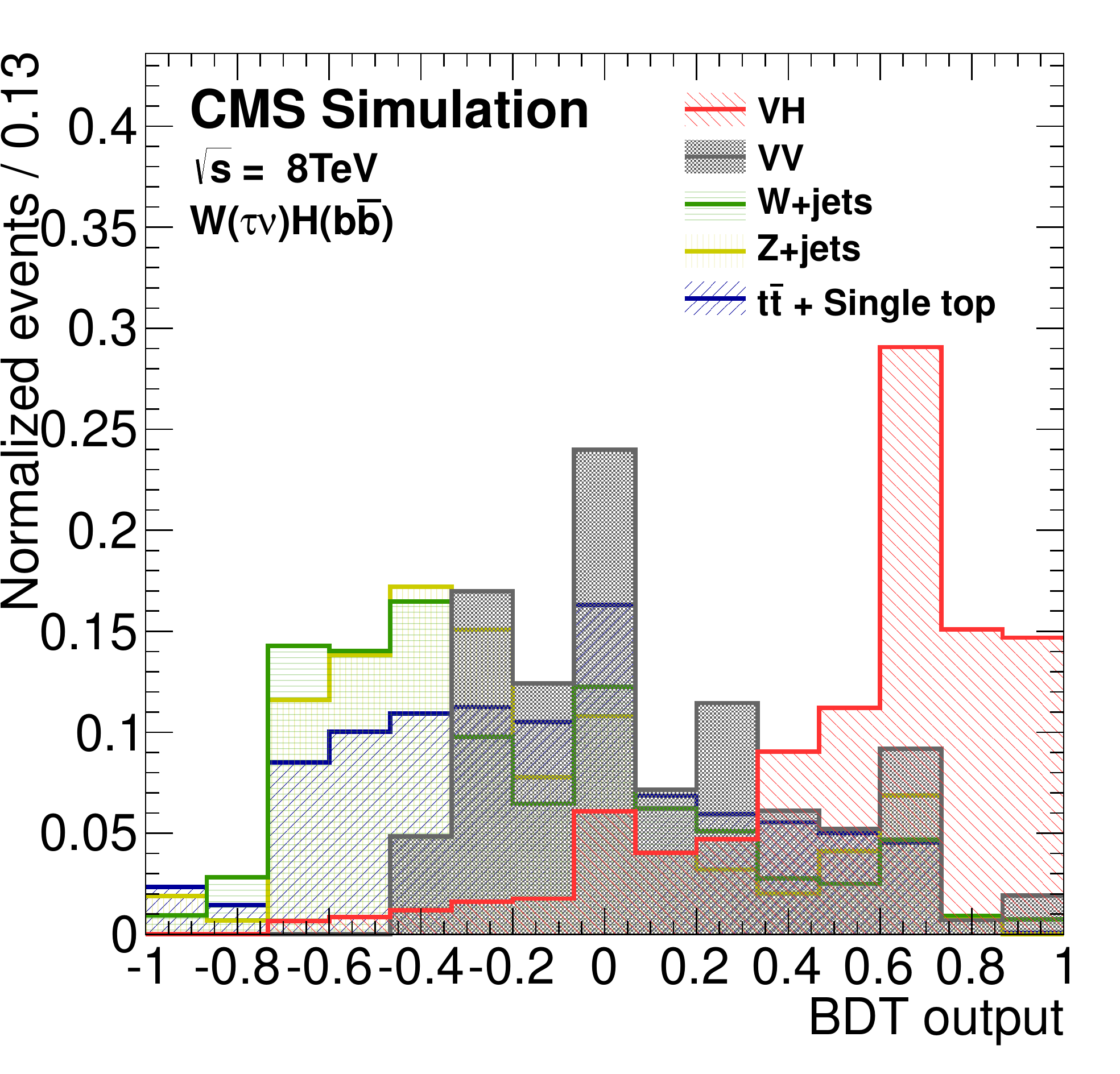}
    \includegraphics[width=0.4\textwidth]{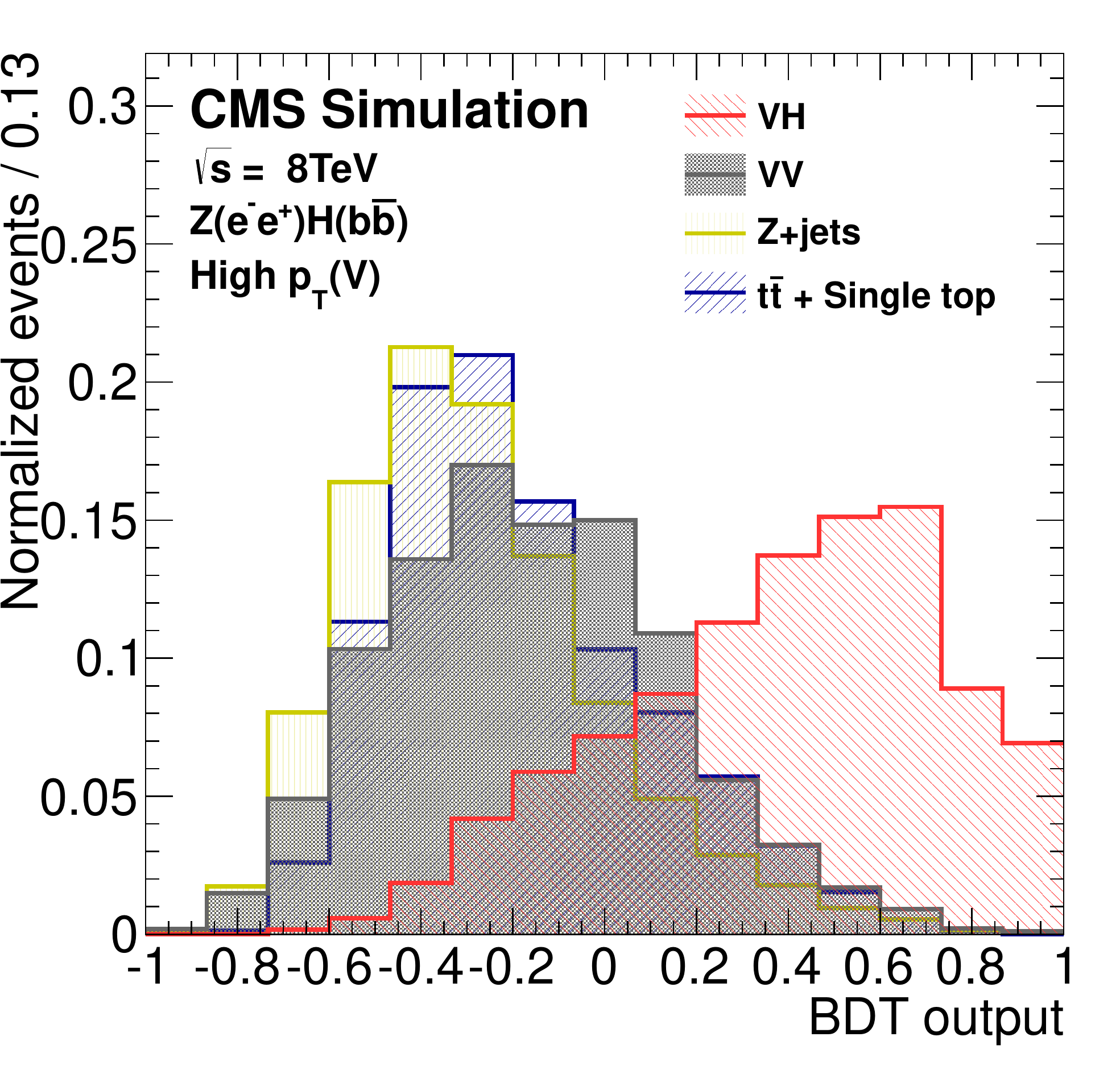}
    \includegraphics[width=0.4\textwidth]{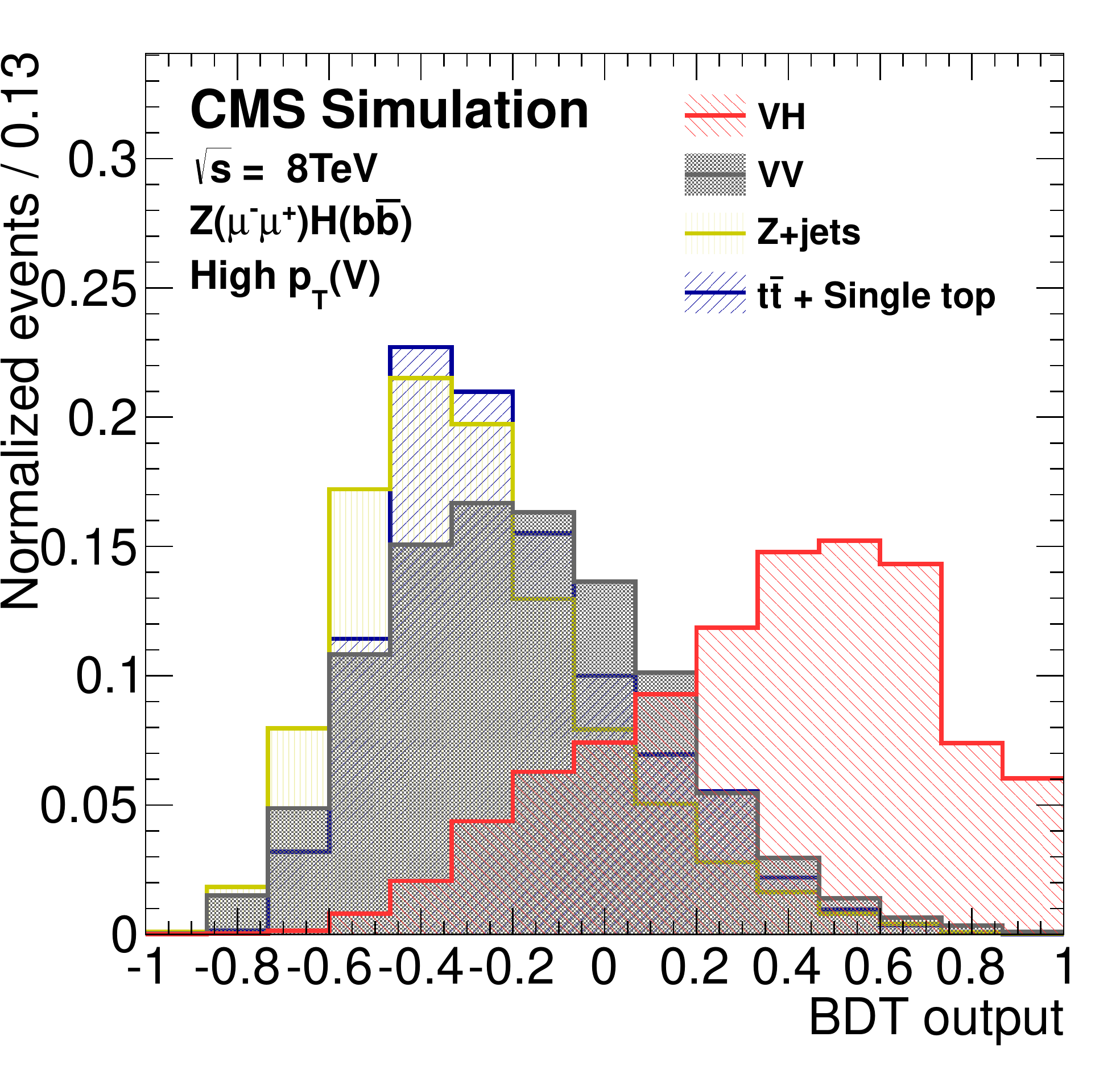}
    \includegraphics[width=0.4\textwidth]{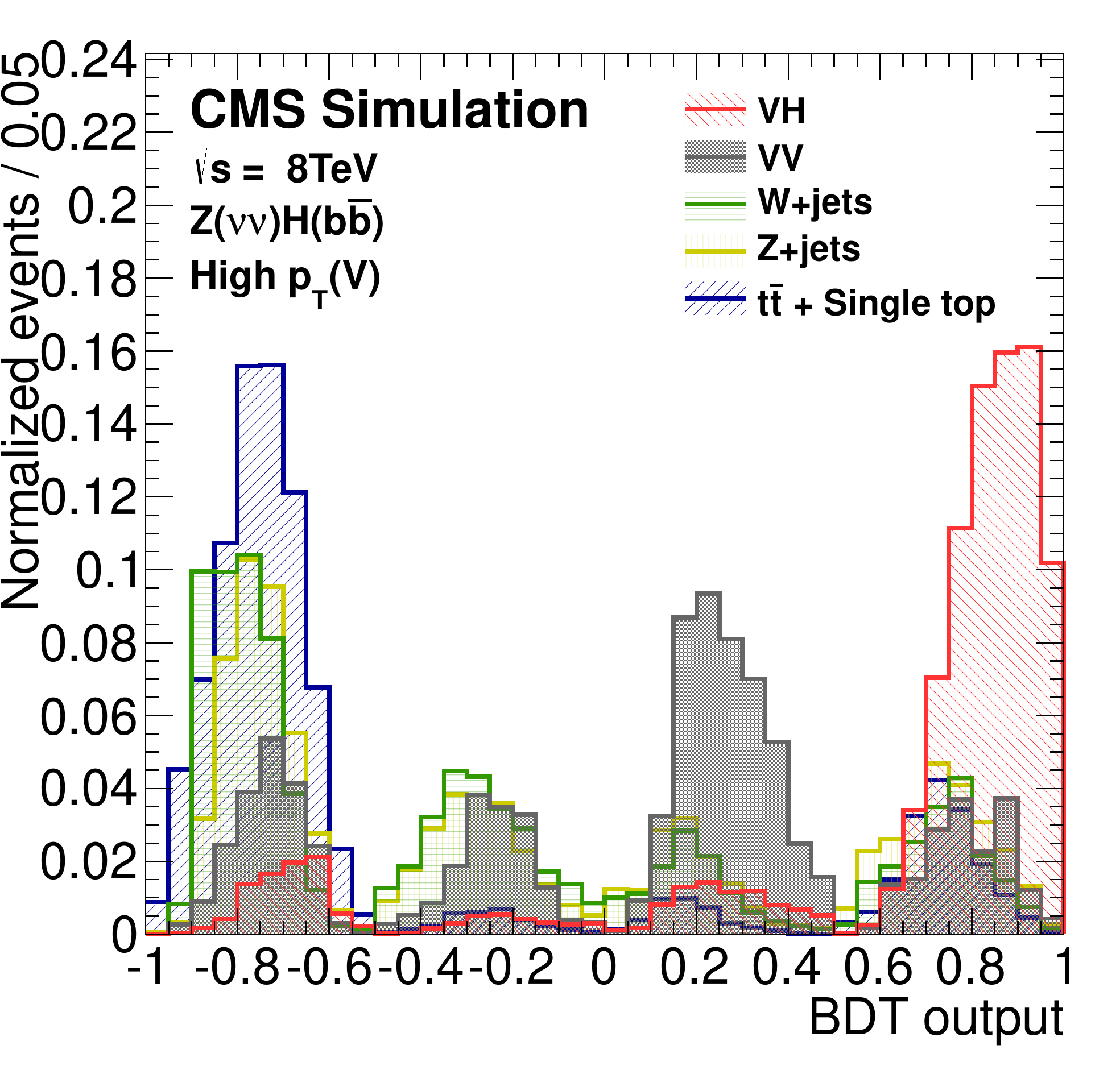}
    \caption{BDT output distributions, normalized to unity, for the highest-boost region in each
    channel, for all backgrounds and signal, after all selection
    criteria have been applied.}
    \label{fig:BDT_norm}
\end{figure*}

Figures~\ref{fig:BDTWln8TeV_mu}--\ref{fig:BDTZnn8TeV} show all the post-fit
BDT distributions, for the  \mH = 125\GeV training, for all channels
and for all boost regions. In order to better display the
different shapes of the signal and background BDT distributions, Fig.~\ref{fig:BDT_norm} shows these
distributions for the highest-boost region in each
channel, normalized to unity. See Section~\ref{sec:hbb_Results} for more details.

\cleardoublepage \appendix\section{The CMS Collaboration \label{app:collab}}\begin{sloppypar}\hyphenpenalty=5000\widowpenalty=500\clubpenalty=5000\textbf{Yerevan Physics Institute,  Yerevan,  Armenia}\\*[0pt]
S.~Chatrchyan, V.~Khachatryan, A.M.~Sirunyan, A.~Tumasyan
\vskip\cmsinstskip
\textbf{Institut f\"{u}r Hochenergiephysik der OeAW,  Wien,  Austria}\\*[0pt]
W.~Adam, T.~Bergauer, M.~Dragicevic, J.~Er\"{o}, C.~Fabjan\cmsAuthorMark{1}, M.~Friedl, R.~Fr\"{u}hwirth\cmsAuthorMark{1}, V.M.~Ghete, N.~H\"{o}rmann, J.~Hrubec, M.~Jeitler\cmsAuthorMark{1}, W.~Kiesenhofer, V.~Kn\"{u}nz, M.~Krammer\cmsAuthorMark{1}, I.~Kr\"{a}tschmer, D.~Liko, I.~Mikulec, D.~Rabady\cmsAuthorMark{2}, B.~Rahbaran, C.~Rohringer, H.~Rohringer, R.~Sch\"{o}fbeck, J.~Strauss, A.~Taurok, W.~Treberer-Treberspurg, W.~Waltenberger, C.-E.~Wulz\cmsAuthorMark{1}
% ...
\vskip\cmsinstskip
% ...
1:~~Also at Vienna University of Technology, Vienna, Austria\\
2:~~Also at CERN, European Organization for Nuclear Research, Geneva, Switzerland\\
% ...

\end{sloppypar}
\end{document}